\documentclass[10pt,letterpaper]{article}
\usepackage[T1]{fontenc}
\usepackage{lmodern,amsmath,amssymb,amsthm,mathtools,bm}
\usepackage{algorithm}
\usepackage{algpseudocode}
\usepackage[margin=0.8in]{geometry}
\usepackage{microtype,booktabs,tabularx,array,graphicx,xcolor,enumitem}
\usepackage{subcaption}
\usepackage[most]{tcolorbox}
\usepackage{tikz}\usetikzlibrary{positioning,arrows.meta}\usepackage{pgfplots}\usepgfplotslibrary{groupplots}\pgfplotsset{compat=1.18}
\usepackage{needspace,fancyhdr,placeins,float}

\newsavebox{\krylovfigurebox}
\definecolor{paperblue}{HTML}{1764AB}
\definecolor{papergreen}{HTML}{14836B}
\definecolor{vibrant_magenta}{HTML}{D1006C}
\definecolor{palegreen}{HTML}{EBF6F2}
\definecolor{paperpurple}{HTML}{7650A6}
\definecolor{palepurple}{HTML}{F3EDF8}
\usepackage[colorlinks=true,linkcolor=paperblue,citecolor=papergreen,urlcolor=paperblue]{hyperref}
\definecolor{ink}{HTML}{173449}
\definecolor{pale}{HTML}{F2F6F8}
\newtheorem{theorem}{Theorem}[section]

\newtheorem{corollary}[theorem]{Corollary}
\newtheorem{lemma}[theorem]{Lemma}
\theoremstyle{remark}
\newcommand{\E}{\mathbb E}
\newcommand{\Prb}{\mathbb P}
\newcommand{\Var}{\operatorname{Var}}
\newcommand{\Cov}{\operatorname{Cov}}
\newcommand{\MSE}{\operatorname{MSE}}
\newcommand{\tr}{\operatorname{tr}}

\newcommand{\R}{\mathbb R}
\newcommand{\C}{\mathcal C}
\newcommand{\HH}{\mathcal H}

\newcolumntype{Y}{>{\raggedright\arraybackslash}X}
\setlist[enumerate]{leftmargin=*,itemsep=3pt,topsep=3pt}
\hypersetup{pdftitle={Efficient Shot Allocation and Spectral Reconstruction for Real-Time Quantum Krylov Diagonalization},pdfsubject={Working manuscript draft},pdfauthor={}}
\definecolor{burntorange}{HTML}{C04400}
\hypersetup{citecolor=burntorange,
  urlcolor=burntorange}
\newtcolorbox{algobox}[1]{enhanced,breakable,colback=white,colframe=black, 
  boxrule=0.6pt,arc=1mm,left=3mm,right=3mm,top=2mm,bottom=2mm,
  title={#1},fonttitle=\bfseries,coltitle=white, before skip=10pt,after skip=10pt}

\title{Efficient Shot Allocation and Spectral Reconstruction\\
for Real-Time Quantum Krylov Diagonalization}

\author{
  Pegah Mohammadipour\thanks{Email:
    \href{mailto:pegahmp@psu.edu}{pegahmp@psu.edu}}
  \qquad
  Xiantao Li\thanks{Email:
    \href{mailto:xiantao.li@psu.edu}{xiantao.li@psu.edu}}
  \\[0.5em]
  Department of Mathematics, \\
  The Pennsylvania State University,
  University Park, PA 16802, USA
}
\date{}

\AtBeginDocument{\setlength{\abovedisplayskip}{7pt plus 2pt minus 2pt}\setlength{\belowdisplayskip}{7pt plus 2pt minus 2pt}}
\usepackage{titlesec}
\titlespacing*{\section}{0pt}{2.0ex plus .5ex minus .2ex}{1ex}
\titlespacing*{\subsection}{0pt}{1.6ex plus .4ex minus .2ex}{.7ex}
\begin{document}
\maketitle

\begin{abstract}
We develop analytical and statistical methods to improve the energy estimate from the real-time quantum Krylov diagonalization under a fixed measurement budget. We first express the leading energy variance as a function of the number of shots assigned to each circuit setting, accounting for both the sensitivity of the energy to the measured moments and the correlations among outcomes recorded in the same shot.We use an initial batch of measurements to choose how to allocate the remaining shots. Under fixed-model regularity assumptions, the resulting generalized least-squares energy estimator attains the minimum leading mean-square error (MSE) coefficient within the prescribed allocation class, with the initial measurements included in the total shot budget. Furthermore, we use the same data to reconstruct the Hamiltonian and overlap moments jointly by fitting them to a common nonnegative spectral measure. For asymptotically Gaussian moment errors, we characterize precisely when this reconstruction lowers the leading energy MSE in the case where two linearly independent inequalities describe the feasible moment perturbations to first order. Numerical experiments on spin and Hubbard models show that learned allocation can improve Ritz energy estimates at fixed total shot cost. The experiments also illustrate that enforcing spectral consistency can further reduce the energy estimation error.
\end{abstract}

\section{Introduction}
\label{sec:intro}
Computing low-energy eigenvalues of quantum Hamiltonians is a central task in quantum chemistry. Due to its capability in 
 comparing molecular energies, predicting excitation spectra, and
studying strongly correlated electrons and quantum magnets
\cite{cao,parrish,stair,ibm,properties}, such computation has been a leading application for quantum computing. Classically, these calculations become difficult
as the dimension of the Hilbert space grows exponentially with system
size. Quantum Krylov methods, on the other hand, approximate the low-energy spectrum by
restricting the Hamiltonian to the span of a small family of trial states.
The quantum processor prepares these states through time evolution and
collects measurement outcomes. Subsequently, a classical computer processes these
outcomes to estimate the projected matrices and solves the resulting
small eigenvalue problem. For a fixed reference state, time step, and Krylov dimension, let
$E_*$ denote the lowest energy in the resulting Krylov subspace. The physical motivation is to approximate the ground-state energy
$E_0=\lambda_{\min}(\mathcal H)$ of the Hamiltonian $\mathcal H$ by $E_*$. 

Motivated by the goal of near-term implemention in quantum Krylov method,  we ask the following question: how does one distribute a total budget of $N$ circuit shots and
use their outcomes to estimate $E_*$ with the smallest mean-square error?  
To make the problem mathematically concrete, we 
consider a normalized reference state $|\bm{\psi}_0\rangle$ and a time step
$\tau>0$, and $m$ trial states
\[
|\bm{\phi}_j\rangle=e^{-ij\tau\HH}|\bm{\psi}_0\rangle,
\qquad j=0,\ldots,m-1.
\]
These states span the Krylov subspace
$\mathcal K=\operatorname{span}\{|\bm{\phi}_0\rangle,\ldots,|\bm{\phi}_{m-1}\rangle\}$.
Instead of minimizing the energy over the entire Hilbert space, we
minimize it over linear combinations
$|\bm{\psi}(\bm c)\rangle=\sum_j c_j|\bm{\phi}_j\rangle$ in $\mathcal K$.

This restricted minimization is described by the projected Hamiltonian
$H$ and overlap matrix $S$, defined by
\[
H_{ij}=\langle\bm{\phi}_i|\HH|\bm{\phi}_j\rangle,
\qquad
S_{ij}=\langle\bm{\phi}_i|\bm{\phi}_j\rangle.
\]
For the state $|\bm{\psi}(\bm c)\rangle$, the energy numerator is
$\bm c^\dagger H\bm c$ and the squared norm is $\bm c^\dagger S\bm c$.
The lowest energy attainable in $\mathcal K$ is therefore
\[
E_*
=\min_{\bm c:\,\bm c^\dagger S\bm c>0}
\frac{\bm c^\dagger H\bm c}{\bm c^\dagger S\bm c}.
\]
After removing any exact linear dependencies among the trial states,
this minimum is the smallest generalized eigenvalue of $H\bm c=ES\bm c$. Because the search is restricted to $\mathcal K$, we have
$E_*\ge E_0$, with equality when $\mathcal K$ contains a
ground state. 

The key departure from a classical Krylov subspace projection is that the matrix elements are estimated from quantum measurements, which are subject to statistical error from the measurement. But once these matrices are assembled, 
this generalized eigenvalue problem is solved classically. 
Another difference from classical treatment is that the trial states are usually non-orthogonal: the prepared states are generally
nonorthogonal, and performing a Lanczos orthogonalization on the circuit is still non-trivial. 
An immediate consequence is that the computed $E_*$ comes with statistical error, and it can also be sensitive to the conditioning of overlap mamtrix. For example,  small positive eigenvalues of $S$ indicate nearly
linearly dependent trial states and can make the computed energy highly
sensitive to measurement errors. 

It is precisely these observations that motivated this work. To be specific, 
 our analysis concerns estimating $E_*$ from
noisy measurements, while the difference $E_*-E_0$ is a separate
subspace approximation error, the can be improved by increasing $m$ or utilizing multiple reference states $|\bm \psi_0\rangle$.

To trace the error all the way to measurement error, we notice that both $H$ and $S$ possess  Toeplitz structure: entries along each
diagonal are equal. We call these diagonal values \emph{expectation values}.
Reusing an expectation-value estimate across several matrix entries correlates their
errors. Additional correlations arise when the same circuit outcomes
estimate both an overlap and a Hamiltonian component. Thus, an accurate and thorough 
assessment of energy uncertainty must account for the matrix structure and these inherent correlations. 
These observations lead to three questions at a \emph{fixed} measurement budget:
\begin{enumerate}
    \item Which measured components most affect energy accuracy and stability?
    \item How should we distribute the  budget among the  measured components?
    \item Can we obtain a better energy estimate by restructuring the collected data?
\end{enumerate}

We first take a small batch of shots, called the pilot, at every
available circuit setting. From these outcomes, we estimate how
strongly each measured moment affects the energy and how the
outcomes from the same shot are correlated. The pilot determines
how many times the quantum device repeats each setting during the
remaining, or production, shots.

Secondly, we combine the production
measurements by generalized least squares (GLS), fit the overlap
and Hamiltonian moments to a common energy distribution, and solve
the resulting generalized eigenvalue problem. We call the fitting
step \emph{spectral reconstruction}. 

Overall, our analysis examines when shot allocation and spectral reconstruction
improve the computed energy, accounting for uncertainty in the pilot
estimates and sensitivity to errors in the overlap matrix.
This requires more than controlling the expectation-value errors, since smaller
expectation-value errors do not necessarily produce a more accurate energy estimate, and more importantly, we aim at a fixed total budget.

\subsection{Main Results}

Our results concern estimation of the ideal fixed-subspace
Ritz energy $E_*$. 

\begin{enumerate}
\item \textbf{Variance-optimal shot allocation for energy estimation.}
We use the energy derivative and joint measurement covariance to
minimize the variance of the first-order energy error, subject to
a minimum shot fraction at each setting. We can bound how much
further any allowed allocation could reduce the predicted variance.
Under fixed-model regularity assumptions, the allocation learned
from the pilot attains the smallest leading MSE coefficient in this
class for the energy estimate before reconstruction. The total shot
budget includes the pilot.

\item \textbf{Energy improvement from spectral reconstruction.}
We fit the overlap and Hamiltonian moments to a single nonnegative
energy distribution. If the true moments satisfy the constraints,
each nonzero correction brings the measured moments closer to the
true moments in the weighted distance used for the fit.
We show that when two independent inequalities describe the feasible first-order
changes, reconstruction also lowers the leading energy MSE coefficient, 
provided that the energy has a nonzero derivative along either correction direction.

\item \textbf{Algorithm and numerical comparisons.}
Our algorithm uses pilot data to allocate shots, combines production
measurements by GLS, and reconstructs the moments before solving
for $E_*$ with reporting safeguards.
A product-state example tests the predicted MSE coefficient for
learned allocation, and a four-spin example tests the predicted
reduction from reconstruction. Across seven spin and Hubbard cases,
equal shot budgets, including pilot shots, we compare the two
steps separately and together.
\end{enumerate}

\subsection{Related Work}

Classical Krylov and Lanczos methods approximate eigenvalues by
restricting the eigenvalue problem to a low-dimensional trial space
\cite{krylov,lanczos,parlett}. Quantum subspace expansion uses a
related construction to estimate excited-state energies and mitigate
errors \cite{mcclean}. Real-time quantum methods include filter
diagonalization, multireference constructions, compressed Hamiltonian
factorizations, and variational phase estimation
\cite{parrish,stair,cohn,klymko}.

Quantum phase estimation (QPE) provides another way to estimate
Hamiltonian eigenvalues and is a central algorithm for fault-tolerant
quantum computing \cite{abramslloyd1999,reiher2017}. For many
many-body applications, high-precision QPE requires long sequences
of coherent controlled evolutions, motivating the use of quantum
error correction \cite{reiher2017}. Methods with lower quantum
resource requirements have also been developed for early
fault-tolerant devices \cite{lintong2022}. Our work focuses on
measurement-based Krylov methods, in which a quantum device supplies
data for the projected matrices and a classical computer solves the
resulting eigenvalue problem. The experiments of Yoshioka et al.\
illustrate this approach on near-term hardware \cite{ibm}.

Two statistical questions arise in measurement-based Krylov methods.
The first is how errors in the measured matrices affect the estimated
energy. Epperly, Lin, and Nakatsukasa analyze thresholding, which
discards directions associated with small eigenvalues of the measured
overlap matrix \cite{epperly}. Analyses of sampling error and
real-time Krylov approximation further explain how the conditioning
of the overlap matrix affects energy accuracy \cite{sampling,kirby}.
Perrin et al.\ address conditioning when constructing the basis.
They optimize each new trial state with a penalty on its overlap
with the states already selected \cite{perrin2026variational}.

The second question is how to distribute a fixed measurement budget.
Measurement-efficient Krylov constructions \cite{zhang} and
Hamiltonian shifting, decomposition, and coefficient splitting
\cite{lee} address different aspects of this cost. In quantum
subspace expansion, Nakamura, Yano, and Yoshioka use an estimated
generalized eigenvector to update the probabilities of Pauli
measurement bases \cite{nakamura}. We consider a fixed
family of circuit settings in which a single shot can contribute to
both an overlap estimate and a Hamiltonian estimate. From pilot
shots, we estimate the covariance of these joint outcomes and the
sensitivity of the lowest Ritz energy to the measured moments.
These estimates determine how often to repeat each setting. Under
our assumptions, the learned allocation attains the smallest leading
mean-square error coefficient for the lowest Ritz energy before
reconstruction, among allocations that assign a prescribed minimum
fraction of shots to every setting. The total budget includes the
pilot shots.

Several approaches use structure in measured data to improve spectral
estimates. Examples include positivity-based denoising of response
functions \cite{kemper} and observable dynamic mode decomposition
with Fourier denoising \cite{odmd,fdodmd}. Other approaches include
partitioned subspace expansion \cite{partitioned}, super-Krylov
methods \cite{superkrylov}, and sample-based Krylov methods, which
diagonalize the Hamiltonian in a sampled configuration space
\cite{samplekrylov}. Related developments include rotated thresholding,
filter-based diagnostics, finite-temperature Lanczos regularization,
mirror subspace diagonalization, and Krylov learning
\cite{rotated,filters,thermal,mirror,qkl}.

Joint spectral fitting is particularly relevant to our reconstruction
step. Krilov, Sim, and Berne improved their estimate of the dominant
transition of a quartic oscillator by fitting a single spectrum to
two correlation functions \cite{krilov2001}. Shen et al.\ used
frequencies shared by several observables to improve estimates of
low-lying energies in lithium hydride \cite{shen2026modmd}. In
real-time Krylov diagonalization, the overlap and Hamiltonian moments
share both the energies and a known relation between their weights.
If the reference state has probability $p_j$ at energy $E_j$, that
energy contributes with weight $p_j$ to the overlap moments and
$E_jp_j$ to the Hamiltonian moments. We combine the production
measurements by generalized least squares, then use convex
optimization \cite{boyd} to fit both moment sequences to a single
energy distribution. When two independent inequalities describe the
feasible moment perturbations to first order, our analysis gives a
condition under which this fit strictly reduces the leading
mean-square error of the lowest Ritz energy estimate.
\subsection{Notation}
Throughout, $\HH$ denotes the physical Hamiltonian. We reserve $H$ and $S$
for its exact projected Hamiltonian and overlap matrices, and
$\widehat H,\widehat S$ for their production estimates. The number of
trial states is $m$, their time step is $\tau$, and $N$ is the total shot
budget. A shot is one state preparation, circuit execution and readout.
We write vectors in bold.
The coefficient vector is written $\bm c=(c_0,\ldots,c_{m-1})^T$;
scalar components such as $c_j$ are not bold. Other vectors and matrices
are identified by their definitions and dimensions.

A dagger denotes conjugate transpose, an overline on a complex scalar
denotes conjugation, and $\Re,\Im$ denote real and imaginary parts, which emerge from Hardmard type of gate operations toward measurements. 
Overbars on recorded outcomes denote sample means instead. We use
$i=\sqrt{-1}$ in complex exponentials and units with $\hbar=1$.
For Hermitian matrices, $S\succeq0$ and $S\succ0$ mean positive
semidefinite and positive definite. The norms $\|\cdot\|_{\rm op}$ and
$\|\cdot\|_F$ are the operator and Frobenius matrix norms;
$\|\cdot\|$ is the Euclidean norm of a vector.

Of particular importance is the vector $\bm{z}_*$ containing the ideal real expectation-value coordinates, and
$\widehat{\bm{z}}$ is their estimate from the production measurements.
The assembly maps introduced below satisfy $H=H(\bm{z}_*)$, $S=S(\bm{z}_*)$,
$\widehat H=H(\widehat{\bm{z}})$, and $\widehat S=S(\widehat{\bm{z}})$.
The ideal fixed-subspace energy is $E_*=E(\bm{z}_*)$.
Expectations refer to repeated measurements; $\E$, $\Var$, $\Cov$ and
$\MSE$ denote expectation, variance, covariance and mean-square error, respectively.
We use $\tr$ for the trace and $I$ for an identity matrix of the required
dimension. The notation $O(a_n)$ means bounded in magnitude by a fixed
multiple of $a_n$; $o(a_n)$ means that the ratio tends to zero.
A subscript $p$ denotes the corresponding statement in probability.
The relation $a_n\asymp b_n$ means that their positive ratio stays bounded
above and away from zero.

\section{Problem Statement: From Measurements to the Eigenvalue Problem}
\label{sec:problem}

We fix the Hamiltonian $\HH$, reference state $|\bm{\psi}_0\rangle$, time step
$\tau$, and $m$ trial states introduced above. Our target is the lowest
energy in their span, the Ritz energy $E_*\ge E_0$, where $E_0$ is the
physical ground-state energy. Enlarging the trial space can reduce
$E_*-E_0$, but nearly dependent trial states can make the estimate more
sensitive to measurement errors. We therefore hold the trial space fixed
and study how the measurement budget and postprocessing affect the
estimate of $E_*$.

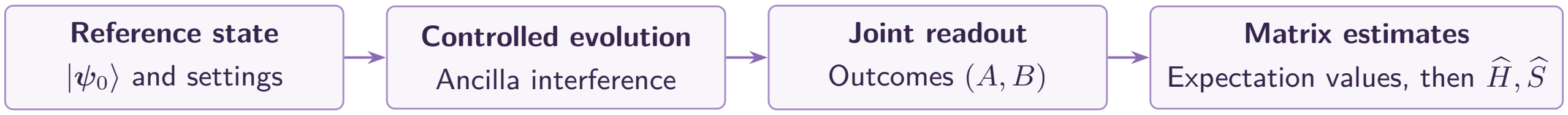
\begin{figure}[!htbp]\centering
\begin{tikzpicture}[
 node distance=4.5mm,
 card/.style={draw=paperpurple!65,fill=palepurple!55,rounded corners=2.5pt,
   line width=.55pt,minimum width=3.65cm,minimum height=1.1cm,
   inner sep=5pt,align=center,text=paperpurple!45!black,font=\sffamily\small},
 link/.style={-{Stealth[length=2mm,width=1.4mm]},draw=paperpurple!85,line width=.75pt},
 stage/.style={circle,fill=paperpurple,text=white,font=\sffamily\bfseries\scriptsize,
   minimum size=4.5mm,inner sep=0pt}]
\node[card] (i) {\textbf{Reference state}\\[2pt]$|\bm{\psi}_0\rangle$ and settings};
\node[card,right=of i] (c) {\textbf{Controlled evolution}\\[2pt]Ancilla interference};
\node[card,right=of c] (o) {\textbf{Joint readout}\\[2pt]Outcomes $(A,B)$};
\node[card,right=of o] (m) {\textbf{Matrix estimates}\\[2pt]Expectation values, then $\widehat H,\widehat S$};

\draw[link] (i.east)--(c.west);
\draw[link] (c.east)--(o.west);
\draw[link] (o.east)--(m.west);
\end{tikzpicture}
\caption{Standard procedure for estimating the projected Hamiltonian
and overlap matrices. Each shot gives
an ancilla outcome $A$ and a system outcome $B$. Averages of $A$ and
$AB$ are used to estimate entries of the overlap and Hamiltonian
matrices, respectively.}
\label{fig:workflow}\end{figure}

The measurement circuit uses an auxiliary qubit, called the ancilla, to
prepare a superposition of the reference state and its time-evolved state.
Interference between these two branches allows us to estimate real and
imaginary parts of the required expectation values. Figure~\ref{fig:workflow}
shows how the measurement outcomes are used to estimate $H$ and $S$;
Appendix~\ref{app:measurement-details} derives the measurement identities.

We continue the procedure in Figure \ref{fig:workflow} in two steps.
First, we use an initial batch of measurements, called the pilot,
to allocate the remaining measurement shots across settings. Second,
we combine the new measurements by generalized least squares and fit
a single spectral distribution to the estimated overlap and Hamiltonian
moments to obtain the reconstructed energy estimate.

\subsection{Measurement records and the statistical model}
We study real-time quantum Krylov diagonalization \cite{kirby}, using
ancilla-interference measurements as illustrated in \cite{ibm}.
 To start, we write the Hamiltonian as 
\[
\HH=\sum_{k=1}^{J}\alpha_kP_k, 
\]
after
subtracting any known identity term, which is added back to the reported
energies. Here $P_k$ is a tensor product of single-qubit Pauli matrices
or identities, $\alpha_k$ is its real coefficient, and $k=1,\ldots,J$
labels the terms.

A measurement setting specifies the circuit and readout used for a group
of shots. For trial states with indices $0\le i<j\le m-1$, their index
difference $\ell=j-i$ is the \emph{lag}, and their time separation is
$\ell\tau$. Under exact evolution, the overlap and Hamiltonian matrix
elements depend only on this lag. We estimate them using a circuit with
$|\bm{\psi}_0\rangle$ in one ancilla branch and
$U_\ell|\bm{\psi}_0\rangle$ in the other, where
$U_\ell=e^{-i\ell\tau\HH}$ and $\ell=1,\ldots,m-1$
\cite{kirby,ibm}.

We label each setting by $q=(\ell,k,Q)$: $\ell$ specifies the evolution
time, $k$ specifies the system observable $P_k$, and $Q\in\{X,Y\}$
specifies the ancilla observable. Changing any of these choices gives a
different setting. Each Pauli measurement returns an eigenvalue $+1$ or
$-1$. Appendix~\ref{app:lag-example} gives a two-qubit example showing
how the expectation values at each lag determine the projected
Hamiltonian and overlap matrices.

Each shot at a positive-lag setting gives the measurement
outcomes $(A,B)\in\{-1,1\}^2$. The ancilla outcome is $A$, measured in
basis $Q$. Meanwhile, the system outcome is $B$, the eigenvalue of $P_k$, obtained
by multiplying the system-qubit outcomes on its nonidentity factors.
We also compute $AB$ from the same shot. With exact evolution, the joint
state before readout is
\[
 |\bm{\Psi}_\ell\rangle=\frac{|0\rangle|\bm{\psi}_0\rangle+
 |1\rangle e^{-i\ell\tau\HH}|\bm{\psi}_0\rangle}{\sqrt2}.
\]
The expectation values needed for the overlap and Hamiltonian matrices are therefore given by, 
\begin{equation}
 s_\ell=\langle\bm{\psi}_0|U_\ell|\bm{\psi}_0\rangle,
 \qquad
 h_\ell=\langle\bm{\psi}_0|\HH U_\ell|\bm{\psi}_0\rangle
       =\sum_k\alpha_k t_{\ell k},
 \label{eq:moments}
\end{equation}
where $t_{\ell k}=\langle\bm{\psi}_0|P_kU_\ell|\bm{\psi}_0\rangle$ is the
contribution associated with $P_k$. The means of the measured outcomes
give their real and imaginary parts:
\begin{equation}
\begin{array}{c|cc}
Q&\E A&\E(AB)\\\hline
X&\Re s_\ell&\Re t_{\ell k}\\
Y&\Im s_\ell&\Im t_{\ell k}
\end{array}\label{eq:outcomes}
\end{equation}
For fixed $\ell$ and $Q$, every Pauli setting measures the same overlap
component through $A$. \emph{Direct pooling} combines these ancilla
outcomes, weighting each setting's mean by its shot count. It estimates
the corresponding Hamiltonian component by summing the product means
with coefficients $\alpha_k$. Appendix~\ref{app:pooling} gives the
estimators and shot counts. Our main estimator uses generalized least
squares (GLS) to account for all shared-mean identities and correlations,
as described in Section~\ref{sec:gls-design}. We use direct pooling as a
comparison. At lag zero, $s_0=1$ is known, and ordinary measurements of
$P_k$ on $|\bm{\psi}_0\rangle$ estimate $h_0$.

\medskip 

Finally, we collect the ideal real components in the vector
\begin{equation} \label{eq:real_compo}
 \bm{z}_*=(h_0,\Re s_1,\Re h_1,\Im s_1,\Im h_1,\ldots)^T\in\R^d,
 \qquad d=4m-3.
\end{equation}
Each positive lag contributes four components, and $h_0$ contributes
one. The known value $s_0=1$ is omitted. Here ``ideal'' refers to exact
Hamiltonian evolution and exact expectation values, with no finite-shot
error or hardware noise.

An initial batch of $N_p$ shots, called the \emph{pilot}, estimates the
measurement covariances and the sensitivity of the energy to errors in
the expectation values. These estimates determine how we distribute
fresh production shots across settings. We write their counts as
$\boldsymbol n=(n_q)_q$, where $n_q$ is the count for setting $q$.
The total budget is
\begin{equation}N=N_p+n,\qquad n=\sum_q n_q.\label{eq:budget}\end{equation}
The index $q$ includes ordinary measurements for $h_0$ as well as the
positive-lag joint measurements. The pilot and production batches use
independent circuit runs. For the production estimate $\widehat{\bm{z}}$, we write, for the clarity of the following discussions,  assuming unbiased estimations,
\begin{equation}
\begin{aligned}
 \widehat{\bm{z}}&=\bm{z}_*+\bm{\varepsilon},\\
 \bm{\varepsilon}&=\widehat{\bm{z}}-\E[\widehat{\bm{z}}\mid\text{pilot}],\\
 \Sigma(\boldsymbol n)&=\Cov(\widehat{\bm{z}}\mid\text{pilot}).
\end{aligned}
\label{eq:noise}
\end{equation}
The sampling error $\bm{\varepsilon}$ has zero mean conditional on
the pilot. Once the pilot fixes the production counts and
estimation weights, $\Sigma(\boldsymbol n)$ is the covariance
of the production estimate $\widehat{\bm{z}}$.

\subsection{Toeplitz structure and shared matrix entries}
The expectation values in \eqref{eq:moments} determine the exact projected overlap and Hamiltonian matrices :
\begin{equation}
 S_{ij}=\langle\bm{\phi}_i|\bm{\phi}_j\rangle=s_{j-i},\qquad
 H_{ij}=\langle\bm{\phi}_i|\HH|\bm{\phi}_j\rangle=h_{j-i}.
 \label{eq:matrices}
\end{equation}
Both matrices enjoy the Hermitian Toeplitz structure: entries on the same diagonal have
the same value because they share the lag $j-i$, which can reduce the needed quantum measurements.  To write this dependence
explicitly, let $T_\ell\in\R^{m\times m}$ have ones on its $\ell$th
upper diagonal and zeros elsewhere. Then

\begin{equation}\label{eq:toeplitz-assembly}
\begin{aligned}
 S&=I+\sum_{\ell=1}^{m-1}
       (s_\ell T_\ell+\overline{s_\ell}T_\ell^\dagger),\\
 H&=h_0I+\sum_{\ell=1}^{m-1}
       (h_\ell T_\ell+\overline{h_\ell}T_\ell^\dagger).
\end{aligned}
\end{equation}
For a real expectation-value vector $\bm{z}$, we write $H(\bm{z})$ and $S(\bm{z})$
for the matrices assembled by these formulas. Thus $H=H(\bm{z}_*)$ and
$S=S(\bm{z}_*)$, while the production estimates are $\widehat H=H(\widehat{\bm{z}})$ and
$\widehat S=S(\widehat{\bm{z}})$. Appendix~\ref{app:lag-example} illustrates this
construction, and Appendix~\ref{app:matrices} derives it from exact
evolution. The Toeplitz property applies to the full projected
Hamiltonian; individual projected Pauli terms need not have this
property because $P_k$ need not commute with $\HH$.

An error in one expectation value therefore appears in every entry on
the corresponding diagonal. The overlap and Hamiltonian estimates are
also correlated because they use the same measurement records. The
allocation in Section~\ref{sec:allocation} accounts for both effects.
Section~\ref{sec:spectral} imposes a further constraint: the overlap
and Hamiltonian expectation values must come from the same spectral
distribution.

\subsection{The energy target and its mean-square error}
We assume the trial states are linearly independent, which is equivalent
to $S\succ0$. 
For an expectation-value vector $\bm{z}$ with $S(\bm{z})\succ0$, the lowest Ritz
energy is
\begin{equation}
 E(\bm{z})=\min_{\bm c\ne\bm{0}}\frac{\bm c^\dagger H(\bm{z})\bm c}{\bm c^\dagger S(\bm{z})\bm c}
     =\lambda_{\min}(H(\bm{z}),S(\bm{z})),\qquad E_*=E(\bm{z}_*).
 \label{eq:target}
\end{equation}
Here $\lambda_{\min}(H(\bm{z}),S(\bm{z}))$ denotes the smallest generalized
eigenvalue of the matrix pair. At the ideal expectation values, the
minimizing coefficients satisfy $H\bm c=E_*S\bm c$.
Appendix~\ref{app:matrices} derives this eigenvalue problem and shows
how to reduce it to an ordinary Hermitian problem.

The production energy estimate is
\[
 \widehat E=E_{\rm rep}(\widehat{\bm{z}}).
\]
Its mean-square error is
\[
 \MSE(\widehat E;E_*)=\E[(\widehat E-E_*)^2],
\]
including all clipped outputs and fallback values.
The results below assume that the reporting rule agrees
with $E(\bm{z})$ in a neighborhood of $\bm{z}_*$.

\subsection{ Spectral intervals} 

Spectral reconstruction requires a known interval $[L,U]$ containing
the spectrum of $\HH$. This interval is needed to choose the reconstruction
grid and to bound the reported energy estimates. Since every Pauli
string has operator norm one,
\[
 \|\HH\|_{\rm op}\le W_H:=\sum_k|\alpha_k|,
 \qquad [L,U]=[-W_H,W_H]
\]
gives a valid interval for the shifted Hamiltonian, using only its known
coefficients. Adding back a removed identity term shifts both endpoints
by the same constant. Tighter certified bounds may also be used. In the
numerical models, we divide the Hamiltonian by $W_H$, so the interval
becomes $[-1,1]$ in the normalized energy units.

We also  use a fixed reporting rule to include failed calculations
in the error assessment. Given moments $\bm{z}$, first check
$\lambda_{\min}(S(\bm{z}))\ge s_{\rm cut}$, where
$s_{\rm cut}>0$ is a numerical cutoff.
If this check and the eigenvalue calculation succeed,
$E_{\rm rep}(\bm{z})$ is the lowest Ritz value clipped to $[L,U]$.
Otherwise, it is a fixed fallback value
$E_{\rm default}\in[L,U]$.

\section{Shot Allocation and Energy Sensitivity}
\label{sec:allocation}

In this section, we first determine how the shot counts at each measurement setting
affect the covariance of the estimated moments. We then derive the
sensitivity of the lowest Ritz energy to each moment. Together,
the moment covariance and the energy sensitivity give the variance
of the first-order energy error. We use the pilot data to estimate
this variance and allocate the remaining measurement shots to
minimize it.

\subsection{Moment estimation and covariance}
\label{sec:gls-design}

To predict how much a setting gains from additional shots, we first
estimate the covariance of a single shot at that setting. At a joint
measurement setting $q$, the ancilla outcome $A$, the system outcome
$B$, and their product $AB$ come from the same shot. We write the
record and its covariance as
\begin{equation}
 \bm{X}_q=(A,B,AB)^T,\qquad \Gamma_q=\Cov(\bm{X}_q).
 \label{eq:gls-record-summary}
\end{equation}
The diagonal entries of $\Gamma_q$ are the variances of the three
recorded quantities, and the off-diagonal entries are their pairwise
covariances. For a measurement used to estimate $h_0$, the record
contains only $B$, so $X_q=B$ and $\Gamma_q=\Var(B)$.

We distribute the pilot shots as evenly as possible across the
$Q_{\rm tot}$ settings, with at least two shots per setting, and
compute the sample covariance at each setting. We clip its
eigenvalues to the interval $[\kappa,b]$, where $0<\kappa<b$, to
obtain $\widehat\Gamma_q$. Thus
\[
 \kappa I\preceq\widehat\Gamma_q\preceq bI.
\]
The positive lower bound ensures that the covariance estimate can
be inverted when forming the GLS weights.
Appendix~\ref{app:learned-policy} describes the choice of $\kappa$
and $b$.

\paragraph{Constructing the moment estimator.}
Suppose we assign $n_q$ fresh production shots to setting $q$.
Conditional on the pilot, the shots are independent, so their
sample mean has covariance
\[
 \overline{\bm{X}}_q=\frac{1}{n_q}\sum_{r=1}^{n_q}\bm{X}_{q,r},
 \qquad
 \Cov(\overline{\bm{X}}_q\mid\text{pilot})=\frac{\Gamma_q}{n_q}.
\]
We therefore use $\widehat\Gamma_q/n_q$ as the estimated covariance
of this mean. GLS uses these estimates to combine measurements of
the same quantity across settings while accounting for correlations
within each record.

The ancilla means give the overlap moments. For each positive
lag, we sum the fitted product means with the Hamiltonian coefficients
$\alpha_k$ to obtain the corresponding Hamiltonian moment. At lag
zero, we estimate $h_0$ by measuring each $P_k$ on the reference
state and summing its sample mean with the same coefficients.

To express these steps explicitly, for production counts $\boldsymbol n=(n_q)_q$, the GLS estimate of
the real moment vector $\bm{z}_*$ is written as
\begin{equation}
 \widehat{\bm{z}}=\sum_q R_q(\boldsymbol n)\overline{\bm{X}}_q.
 \label{eq:gls-production}
\end{equation}
The matrix $R_q(\boldsymbol n)$ maps the contribution of the sample
mean at setting $q$ to the fitted moments. Its coefficients depend
on the shot counts, the pilot covariance estimates, and which means
are shared across settings. Appendix~\ref{app:allocation-gls-details}
gives their construction. Applying the same fit to the pilot means
and counts gives the pilot moment estimate $\widehat{\bm{z}}_p$, which we
use below to estimate the energy sensitivity.

The pilot fixes the production counts and GLS weights before we
collect the fresh measurements. Under the assumption that the
measurement means equal their ideal values, the production estimate
satisfies $\E[\widehat{\bm{z}}\mid\text{pilot}]=\bm{z}_*$. Because the production
means at different settings are conditionally independent, their
contributions to the covariance in \eqref{eq:gls-production} add:
\begin{equation}
 \Sigma(\boldsymbol n)
 :=\Cov(\widehat{\bm{z}}\mid\text{pilot})
 =\sum_q\frac{R_q(\boldsymbol n)\Gamma_q
                  R_q(\boldsymbol n)^T}{n_q}.
 \label{eq:sandwich}
\end{equation}
Replacing each unknown single-shot covariance $\Gamma_q$ by its
pilot estimate gives
\begin{equation}
 \widehat\Sigma(\boldsymbol n)
 =\sum_q\frac{R_q(\boldsymbol n)\widehat\Gamma_q
                  R_q(\boldsymbol n)^T}{n_q}.
 \label{eq:gls-fit-covariance}
\end{equation}
We can evaluate $\widehat\Sigma(\boldsymbol n)$ for any proposed
allocation before taking the production shots. Crucially, to compare these
allocations by their effect on the energy, we next determine how
an error in each moment changes the lowest Ritz energy.

\subsection{Energy sensitivity}
\label{sec:sensitivity}

To quantify how moment errors affect the computed energy,
assume that $S\succ0$ and that the lowest Ritz value $E_*$
is simple. Let $\bm c$ be its eigenvector, normalized by
\[
 H\bm c=E_*S\bm c,\qquad \bm c^\dagger S\bm c=1.
\]
The estimated matrices are
$\widehat H=H(\widehat{\bm{z}})$ and
$\widehat S=S(\widehat{\bm{z}})$.
Write their errors as
$\delta H=\widehat H-H$ and
$\delta S=\widehat S-S$.

Under the unbiased production model, the moment error
$\bm{\varepsilon}=\widehat{\bm{z}}-\bm{z}_*$ has conditional
mean zero. For sufficiently small $\bm{\varepsilon}$,
\begin{equation}
 \begin{aligned}
 E(\widehat{\bm{z}})-E_*
 &=\bm{g}^T\bm{\varepsilon}+O(\|\bm{\varepsilon}\|^2),\\
 \bm{g}^T\bm{\varepsilon}
 &=\bm c^\dagger(\delta H-E_*\delta S)\bm c,
 \qquad \bm{g}:=\nabla E(\bm{z}_*).
 \end{aligned}
 \label{eq:gradient}
\end{equation}

To obtain this derivative, we differentiated the generalized
eigenvalue equation and multiply by $\bm c^\dagger$.
Hermiticity gives
$\bm c^\dagger H=E_*\bm c^\dagger S$, so the terms
containing derivatives of the eigenvector cancel.
Positive overlap and simplicity make $E(\bm{z})$ smooth near
$\bm{z}_*$, giving the quadratic remainder
\cite{epperly,parlett,silverman1984}.

Importantly, the Toeplitz structure lets us collect the contribution
from each lag. Define
\begin{equation}
 a_\ell(\bm c)
 =\sum_{j=0}^{m-\ell-1}\overline{c_j}c_{j+\ell},
 \qquad \ell=1,\ldots,m-1.
 \label{eq:lag-weights}
\end{equation}
Then we arrive at an estimation in terms of the moment errors,
\begin{equation}
 \bm{g}^T\bm{\varepsilon}
 =\|\bm c\|^2\delta h_0+
 2\Re\sum_{\ell=1}^{m-1}
 a_\ell(\bm c)(\delta h_\ell-E_*\delta s_\ell),
 \label{eq:lag-gradient}
\end{equation}
where $\delta h_0,\delta s_\ell,\delta h_\ell$ are the
corresponding moment errors.
Hamiltonian and overlap errors enter the same expression,
so their correlations affect the first-order energy variance:
\begin{equation}
 \Var(\bm{g}^T\bm{\varepsilon}\mid\text{pilot})
 =\bm{g}^T\Sigma(\boldsymbol n)\bm{g}.
 \label{eq:variance}
\end{equation}

\subsection{Optimizing the production shot counts.} \label{sec:learned-allocation}
With a pilot run to determine $\bm g$ and $\Sigma$, together with the perturbation analysis in \eqref{eq:lag-gradient} and \eqref{eq:gradient} it is now possible to determine an optimized shot allocation
to  reduce the predicted
energy variance as much as possible.
In terms of the desired counts before integer rounding,
we solve
\begin{equation}
 \boxed{
 \begin{aligned}
 \underset{\boldsymbol n\in\R^{Q_{\rm tot}}}
 {\operatorname{minimize}}\quad&
 \widehat V(\boldsymbol n) = 
 \widehat{\bm{g}}^T\widehat\Sigma(\boldsymbol n)\widehat{\bm{g}}\\
 \text{subject to}\quad&
 \sum_q n_q=n,\\
 &n_q\ge\frac{n\epsilon}{Q_{\rm tot}}
 \quad\text{for every setting }q.
 \end{aligned}}
 \label{eq:design}
\end{equation}

The budget constraint fixes the total number of production
shots, while the lower bounds preserves a minimum number of measurements at every
setting. The allocation has a simple form when each setting measures one
moment independently. If moment $a$ has single-shot standard
deviation $\sigma_a$ and energy sensitivity $g_a$, it contributes
$g_a^2\sigma_a^2/n_a$ to the first-order energy variance. When the
lower bounds in \eqref{eq:design} are inactive, the minimizing
counts before rounding satisfy $n_a\propto |g_a|\sigma_a$. A moment
therefore receives more shots when its measurements are more variable
or the energy is more sensitive to it. In the Krylov measurements,
one shot can contribute to both an overlap moment and a Hamiltonian
moment. Equation~\eqref{eq:design} accounts for the covariance
between those contributions.

We solve this convex problem in the fractions $w_q=n_q/n$.
Starting from uniform fractions, each iteration moves toward
an allocation that minimizes the linear approximation
of $\widehat V(\bm{w})$ over $\mathcal W_\epsilon$.
Appendix~\ref{app:learned-policy} gives the update rule and
a computable upper bound on the remaining objective error.
We stop when this bound is at most $N_p^{-1/2}$.
If numerical optimization fails to reach this tolerance,
we use uniform fractions.

For the selected fractions $\widehat{\bm{w}}$, assign
$n_q=\lfloor n\widehat w_q\rfloor$ and distribute the
remaining shots to the settings with the largest fractional
remainders, using a fixed index order for ties.
This gives
\begin{equation}
     \sum_qn_q=n,\qquad
 |n_q-n\widehat w_q|<1,\qquad
 n_q\ge\left\lfloor\frac{n\epsilon}{Q_{\rm tot}}\right\rfloor.
\end{equation}
We require the last lower bound to be at least one.
The optimization tolerance applies before rounding;
the theorem accounts for the resulting integer counts.
We recompute the GLS weights using these counts, then hold
the counts and covariance weights fixed while collecting
the production data.

\subsection{Learning the allocation from pilot measurements}

The allocation rule in the pevious section uses
the energy gradient and the covariance of the measurement outcomes
at each setting. Now we ask: can a pilot that uses a vanishing fraction of the
total shot budget attain the same leading energy accuracy as an
allocation chosen with exact knowledge of these quantities?
Section~\ref{sec:gls-design} gives the pilot covariance estimates.
It remains to estimate the energy gradient and check whether the
pilot eigenpair is reliable enough to guide the allocation.

To see what the variance in \eqref{eq:variance} implies for shot
allocation, first suppose that each moment $z_a$ is measured
independently at its own setting. Let $\sigma_a^2$ be its single-shot
variance, let $n_a$ be its shot count, and let
$g_a=\partial E/\partial z_a$ at $\bm z_*$. The variance of the
first-order energy error is then
\[
 \Var\!\left(\bm g^T(\widehat{\bm z}-\bm z_*)\right)
 =\sum_a\frac{g_a^2\sigma_a^2}{n_a}.
\]
For equal-cost shots, positive $|g_a|\sigma_a$, and a fixed total
$\sum_a n_a=n$, this expression is minimized over positive real
counts by $n_a\propto |g_a|\sigma_a$. Thus a noisy measurement
deserves more shots only to the extent that its error affects the
target energy.

In our circuits, the same shot supplies the ancilla outcome $A$
and the product $AB$. These estimate an overlap component and a
Hamiltonian component, respectively. Their errors enter the
combination $\delta H-E_*\delta S$, so their covariance also
affects the energy variance. We therefore use the full covariance
in \eqref{eq:variance} when choosing the production counts.

The pilot moment estimate gives the matrices
$H(\widehat{\bm{z}}_p)$ and $\widehat S_p=S(\widehat{\bm{z}}_p)$.
When $\widehat S_p\succ0$ and the lowest pilot Ritz value
is simple, let $(\widehat E_p,\widehat{\bm c}_p)$ be the
corresponding eigenpair, normalized by
$\widehat{\bm c}_p^\dagger\widehat S_p\widehat{\bm c}_p=1$.
Evaluating the derivative formula \eqref{eq:gradient}
at these pilot quantities gives
\[
 \widehat g_i
 =\widehat{\bm c}_p^\dagger
   \bigl(\partial_iH-\widehat E_p\partial_iS\bigr)
   \widehat{\bm c}_p,
 \qquad
 \widehat{\bm{g}}=\nabla E(\widehat{\bm{z}}_p).
\]
Here $\partial_i$ denotes differentiation with respect to
the $i$th moment coordinate.
The entries of $H(\bm{z})$ and $S(\bm{z})$ are constants or linear
combinations of the real moment coordinates.
Their derivatives are therefore obtained directly from
the coefficients in \eqref{eq:toeplitz-assembly},
without numerical differentiation.

In practice, small errors in the pilot moments can strongly affect
this gradient estimate when the pilot overlap is nearly
singular or the two lowest Ritz values are close.
We therefore check that the pilot is suitable for choosing
the production counts.

Set $t_p=1/\log(N_p+\mathrm e)$, and for $m\ge2$ let
$\widehat E_{p,2}$ be the second-lowest pilot Ritz value.
We accept the pilot if all required calculations are finite
and
\begin{equation}
 \lambda_{\min}(\widehat S_p)>t_p,\qquad
 \widehat E_{p,2}-\widehat E_p>t_p,\qquad
 \|\widehat{\bm{g}}\|_2\le t_p^{-1}.
 \label{eq:learned-pilot-acceptance}
\end{equation}
Here $\|\cdot\|_2$ denotes the Euclidean norm.
We check the overlap before solving the eigenproblem,
then check the Ritz gap before computing the gradient.
For $m=1$, there is no second Ritz value, so we omit
the gap check.
The final bound prevents an excessively large estimated
gradient from determining the allocation.
We use this logarithmic threshold in the product-state allocation
experiment. For the pilot acceptance argument, any positive
sequence $t_p\to0$ suffices. Under the assumptions of
Theorem~\ref{thm:pilot}, the smallest pilot overlap eigenvalue
and, for $m\ge2$, the lowest Ritz gap converge to positive limits.
The estimated gradient converges to a finite vector. Thus the
pilot passes the checks in \eqref{eq:learned-pilot-acceptance}
with probability tending to one.

If any check or calculation fails, we use uniform production
fractions. Otherwise, we combine $\widehat{\bm{g}}$ with the pilot
covariance estimates to predict the first-order energy
variance for production counts $\boldsymbol n$:
\begin{equation}
 \widehat V(\boldsymbol n)
 =\widehat{\bm{g}}^T\widehat\Sigma(\boldsymbol n)\widehat{\bm{g}}.
 \label{eq:gls-objective}
\end{equation}
To choose the counts, first let $w_q$ be the desired fraction
of the production budget assigned to setting $q$.
Before integer rounding, the counts are $n_q=nw_q$.
The GLS formulas can be evaluated at positive real counts,
and scaling all counts leaves the GLS weights unchanged.
Consequently,
\begin{equation}
 \widehat\Sigma(n\bm{w})=\frac{\widehat\Sigma(\bm{w})}n,\qquad
 \widehat V(n\bm{w})=\frac{\widehat V(\bm{w})}n.
 \label{eq:learned-design-objective}
\end{equation}
For fixed production budget $n$, it therefore suffices
to minimize $\widehat V(\bm{w})$.

We retain measurements at every setting by restricting
the fractions to
\begin{equation}
 \mathcal W_\epsilon
 =\left\{\bm{w}\in\R^{Q_{\rm tot}}:
 \sum_qw_q=1,\quad w_q\ge\epsilon/Q_{\rm tot}\right\},
 \qquad 0<\epsilon<1.
 \label{eq:learned-allocation-class}
\end{equation}
Each setting receives at least the fraction
$\epsilon/Q_{\rm tot}$ of the production budget before rounding.
The remaining fraction $1-\epsilon$ is distributed
according to the optimization, generally unequally,
to reduce the predicted energy variance.
We minimize $\widehat V(\bm{w})$ over this class and convert
the selected fractions to integer counts using the procedure
at the end of this subsection.

The following theorem compares in an asymptotic regime, this learned allocation
with an oracle that knows the true gradient and record
covariances.
To make the pilot estimates consistent while limiting
their cost, we choose
\begin{equation}
 N_p\to\infty,\qquad N_p/N\to0
 \qquad(N\to\infty).
 \label{eq:learned-pilot-growth}
\end{equation}
Thus the pilot grows but uses a vanishing fraction of the
total budget.
Let $V(\bm{w})$ be the first-order energy-variance coefficient
for allocation $\bm{w}$ when GLS uses the true record covariances
and the energy gradient is $\bm{g}$.
Its formula is given in Appendix~\ref{app:gls}.
The oracle minimizes this coefficient over the same
allocation class $\mathcal W_\epsilon$.

The following theorem shows that, under fixed-model regularity
assumptions, the allocation learned from the pilot attains the same
leading MSE coefficient for the production GLS energy estimate as
the best allocation in the prescribed class chosen with exact model
information. The total shot budget includes the pilot measurements.

\begin{theorem}[Accuracy of the learned allocation]
\label{thm:pilot}
For the allocation procedure described in this subsection,
fix the measurement model, Krylov dimension $m$, number
of settings $Q_{\rm tot}$, allocation parameter $\epsilon$,
and covariance clipping bounds $\kappa,b$.

Assume independent bounded pilot records and, conditional
on the pilot, independent production records from the same
fixed setting-specific distributions.
Assume the shared-mean model is correct and identifiable,
$S\succ0$, $E_*$ is simple, and
\begin{equation}
 2\kappa I\preceq\Gamma_q\preceq bI
 \quad\text{for every setting }q.
 \label{eq:learned-allocation-assumptions}
\end{equation}
Assume the bounded energy-reporting rule $E_{\rm rep}$
agrees with $E(\bm{z})$ near $\bm{z}_*$.

If $N_p\to\infty$, $N_p/N\to0$, and numerical allocation
failures have probability $o(1)$, then the production
energy estimate $\widehat E=E_{\rm rep}(\widehat{\bm{z}})$ satisfies
\begin{equation}
 \MSE(\widehat E;E_*)
 =\frac{V_*}{N}+o(N^{-1}),\qquad
 V_*=\min_{\bm{w}\in\mathcal W_\epsilon}V(\bm{w}).
 \label{eq:pilot-oracle}
\end{equation}
\end{theorem}

Appendix~\ref{app:allocation-proof} proves the theorem.
Because $N$ includes the pilot shots, the result accounts
for the cost of learning the allocation.
It shows that the production energy estimate attains the
oracle's leading MSE coefficient within the prescribed
allocation class.
Since uniform fractions belong to this class,
\begin{equation}
 V_*\le V(\boldsymbol1/Q_{\rm tot}),
 \label{eq:learned-uniform-comparison}
\end{equation}
with strict inequality if uniform allocation is not a minimizer.
This comparison is asymptotic; finite-budget improvemnets depend
on the accuracy and cost of the pilot.

\section{Improved Energy Estimation by Joint Spectral Reconstruction}
\label{sec:spectral}

Under exact time evolution, the overlap matrix $S$ and the projected
Hamiltonian matrix $H$ are determined by the same energy distribution
of the reference state. However, sampling error can make the measured moments
of the two matrices inconsistent with any common distribution.  To regain the structure, we form an
 estimated joint covariance and find the closest moment vector
that one distribution can generate. We call this procedure
\emph{joint spectral reconstruction}. We  then form both matrices from the
fitted moments and use their lowest generalized eigenvalue to estimate
the lowest Ritz energy $E_*$. The section studies when this fit reduces
the leading mean-square error of the energy estimate.

\subsection{Fitting the shared distribution to the generating sequences}

Let $E_1,\ldots,E_r$ be the distinct energies that occur with
nonzero probability in the normalized reference state
$|\bm{\psi}_0\rangle$. For each $E_j$, let $\Pi_j$ denote the
projector onto its eigenspace and let
$p_j=\langle\bm{\psi}_0|\Pi_j|\bm{\psi}_0\rangle$ be the
probability of measuring that energy. At lag $\ell$, spectral
decomposition gives the overlap moment $s_\ell$ and the Hamiltonian
moment $h_\ell$ as
\begin{equation}
 \begin{aligned}
 s_\ell&=\sum_{j=1}^r p_j e^{-i\ell\tau E_j},&
 h_\ell&=\sum_{j=1}^r p_j E_j e^{-i\ell\tau E_j},\\
 p_j&>0,&
 \sum_{j=1}^r p_j&=1,
 \end{aligned}
 \label{eq:common-spectrum}
\end{equation}
where $\tau>0$ is the time step. The same probabilities $p_j$
generate both sequences. An energy $E_j$ contributes with weight
$p_j$ to the overlap moment and with weight $p_jE_j$ to the
Hamiltonian moment. In particular, $s_0=1$ and
$h_0=\sum_{j=1}^r p_jE_j$.

The moments $(s_\ell)$ and $(h_\ell)$ are the \emph{generating
sequences} of $S$ and $H$. Each sequence gives one value for each
diagonal through $S_{ij}=s_{j-i}$ and $H_{ij}=h_{j-i}$, as in
\eqref{eq:matrices}. The Toeplitz structure therefore constrains
entries within each matrix. Equation~\eqref{eq:common-spectrum}
also links the two matrices by requiring one energy distribution
to generate both sequences.

To impose this relation on the measured estimates, we collect
the required moments into a real vector. For $m$ Krylov states,
this vector contains $h_0$ and the real and imaginary parts of
$s_\ell,h_\ell$ for $\ell=1,\ldots,m-1$, giving
$d=4m-3$ coordinates. In the order of
Eq.~\eqref{eq:real_compo}, a distribution concentrated at one
energy $x$ produces
\begin{equation}
 \begin{aligned}
 \bm{v}(x)&=\bigl(x,\bm{v}_1(x)^T,\ldots,\bm{v}_{m-1}(x)^T\bigr)^T,\\
 \bm{v}_\ell(x)&=
 \bigl(\cos(\ell\tau x),\,x\cos(\ell\tau x),
       -\sin(\ell\tau x),\,-x\sin(\ell\tau x)\bigr)^T.
 \end{aligned}
 \label{eq:single-energy-vector}
\end{equation}
The exact moment vector is therefore
$\bm{z}_*=\sum_{j=1}^r p_j\bm{v}(E_j)$.

Suppose that a known interval $[L,U]$, with $L<U$, contains
all the energies $E_j$.
Since the locations of the energies $E_j$ and their probabilities
are unknown, we allow any probability distribution on this
interval. Its moment vector is a weighted average of the
vectors $\bm{v}(x)$, with nonnegative weights that sum to one.
The set of compatible moment vectors is the convex hull
\begin{equation}
 \begin{aligned}
 \C
 &=\operatorname{conv}\{\bm{v}(x):x\in[L,U]\}
 =\left\{
 \sum_{a=1}^{d+1}q_a\bm{v}(x_a):
 x_a\in[L,U],\ q_a\ge0,\ 
 \sum_{a=1}^{d+1}q_a=1
 \right\}.
 \end{aligned}
 \label{eq:feasible}
\end{equation}
Carath\'eodory's theorem gives the second representation. Every
vector in $\C$ can be represented using at most $d+1$ energies.
The true distribution may contain more energies, since different
distributions can produce the same stored moments. The set $\C$
is convex by construction and compact because $\bm{v}$ is continuous
on the compact interval $[L,U]$. It contains
$\bm{z}_*=\sum_{j=1}^r p_j\bm{v}(E_j)$, and defining it requires
only $\bm{v}$ and the interval.

The generalized least-squares estimate $\widehat{\bm{z}}$ obtained
from the production measurements need not belong to $\C$. We
therefore adjust it to satisfy the common-distribution constraint,
with the adjustment weighted by its estimated uncertainty.
The pilot covariance estimates, production shot counts, and
generalized least-squares weights give an estimate of the covariance
of $\widehat{\bm{z}}$. Denote the positive-definite covariance
estimate used in the fit by $\widehat\Sigma_{\rm fit}$, and set
\[
 M=\widehat\Sigma_{\rm fit}^{-1},
 \qquad
 \|\bm{u}\|_M^2=\bm{u}^TM\bm{u}.
\]
The inverse covariance measures the adjustment
$\bm{z}-\widehat{\bm{z}}$ relative to its estimated uncertainty.
It accounts for the variances of the individual moment estimates
and for correlations between them.

Using this weight, we define the reconstructed moment vector by
\begin{equation}
 \widehat{\bm{z}}_{\rm rec}=\Pi_\C^M(\widehat{\bm{z}})
 :=\underset{\bm{z}\in\C}{\operatorname{argmin}}
 \ \tfrac12\|\widehat{\bm{z}}-\bm{z}\|_M^2.
 \label{eq:projection}
\end{equation}
The fitted moment vector is unique because $\C$ is nonempty,
compact, and convex, and $M\succ0$ makes the objective strictly
convex in $\bm{z}$. Several energy distributions may produce
this vector. If $\widehat{\bm{z}}\in\C$, the projection gives
$\widehat{\bm{z}}_{\rm rec}=\widehat{\bm{z}}$.

The projection in \eqref{eq:projection} is convex in the moment
vector $\bm{z}$, even though $\C$ allows every energy in $[L,U]$.
If we use the representation in \eqref{eq:feasible} and vary both
the locations $x_a$ and the weights $q_a$, the dependence on the
locations through $\bm{v}(x_a)$ is nonlinear and the resulting
problem is generally nonconvex. We obtain a finite convex problem
by fixing the candidate energies on a grid and fitting their
probabilities.

\paragraph{Grid-based discretization.}
Choose $G\ge2$ trial energies on a uniform grid,
\begin{equation}
 x_a=L+(a-1)h,\qquad
 h=\frac{U-L}{G-1},\qquad a=1,\ldots,G.
 \label{eq:spectral-grid-points}
\end{equation}
The grid includes both endpoints, but its points need not
be true eigenvalues. We reserve $p_j$, $j=1,\ldots,r$, for
the true probabilities and use
$\boldsymbol q=(q_1,\ldots,q_G)^T$ for the grid probabilities.

For $q_a\ge0$ and $\sum_{a=1}^Gq_a=1$, the grid distribution
predicts both sequences through
\begin{equation}
 \begin{aligned}
 s_\ell(\boldsymbol q)
 &=\sum_{a=1}^G q_a e^{-i\ell\tau x_a},\\
 h_\ell(\boldsymbol q)
 &=\sum_{a=1}^G q_a x_a e^{-i\ell\tau x_a},
 \qquad \ell=0,\ldots,m-1.
 \end{aligned}
 \label{eq:grid-common-spectrum}
\end{equation}
The corresponding moment vector is $D_G\boldsymbol q$, where
$D_G=[\bm{v}(x_1)\ \cdots\ \bm{v}(x_G)]$. We choose the probabilities
so that the predicted moments are as close as possible to the
measured moments $\widehat{\bm{z}}$ in the metric defined by $M$.
Holding the grid, $\widehat{\bm{z}}$, and $M$ fixed, we solve
\begin{equation}
 \begin{aligned}
 \widehat{\boldsymbol q}
 &\in\underset{\boldsymbol q\ge0,\,
                   \boldsymbol1^T\boldsymbol q=1}
          {\operatorname{argmin}}
 \ \tfrac12(D_G\boldsymbol q-\widehat{\bm{z}})^TM
           (D_G\boldsymbol q-\widehat{\bm{z}}),\\
 \widehat{\bm{z}}_{{\rm rec},G}&=D_G\widehat{\boldsymbol q}.
 \end{aligned}
 \label{eq:grid-fit}
\end{equation}
Here $\boldsymbol1$ is the all-ones vector. The constraints require
the weights to be nonnegative and sum to one, so they define a
probability distribution on the grid.

This is a convex quadratic program with linear constraints. Its
optimality conditions are linear equations and inequalities, but
which weights vanish depends on the data. We solve the problem
numerically. Different probability vectors can produce the same
moments, so $\widehat{\boldsymbol q}$ need not be unique. Every
minimizer gives the same fitted moment vector
$\widehat{\bm{z}}_{{\rm rec},G}$ because the objective is strictly
convex in $D_G\boldsymbol q$.

Restricting the candidate energies to the grid replaces $\C$ by
\begin{equation}
 \C_G=
 \{D_G\boldsymbol q:\boldsymbol q\ge0,\;
                       \boldsymbol1^T\boldsymbol q=1\}
 \subseteq\C.
 \label{eq:grid-feasible-set}
\end{equation}
The true vector $\bm{z}_*$ need not belong to $\C_G$.
Its distance to $\C_G$ measures the approximation error
introduced by the grid.
If $h_0$ is known exactly, we restrict the continuous fit
to vectors whose first coordinate equals $h_0$ and add
\[
 \sum_{a=1}^G q_ax_a=h_0
\]
to the grid fit. In that case, we form the fitting objective
and covariance from the remaining, uncertain coordinates.

\paragraph{Computing the reconstructed energy.}
The fitted grid moments give the reconstructed matrices
\begin{equation}
 \widehat H_{\rm rec}=H(\widehat{\bm{z}}_{{\rm rec},G}),
 \qquad
 \widehat S_{\rm rec}=S(\widehat{\bm{z}}_{{\rm rec},G}).
 \label{eq:reconstructed-matrices}
\end{equation}
We compute their lowest generalized eigenvalue using the same
overlap check, clipping, and fallback as for the production
estimate. With the reporting rule of
Section~\ref{sec:algorithm-energy}, the reconstructed energy is
\begin{equation}
 \widehat E_{{\rm rec},G}
 =E_{\rm rep}(\widehat{\bm{z}}_{{\rm rec},G}).
 \label{eq:reconstructed-energy}
\end{equation}
The overlap must be checked again because the fit changes its
entries. When the fitted overlap is positive definite, the common
spectral representation places its exact Ritz values in $[L,U]$.
Clipping also guards against numerical roundoff.

\subsection{Lower energy mean squares error with joint reconstruction}
\label{sec:reconstruction-accuracy}

Reconstruction adjusts the measured moments so that
they are compatible with a single energy distribution.
We now examine whether this adjustment also improves
the energy estimate. We first analyze exact projection
onto the continuous compatible set $\C$, then show
how the computed fit retains the same improvement.

\begin{lemma}[Weighted projection bound]
\label{lem:projection-bound}
Let $\C\subset\mathbb R^d$ be closed and convex, and let
$\bm{z}_*\in\C$ be the ideal moment vector. For a symmetric
positive-definite matrix $M$, write
$\|\bm{u}\|_M^2=\bm{u}^T M\bm{u}$.
Given a moment estimate $\widehat{\bm{z}}$, let
$\widehat{\bm{z}}_{\rm rec}=\Pi_\C^M(\widehat{\bm{z}})$ be its
projection onto $\C$ in this norm. Then
\begin{equation}
 \|\widehat{\bm{z}}_{\rm rec}-\bm{z}_*\|_M^2
 \le \|\widehat{\bm{z}}-\bm{z}_*\|_M^2
    -\|\widehat{\bm{z}}-\widehat{\bm{z}}_{\rm rec}\|_M^2.
 \label{eq:contraction}
\end{equation}
\end{lemma}

Appendix~\ref{app:proj_ineq} proves the lemma.
Thus, whenever
reconstruction changes the measured moments, it reduces
their squared error in the $M$-norm.

The energy depends on a particular combination of
these errors. For $S(\bm{z})\succ0$, let $E(\bm{z})$ be the
smallest generalized eigenvalue of $H(\bm{z}),S(\bm{z})$,
and write $E_*=E(\bm{z}_*)$.
Assume $S(\bm{z}_*)\succ0$ and that $E_*$ is simple.
Then $E$ is differentiable near $\bm{z}_*$, with
\[
 E(\bm{z}_*+\bm{e})-E_*=\bm{g}^T\bm{e}+o(\|\bm{e}\|),
 \qquad \bm{g}=\nabla E(\bm{z}_*).
\]
Errors in different moments can cancel in $\bm{g}^T\bm{e}$,
and reconstruction can alter those cancellations.
The projection inequality alone therefore does not
guarantee a smaller energy MSE.

To determine when the energy improves, we examine
the directions corrected by the fit. Projection
leaves $\widehat{\bm{z}}$ unchanged when $\widehat{\bm{z}}\in\C$, so we consider
a boundary point $\bm{z}_*$ where the constraints can
affect small measurement errors.
Suppose its tangent cone is
\[
 T=\overline{\{t(\bm{z}-\bm{z}_*):t\ge0,\ \bm{z}\in\C\}}
   =\{\bm{e}:\bm{a}_1^T\bm{e}\ge0,\ \bm{a}_2^T\bm{e}\ge0\},
\]
where $\bm{a}_1,\bm{a}_2$ are linearly independent inward normals.
Each normal satisfies
\[
 \bm{a}_i^T(\bm{z}-\bm{z}_*)\ge0
 \qquad\text{for every }\bm{z}\in\C.
\]
The corresponding supporting plane passes through
$\bm{z}_*$, and $\C$ lies on the side toward which $\bm{a}_i$
points. The tangent-cone assumption says that these
two inequalities describe all feasible changes to
first order.

As the measurement errors shrink, projection onto
$\C$ is described to first order by projection of
the errors onto $T$. This local description lets us
compare the correction directions with the energy
gradient $\bm{g}$. The normals are needed only for the
analysis, not for computing the reconstruction.

We apply the same reporting rule $E_{\rm rep}$ before
and after reconstruction. This rule includes the
overlap check, clipping, and fallback described above.
Assume it is bounded and measurable and agrees with
$E$ near $\bm{z}_*$, so the safeguards leave sufficiently
accurate estimates unchanged.
We measure accuracy relative to the ideal Krylov energy:
\[
 \MSE(\widehat E;E_*)=\E[(\widehat E-E_*)^2],
\]
where the expectation averages over measurement randomness.

\begin{theorem}[Leading energy-MSE reduction]
\label{thm:wedge}
Under the geometric and spectral assumptions above,
let $\widehat{\bm{z}}_n$ be the moment estimate from $n$ production
shots. Suppose
\[
 \sqrt n(\widehat{\bm{z}}_n-\bm{z}_*)\Rightarrow N(\bm{0},\Omega),
 \qquad
 \sup_n\E\|\sqrt n(\widehat{\bm{z}}_n-\bm{z}_*)\|^{2+\eta}<\infty
\]
for some $\Omega\succ0$ and $\eta>0$, where
$\Rightarrow$ denotes convergence in distribution.
Define
\[
 \widehat{\bm{z}}_{{\rm rec},n}
 =\Pi_\C^{\Omega^{-1}}(\widehat{\bm{z}}_n)
 =\underset{\bm{z}\in\C}{\operatorname{argmin}}\,
   (\widehat{\bm{z}}_n-\bm{z})^T\Omega^{-1}(\widehat{\bm{z}}_n-\bm{z}),
\]
and report
\[
 \widehat E=E_{\rm rep}(\widehat{\bm{z}}_n),
 \qquad
 \widehat E_{\rm rec}=E_{\rm rep}(\widehat{\bm{z}}_{{\rm rec},n}).
\]
Here $\widehat E$ uses the measured moments,
while $\widehat E_{\rm rec}$ uses their spectral
reconstruction. Then
\begin{equation}
 \MSE(\widehat E;E_*)
 -
 \MSE(\widehat E_{\rm rec};E_*)
 =\frac{\Delta}{n}+o(n^{-1}),
 \qquad \Delta\ge0.
 \label{eq:gain}
\end{equation}
Moreover,
\[
 \Delta>0
 \quad\Longleftrightarrow\quad
 (\bm{a}_1^T\Omega \bm{g},\ \bm{a}_2^T\Omega \bm{g})\ne(0,0).
\]
\end{theorem}

The strict-improvement condition identifies whether
the corrections made by reconstruction affect the
energy. In the tangent-cone approximation, projection
corrects constraint violations by adding nonnegative
combinations of $\Omega \bm{a}_1$ and $\Omega \bm{a}_2$.
The first-order energy change along $\Omega \bm{a}_i$ is
\[
 \bm{g}^T\Omega \bm{a}_i=\bm{a}_i^T\Omega \bm{g}.
\]
Thus the leading energy MSE decreases when at least
one correction direction affects the energy.
If both inner products vanish, the leading coefficient
is unchanged, and the theorem does not determine the
smaller-order difference. The comparison includes any
bias introduced by reconstruction.
Appendix~\ref{app:wedge} defines $\Delta$ and proves
the result.

The statistical assumptions hold, for example, for
unbiased linear estimates formed from independent
bounded shots with fixed pooling weights, positive
limiting shot fractions, and a positive-definite
limiting covariance. The Gaussian limit follows from
the multivariate central limit theorem, while bounded
outcomes give the required moment bound.

Theorem~\ref{thm:pilot} gives the leading mean-square error of the
GLS energy estimate under learned shot allocation.
Theorem~\ref{thm:wedge} shows when spectral reconstruction reduces
the leading mean-square error for a moment estimator with the
Gaussian limit assumed there.

These assumptions explain the fitting weight:
\[
 n\operatorname{Cov}(\widehat{\bm{z}}_n)\longrightarrow\Omega,
 \qquad
 \sqrt n\,(\E \widehat{\bm{z}}_n-\bm{z}_*)\longrightarrow\bm{0}.
\]
The covariance is approximately $\Omega/n$, and the
bias is negligible on the $n^{-1/2}$ scale.
The corresponding inverse-covariance weight is
approximately $n\Omega^{-1}$; removing the common
factor $n$ leaves the projection unchanged.

The implementation estimates this weight from pilot
data, restricts the fit to an energy grid, and solves
the resulting optimization problem approximately.
We now specify numerical accuracy conditions under
which these steps preserve the MSE reduction.
Write
\[
 M_n=(n\widehat\Sigma_{{\rm fit},n})^{-1},
\]
which gives the same fit as
$\widehat\Sigma_{{\rm fit},n}^{-1}$.
At production budget $n$, choose a grid containing
$L,U$, with maximum spacing $h_n$, and return a
feasible numerical fit $\widehat{\bm{z}}_{{\rm rec},n}^{\,{\rm num}}\in\C_{G_n}$ satisfying
\begin{equation}
 \sup_{\bm{z}\in\C_{G_n}}
 (\widehat{\bm{z}}_n-\widehat{\bm{z}}_{{\rm rec},n}^{\,{\rm num}})^TM_n(\bm{z}-\widehat{\bm{z}}_{{\rm rec},n}^{\,{\rm num}})
 \le\epsilon_n.
 \label{eq:computed-fit-gap}
\end{equation}
The left-hand side is the solver gap: it is
nonnegative and vanishes at the exact grid solution.

\begin{corollary}[MSE reduction for the discretized estimator]
\label{cor:computed-mse}
Under the assumptions of Theorem~\ref{thm:wedge},
keep the number of Krylov trial states $m$, their
time spacing $\tau$, and the spectral interval $[L,U]$
fixed as the production budget $n$ increases, and suppose
\[
 n\widehat\Sigma_{{\rm fit},n}\xrightarrow{p}\Omega,
 \qquad
 c_0I\preceq n\widehat\Sigma_{{\rm fit},n}\preceq C_0I
 \quad\text{almost surely for every }n,
\]
for fixed $0<c_0<C_0<\infty$.
If the grid spacing and solver-gap bounds are
deterministic and satisfy
\[
 h_n=o(n^{-1/4}),
 \qquad
 \epsilon_n=o(n^{-1}),
\]
then the computed report
$\widehat E_{\rm rec}^{\,{\rm num}}=E_{\rm rep}(\widehat{\bm{z}}_{{\rm rec},n}^{\,{\rm num}})$
satisfies
\begin{equation}
 \MSE(\widehat E;E_*)
 -
 \MSE(\widehat E_{\rm rec}^{\,{\rm num}};E_*)
 =\frac{\Delta}{n}+o(n^{-1}),
 \label{eq:computed-mse-gain}
\end{equation}
with the same $\Delta$ as in Theorem~\ref{thm:wedge}.
In particular, if
$(\bm{a}_1^T\Omega \bm{g},\bm{a}_2^T\Omega \bm{g})\ne(0,0)$, the computed
reconstruction has smaller energy MSE than the production
estimator for all sufficiently large $n$.
\end{corollary}

The covariance conditions ensure that the estimated
weight approaches the correct weight and remains
uniformly well conditioned.
The grid and solver conditions make their contribution
negligible at the statistical error scale.

If the solver uses the unnormalized weight
$\widehat\Sigma_{{\rm fit},n}^{-1}$, the corresponding
gap tolerance is $n\epsilon_n$.

Appendix~\ref{app:finite} proves that these choices
make the mean squared difference between the computed
and exact reconstructed energies $o(n^{-1})$.
They therefore preserve the leading improvement
$\Delta/n$, giving an MSE-reduction guarantee for
the computed algorithm.

\section{The Complete Algorithm}
\label{sec:algorithm}

Algorithm~\ref{alg:shot-allocation-reconstruction} combines
the shot allocation of Section~\ref{sec:gls-design}
with the spectral fit of Section~\ref{sec:spectral}.
The subsections below specify the numerical choices,
stopping rules, and fallback procedures.

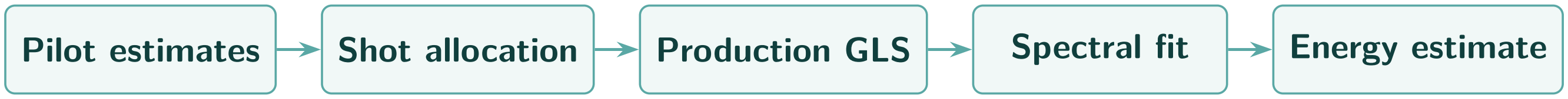
\begin{figure}[!htbp]
\centering
\definecolor{seagreen}{HTML}{238B87}
\definecolor{paleseagreen}{HTML}{E5F3F0}
\resizebox{\linewidth}{!}{%
\begin{tikzpicture}[
 node distance=4mm,
 card/.style={
   draw=seagreen!75,
   fill=paleseagreen!55,
   rounded corners=2.5pt,
   line width=.55pt,
   minimum width=2.3cm,
   minimum height=.8cm,
   inner sep=4pt,
   align=center,
   text=seagreen!45!black,
   font=\sffamily\small\bfseries},
 link/.style={
   -{Stealth[length=2mm,width=1.4mm]},
   draw=seagreen!75,
   line width=.75pt}]
\node[card] (pilot) {Pilot estimates};
\node[card,right=of pilot] (design) {Shot allocation};
\node[card,right=of design] (prod) {Production GLS};
\node[card,right=of prod] (fit) {Spectral fit};
\node[card,right=of fit] (energy) {Energy estimate};
\draw[link] (pilot)--(design);
\draw[link] (design)--(prod);
\draw[link] (prod)--(fit);
\draw[link] (fit)--(energy);
\end{tikzpicture}%
}
\caption{An initial batch of shots (the pilot) lets us estimate how
measurement noise affects the energy and allocate the remaining shots
to minimize the predicted first-order energy variance. We use generalized
least squares (GLS) to estimate the overlap and Hamiltonian moments from
these production shots, then fit both moment sequences to one energy
distribution to estimate the lowest Krylov energy.}
\label{fig:algorithm-flow}
\end{figure}

\begin{algorithm}[tbp]
\caption{Shot allocation and shared spectral reconstruction}
\label{alg:shot-allocation-reconstruction}
\small
\begin{algorithmic}[1]

\Statex \textbf{Inputs:}
Measurement setup, budgets $N,N_p$, grid $D_G$,
and numerical controls specified below.
\Statex \textbf{Outputs:}
Reconstructed energy $\widehat E_{\rm rec}^{\,{\rm num}}$,
production moments $\widehat{\bm{z}}$,
fitted moments $\widehat{\bm{z}}_{\rm rec}^{\,{\rm num}}$,
shot counts, and diagnostics.

\State \textbf{Pilot.}
Estimate the moments and measurement covariances
from $N_p$ shots
(Section~\ref{sec:algorithm-pilot}).

\State \textbf{Allocate.}
Use the pilot to choose the production counts
and fix the GLS weights
(Section~\ref{sec:algorithm-pilot}).

\State \textbf{Measure.}
Collect fresh production data and compute $\widehat{\bm{z}}$
and its covariance estimate $C$
(Section~\ref{sec:algorithm-reconstruction}).

\State \textbf{Reconstruct.}
Fit one distribution on the spectral grid to obtain
$\widehat{\bm{z}}_{\rm rec}^{\,{\rm num}}$
(Section~\ref{sec:algorithm-reconstruction}).

\State \textbf{Compute the energy.}
Return
$\widehat E_{\rm rec}^{\,{\rm num}}
 =E_{\rm rep}(\widehat{\bm{z}}_{\rm rec}^{\,{\rm num}})$,
using the reporting checks and fallback
in Section~\ref{sec:algorithm-energy}.

\end{algorithmic}
\end{algorithm}

\subsection{Pilot and shot allocation}
\label{sec:algorithm-pilot}

Reserve $N_p$ shots for the pilot and $n=N-N_p$ for
production. Distribute the pilot shots as evenly as
possible among the $Q_{\rm tot}$ settings, with at
least two shots per setting.
Require
\[
 \left\lfloor\frac{n\epsilon}{Q_{\rm tot}}\right\rfloor
 \ge1
\]
so every setting also receives a production shot.

Compute the pilot sample means and covariances.
Clip the covariance eigenvalues to the bounds
$[\kappa,b]$ chosen in Section~\ref{sec:gls-design},
obtaining $W_q$, and compute $\widehat{\bm{z}}_p$ using
\eqref{eq:gls-normal} with the pilot counts.

To evaluate the predicted energy variance used for
allocation, we estimate the energy gradient at $\widehat{\bm{z}}_p$.
For the lowest pilot eigenpair $(\widehat E_p,\widehat{\bm c}_p)$, normalized
by $\widehat{\bm c}_p^\dagger S(\widehat{\bm{z}}_p)\widehat{\bm c}_p=1$, compute
\begin{equation}
 \widehat g_i
 =\widehat{\bm c}_p^\dagger
   \bigl(\partial_iH-\widehat E_p\partial_iS\bigr)\widehat{\bm c}_p,
 \label{eq:algorithm-allocation}
\end{equation}
with derivatives evaluated at $\widehat{\bm{z}}_p$.
Substituting this gradient into
\eqref{eq:learned-design-objective} gives
\[
 f_p(\bm{w})
 =\widehat{\bm{g}}^T
   F\mathcal J(\bm{w})^{-1}F^T\widehat{\bm{g}}.
\]
Thus, $\widehat{\bm{g}}$ converts the predicted moment
covariance into the predicted energy variance
$f_p(\bm{w})/n$, which we minimize to choose the shot counts.

The gradient also enters the pilot acceptance test
$\|\widehat{\bm{g}}\|\le t_p^{-1}$.
Proceed with learned allocation only if the eigenvalue
calculation succeeds and all checks in
\eqref{eq:learned-pilot-acceptance} pass.

For an accepted pilot, minimize the predicted variance
objective $f_p(\bm{w})$ in \eqref{eq:learned-design-objective}
over $\mathcal W_\epsilon$.
The gap
\begin{equation}
 \mathcal G_p(\bm{w})
 =\max_{\bm{v}\in\mathcal W_\epsilon}
   \nabla f_p(\bm{w})^T(\bm{w}-\bm{v})
 \label{eq:algorithm-allocation-gap}
\end{equation}
bounds the remaining objective error.
Stop when $\mathcal G_p(\bm{w})\le N_p^{-1/2}$.
If the pilot is rejected or this tolerance is not
certified, use uniform fractions $w_q=1/Q_{\rm tot}$.

Set $n_q=\lfloor nw_q\rfloor$ and assign the remaining
shots to the largest fractional remainders.
Keep these counts and the weights $W_q$ fixed
throughout production.

\subsection{Production estimate and reconstruction}
\label{sec:algorithm-reconstruction}

Collect $n_q$ fresh shots at each setting.
Compute $\widehat{\bm{z}}$ from \eqref{eq:gls-normal} and use
\[
 C=\widehat\Sigma_{\rm fit}
   =F\mathcal J(\boldsymbol n)^{-1}F^T.
\]
This covariance estimate uses the actual production
counts and corresponds to $\widehat\Gamma_q=W_q$
in \eqref{eq:gls-fit-covariance}.

Fit the grid probabilities using \eqref{eq:grid-fit}
with $M=C^{-1}$.
For a feasible fitted moment vector $\widehat{\bm{z}}_{\rm rec}^{\,{\rm num}}$,
compute
\begin{equation}
 \mathcal G_{\rm fit}
 =\max_{\bm{z}\in\C_G}
   (\widehat{\bm{z}}-\widehat{\bm{z}}_{\rm rec}^{\,{\rm num}})^TC^{-1}(\bm{z}-\widehat{\bm{z}}_{\rm rec}^{\,{\rm num}}).
 \label{eq:algorithm-fit-gap}
\end{equation}
This gap bounds the remaining fitting-objective error.
For the reconstruction fit, the maximum is obtained by checking
the grid columns $\bm{z}=\bm{v}(x_a)$.
If $h_0$ is known, impose $\sum_aq_ax_a=h_0$ and
maximize over the restricted feasible set instead.
The fit and covariance then use only the uncertain
coordinates; restore $h_0$ when assembling the matrices.

Stop when $\mathcal G_{\rm fit}\le\delta_{\rm fit}$.
If the solver stops early, keep its last finite
feasible fit and record the gap.
If no such fit is available, retain $\widehat{\bm{z}}$ by setting
$\widehat{\bm{z}}_{\rm rec}^{\,{\rm num}}=\widehat{\bm{z}}$ and flag the skipped reconstruction.

For the large-budget guarantee in
Corollary~\ref{cor:computed-mse}, the maximum grid
spacing must satisfy $h_n=o(n^{-1/4})$ and the
tolerance must satisfy $\delta_{{\rm fit},n}=o(1)$.
The corollary also requires its stated covariance
conditions and a returned feasible fit meeting
the tolerance.

\subsection{Energy reporting and diagnostics}
\label{sec:algorithm-energy}

Use the reconstructed moments to compute the final energy. If
reconstruction was skipped, use the production GLS moments instead.
Let $\bm z$ denote the moments used for this calculation.

Form the projected Hamiltonian $H(\bm z)$ and overlap matrix
$S(\bm z)$. If $\lambda_{\min}(S(\bm z))$ is below the prescribed
cutoff $s_{\rm cut}>0$, report the default energy
$E_{\rm default}\in[L,U]$. Otherwise, compute the lowest
generalized eigenvalue of the pair $(H(\bm z),S(\bm z))$ and
clip it to $[L,U]$. If the eigenvalue calculation fails, report
$E_{\rm default}$. We use the same rule to report the GLS energy
before reconstruction.

Return the reported energy, the production moments, the moments
used for the final energy, and the shot count at each setting.
For the allocation and spectral fit, record the available
optimization gaps and any unmet tolerances. Also record whether
the algorithm used uniform allocation, skipped reconstruction,
or returned the default energy.

\section{Numerical Experiments}
\label{sec:numerics}

We remind you that in this work, shot allocation aims to reduce the statistical error in the lowest
Krylov energy by assigning more measurements to the settings that
matter most for that energy. Joint spectral reconstruction seeks a
further reduction by using the fact that the Hamiltonian and overlap
moments arise from the same spectral measure. We first test whether
learned allocation approaches the best leading MSE coefficient
available with exact model information, and whether reconstruction
gives the reduction predicted by Theorem~\ref{thm:wedge}. We then
compare the methods across several models and reference states. We generate
measurement outcomes from exact real-time evolution, so the
experiments isolate sampling error from evolution error.

\paragraph{Models and evaluation.}
\label{sec:numerical-setup}
We use four- and six-spin XYZ chains and a three-site spinful Hubbard
model on six qubits. Eigenstate superpositions let us control the
spectral support of the reference state, while a product state lets
us test learned allocation with a directly preparable input. The
Hamiltonians, state choices, normalization, and time steps are given
in Appendix~\ref{app:numerical-states}.

For each choice of model, reference state, Krylov dimension, and
shot budget, we run independent trials. The total budget $N$
includes any shots used for the pilot. We measure error relative
to $E_*$, the lowest energy in the chosen Krylov subspace. This
choice isolates the error in estimating $E_*$ from the subspace
approximation error.

We report each trial energy using the rule in
Section~\ref{sec:algorithm-energy}, with overlap cutoff $10^{-8}$
and default energy zero. We clip the computed energies to
$[-0.7,0.4]$ in the controlled reconstruction test and to $[-1,1]$
in the other experiments. The empirical MSE is the average of
$(\widehat E-E_*)^2$ over all trials, including those that return
a clipped or default energy.

\subsection{Testing Learned Shot Allocation}
\label{sec:new-design}

\begin{figure}[!htbp]
\centering
\includegraphics[width=0.92\linewidth]{
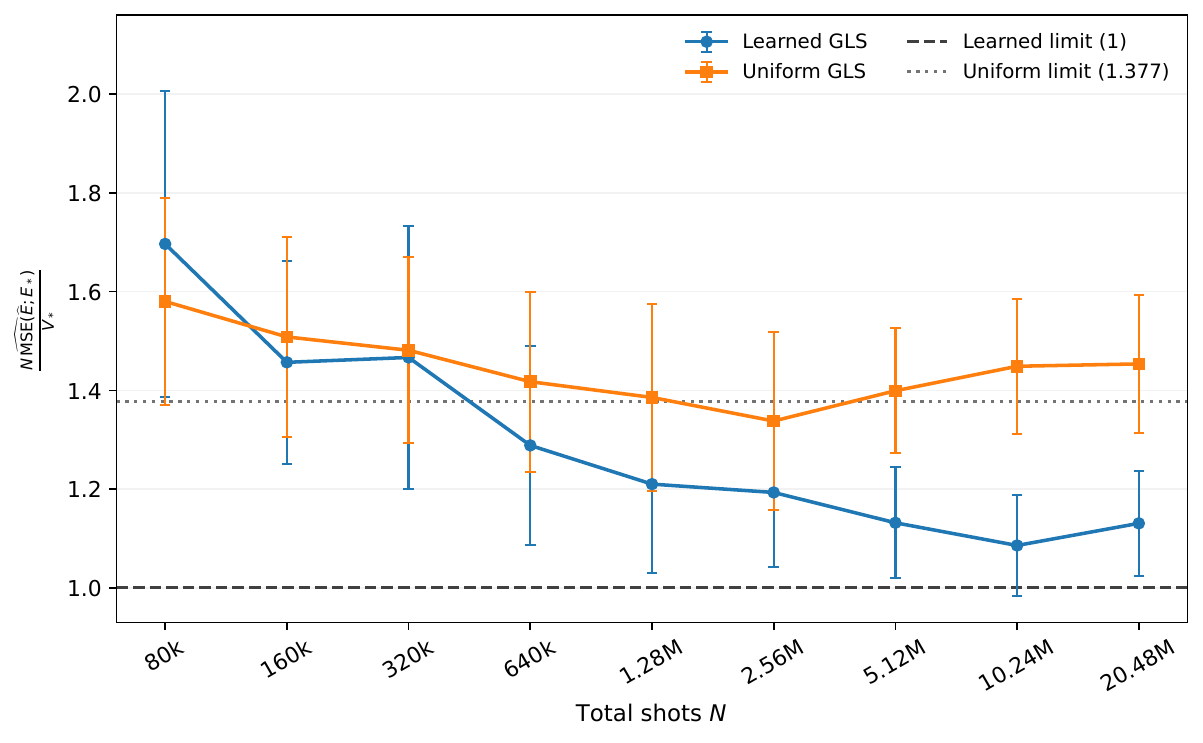}
\caption{Measured energy MSE relative to $V_*/N$,
including the pilot cost. The dashed line at one
marks the learned-allocation theorem's asymptotic
prediction. Error bars show the statistical uncertainty in the estimated MSE due to the finite number of independent Monte Carlo
repetitions. They represent approximate simultaneous $95\%$
confidence intervals across both methods and all nine shot
budgets. }
\label{fig:allocation-convergence}
\end{figure}

\begin{figure}[!htbp]
\centering
\includegraphics[width=0.92\linewidth]{
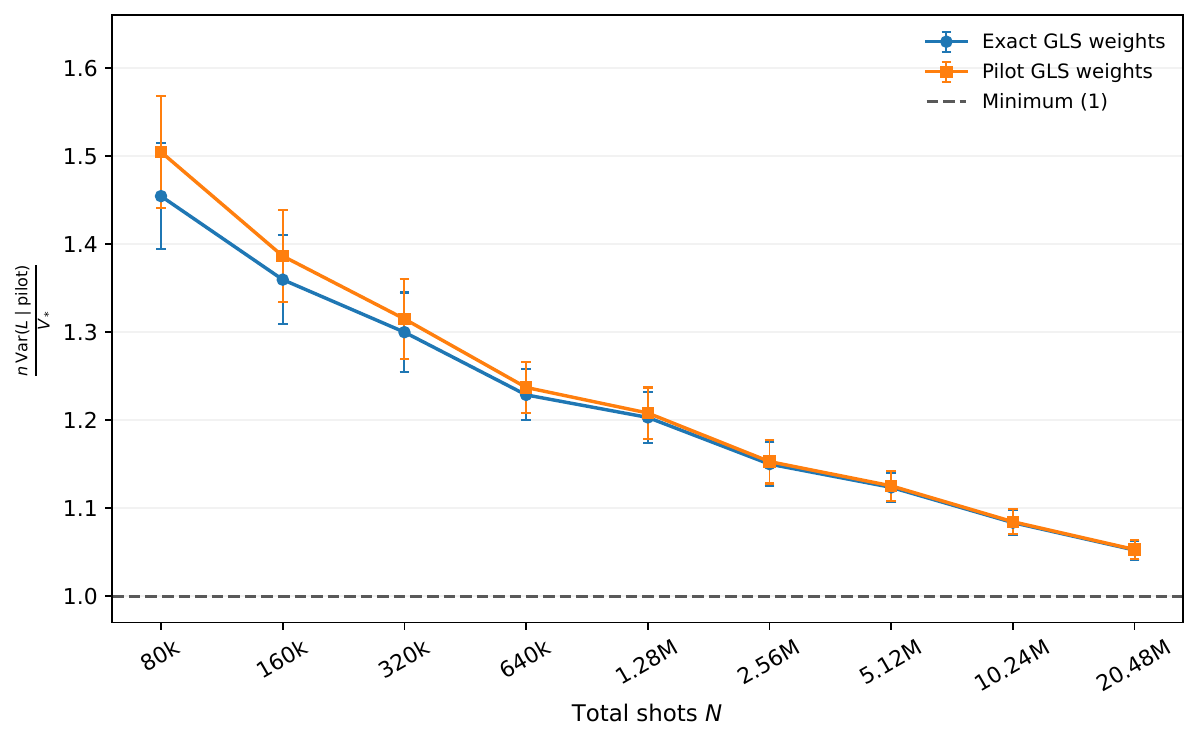}
\caption{Minimum first-order energy variance as a benchmark for
pilot-selected shot allocation. The curves show
$\Var(L\mid\text{pilot})/(V_*/n)$, where
$L=\bm{g}^T(\widehat{\bm{z}}-\bm{z}_*)$ is the first-order energy
error and $\bm{g}=\nabla E(\bm{z}_*)$ is the energy gradient at the
ideal moments. The two curves use exact and pilot-fitted GLS weights.
The dashed line at one marks the minimum variance $V_*/n$ achievable
with the same $n$ production shots. This comparison excludes pilot shot cost.}
\label{fig:allocation-learning}
\end{figure}

We test whether pilot measurements can identify a shot allocation that
achieves the leading energy accuracy predicted by
Theorem~\ref{thm:pilot}. The theorem concerns the energy $\widehat E$
computed from GLS moments before spectral reconstruction.
We showed that for small moment errors, the energy error is
\[
 \widehat E-E_* = \bm{g}^T(\widehat{\bm{z}}-\bm{z}_*)+O(\|\widehat{\bm{z}}-\bm{z}_*\|^2).
\]
With $n$ production shots allocated according to fractions $\bm{w}$,
the variance of the linear term $\bm{g}^T(\widehat{\bm{z}}-\bm{z}_*)$ is $V(\bm{w})/n$ when GLS
uses the true measurement covariances. We calculate the smallest
coefficient within the permitted allocation class,
\[
 V_* = \min_{\bm{w}\in\mathcal W_{0.16}}V(\bm{w})\simeq20.69385,
\]
using the exact covariances and energy gradient. This is a
known-information benchmark, not an additional measurement procedure.
The learned method does not receive these exact quantities.
The theorem predicts that its energy MSE approaches $V_*/N$ as the
total shot budget $N$ grows, even though $N$ includes the pilot.

We use the four-spin model with a product reference state, Krylov
dimension $m=3$, and time step $\tau=6$. The pilot estimates the
covariances and energy gradient and determines the production shot
counts; a rejected pilot or unsuccessful allocation solve uses
uniform counts.
The pilot-budget rule is specified in
Appendix~\ref{sec:numerical-policies}.
For each total shot budget $N$, we independently repeat the
complete measurement and energy-estimation procedure to estimate
its MSE. Each repetition uses $N$ fresh measurement shots and
produces one energy estimate. We use 1,000 repetitions at the
six smaller budgets and 2,000 at the three larger budgets.
Learned and uniform GLS share the pilot and use fresh
production measurements under their respective allocations.

Our first test asks whether the learned allocation gives the energy
accuracy predicted by Theorem~\ref{thm:pilot}. For each total shot
budget $N$, we estimate the energy MSE over independent repetitions
and compare it with $V_*/N$, the theorem's leading-order benchmark.
Figure~\ref{fig:allocation-convergence} plots
\[
 \frac{\widehat{\MSE}(\widehat E;E_*)}{V_*/N}.
\]
A ratio of one means that the measured MSE equals the benchmark;
a ratio above one means that the measured error is larger.
The dashed line marks one. The thin dotted line at $1.377$ is a
separate large-budget reference for uniform GLS, computed as
$V(\bm{w}_{\rm unif})/V_*$ using exact model information.

For learned GLS, the ratio falls from $1.696$ at $N=80{,}000$
to $1.131$ at $N=20{,}480{,}000$, although the values do not
decrease at every intermediate budget. At the largest budget, its
measured MSE is $22.2\%$ below uniform GLS at the same total shot
cost. The paired confidence intervals also favor learned GLS at
the three largest budgets. At the smallest budget, learned GLS
instead has a higher measured MSE; $52.4\%$ of its pilots are
rejected and use uniform allocation. The largest-budget confidence
interval for the learned ratio is $[1.024,1.238]$, still above one.
The results therefore show a finite-budget gain at large $N$,
but do not demonstrate that the asymptotic benchmark has been reached.

Our second test asks whether the pilot actually selects good shot
counts. This can be checked without solving the nonlinear eigenvalue
problem. If $\widehat{\bm{z}}$ denotes the measured moment vector and
$\bm{g}=\nabla E(\bm{z}_*)$, the linear approximation to the energy error is
$L=\bm{g}^T(\widehat{\bm{z}}-\bm{z}_*)$. Its conditional variance is
\[
 \Var(L\mid\text{pilot})=\bm{g}^T\Sigma(\boldsymbol n)\bm{g}.
\]
For each pilot, Figure~\ref{fig:allocation-learning} evaluates this
variance at the shot counts chosen by that pilot. It divides the
variance by $V_*/n$, the smallest first-order variance available
with $n=N-N_p$ production shots, and then averages over pilots.
A plotted value of $1.05$, for example, means that the selected
counts give a linearized variance $5\%$ above this minimum.

The two curves differ in how the production GLS estimator combines
the measured records. One uses GLS weights computed from the exact
covariances; the other uses the weights estimated from the pilot.
The exact covariances and gradient are used only to assess the
resulting variance, never to choose the shots. With exact GLS
weights, the average ratio decreases from $1.454$ to $1.052$;
with pilot-fitted weights, it decreases from $1.504$ to $1.053$.
Thus the pilot-selected counts approach the minimum first-order
variance in this test, and the additional effect of estimating GLS
weights becomes small.

The two figures measure different errors.
Figure~\ref{fig:allocation-learning} examines the linearized
variance of the production estimator at fixed pilot-selected counts.
Figure~\ref{fig:allocation-convergence} measures actual energy MSE
and includes the shots spent on the pilot, nonlinear energy errors,
and fallback outcomes. Together they support the learned-allocation
mechanism and show an advantage over uniform GLS at the larger tested
budgets.
\subsection{Testing the Predicted Reduction in Energy MSE}
\label{sec:theory-validation}

We test whether spectral reconstruction reduces the energy MSE by
the amount predicted by Theorem~\ref{thm:wedge}. Let $n$ be the
actual number of production shots, and let $\widehat E$ and
$\widehat E_{\rm rec}$ be the energy estimates before and after
reconstruction, respectively. The theorem predicts
\[
  n\,\operatorname{MSE}(\widehat E;E_*) \longrightarrow V,
  \qquad
  n\,\operatorname{MSE}(\widehat E_{\rm rec};E_*)
  \longrightarrow V-\Delta.
\]
Thus, $\Delta/V$ is the predicted relative reduction in energy MSE
as the shot budget grows.

We use the four-spin XYZ example with Krylov dimension $m=2$ and
time step $\tau=2.5$. The reference state has probabilities $0.6$
and $0.4$ on the energy eigenvectors indexed by $0$ and $13$,
respectively. In this example, two independent inequalities describe
the tangent cone of the compatible moment set, so the theorem
applies and gives $\Delta/V \approx 45.9\%$.

To isolate the effect of reconstruction, we omit the pilot, allocate
shots uniformly across the 39 settings, and use the exact measurement
covariances to set the GLS and reconstruction weights. We compute
both energy estimates from the same measurement records in each
repetition. To test the theorem's prediction for projection onto the
continuous moment set, we initialize the spectral fit on a grid of
401 energies and then allow the energy locations to vary continuously
over $[-0.7,0.4]$. The occupied energies are unknown to the solver.
In Algorithm~\ref{alg:shot-allocation-reconstruction}, the grid
locations are fixed and the quadratic program~\eqref{eq:grid-fit}
fits the weights on that grid.

We run 5,000 independent repetitions at each of ten nominal shot
budgets from $10^4$ to $10^7$. We apply the bounded energy-reporting
rule and include every repetition in the empirical MSE, including
those for which the solver reports a non-success flag.

\begin{figure}[!htbp]
\centering
\includegraphics[width=0.92\linewidth]{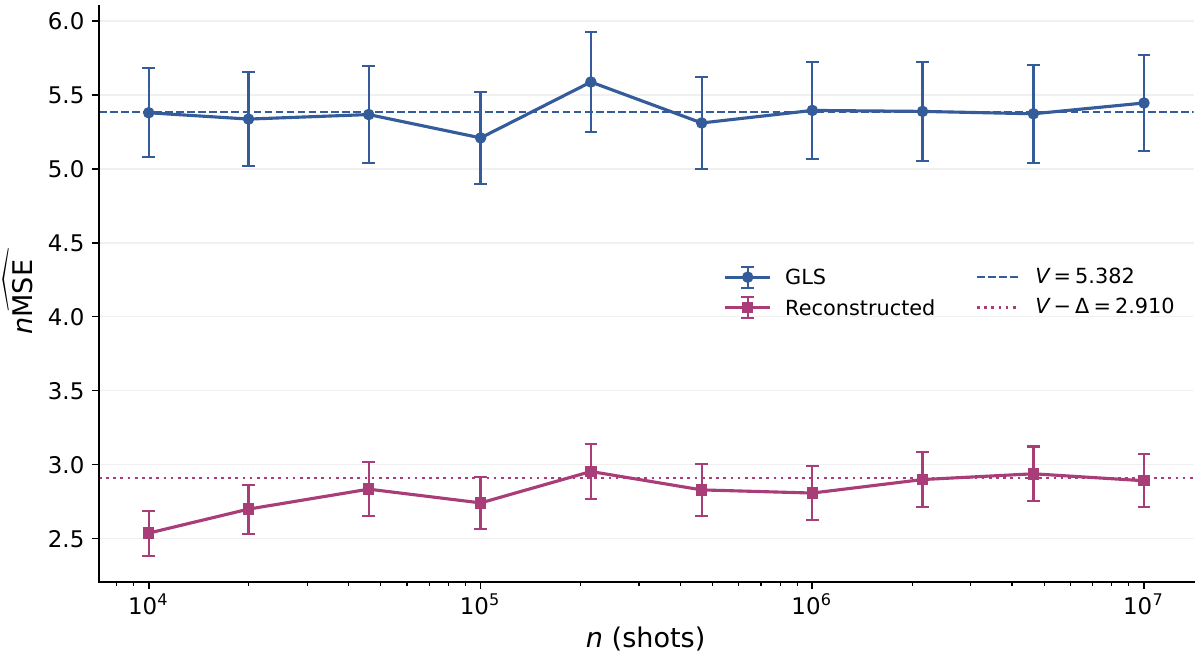}
\caption{Energy MSE before and after spectral reconstruction in the
four-spin XYZ example. Each point shows the empirical MSE from
5,000 repetitions multiplied by the actual shot count $n$. The
horizontal lines mark the predicted limits $V$ before reconstruction
and $V-\Delta$ after reconstruction. The Monte Carlo error bars have
approximately $95\%$ simultaneous coverage across all 20 estimates.}
\label{fig:theory}
\end{figure}

Figure~\ref{fig:theory} compares the scaled empirical MSE with the
two predicted limits. At the two smallest budgets, the scaled MSE
after reconstruction lies below $V-\Delta$. At each of the remaining
eight budgets, the Monte Carlo intervals contain the predicted
limits for both estimates.

At the largest budget, reconstruction reduced the empirical energy
MSE by $46.9\%$, compared with the predicted $45.9\%$. The paired
Monte Carlo interval for this reduction was $[44.1\%,49.6\%]$,
which contains the prediction. 

\subsection{The Complete Algorithm Across Different Models}
\label{sec:factorial}

\begin{figure}[!htbp]
\centering
\includegraphics[width=\linewidth]{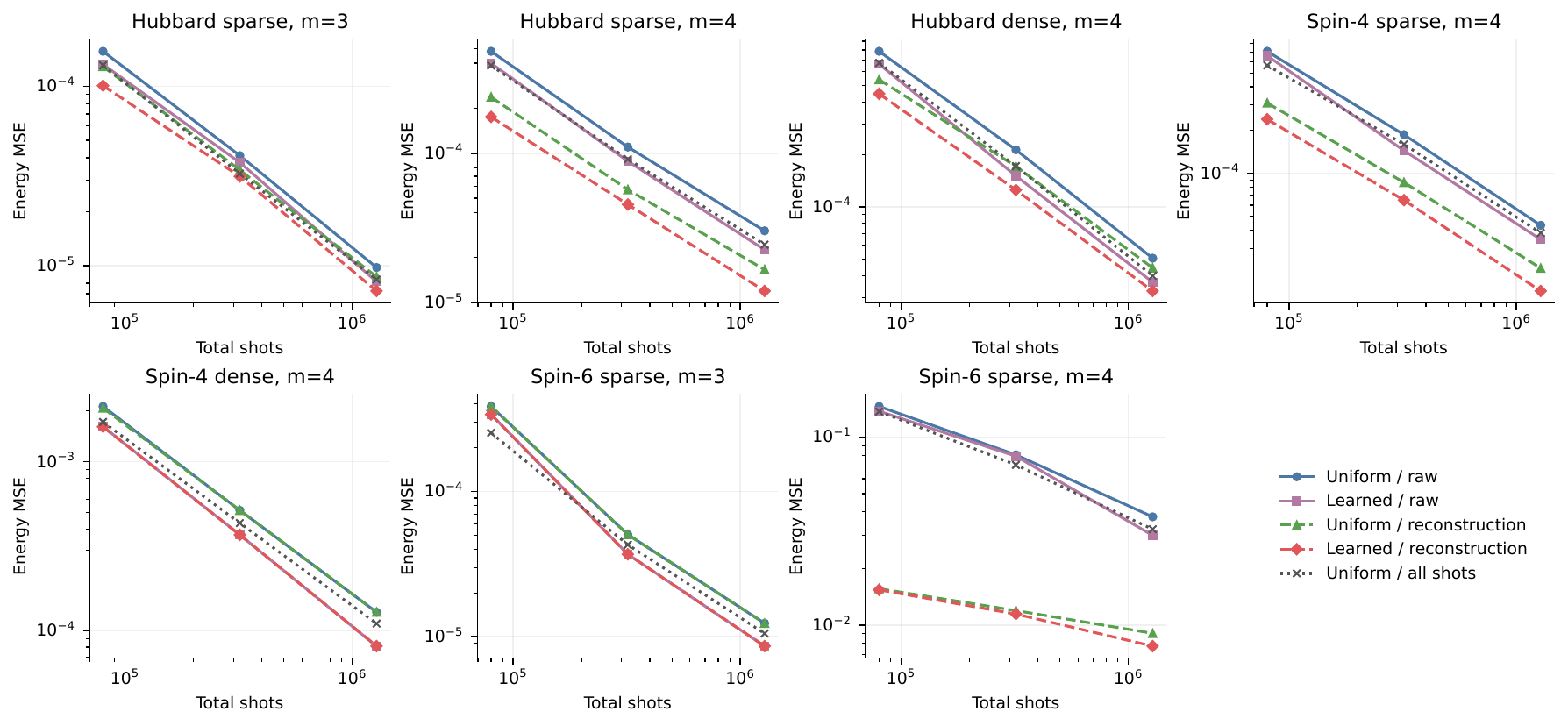}
\caption{Energy MSE at equal total shot budgets for seven
model--state--subspace combinations, with 1,000 independent Monte
Carlo repetitions per case and budget.
Each panel plots five estimators:
$U_{\rm all}$ is the direct-pooling baseline using all $N$ shots
with uniform allocation and no pilot;
$U$ is GLS with a $20\%$ pilot followed by uniform production
allocation;
$A$ is GLS with the same pilot followed by learned production
allocation;
$U_{\rm rec}$ and $A_{\rm rec}$ apply joint spectral reconstruction
to the production records from $U$ and $A$, respectively.
Comparing $U$ with $A$ isolates the effect of learned allocation
within GLS.
Comparing $U$ with $U_{\rm rec}$, or $A$ with $A_{\rm rec}$,
isolates the effect of reconstruction on the same measurement data.
Across the tested cases, learned allocation consistently reduces the
GLS MSE, while reconstruction provides an additional gain in several,
but not all, problems.
Both axes are logarithmic, and the vertical scale is allowed to vary
across panels to keep the trends visible.}
\label{fig:budget-curves}
\end{figure}

The previous experiments test the quantitative predictions of the
allocation and reconstruction theory in controlled settings.
We now ask a broader practical question: across a range of models,
reference states, and Krylov dimensions, how much do learned shot
allocation and joint spectral reconstruction each improve the final
energy estimate?

Figure~\ref{fig:budget-curves} compares five estimators on seven
model--state--subspace combinations, using 1,000 independent Monte
Carlo repetitions at each total shot budget
$N=80{,}000$, $320{,}000$, and $1{,}280{,}000$.
In every case, the error metric is the MSE relative to the ideal
fixed-subspace Ritz energy $E_*$, so the comparison isolates
statistical error rather than subspace approximation error.

Unlike the preceding learned-allocation test, this comparison uses
a fixed pilot fraction and the practical allocation settings
specified in Appendix~\ref{sec:numerical-policies}.
The four GLS-based methods use the same balanced pilot, which
consumes $20\%$ of the total budget.
After the pilot, the remaining shots are allocated either uniformly
or by the learned policy.
The corresponding GLS energies are denoted by $U$ and $A$,
respectively.
Applying joint spectral reconstruction to the same production records
gives the reconstructed estimators $U_{\rm rec}$ and $A_{\rm rec}$.
The reconstruction procedure is described
in Appendix~\ref{sec:numerical-policies}.
For comparison, we also include a direct-pooling baseline
$U_{\rm all}$, which allocates all $N$ shots uniformly and does not
use a pilot.

The figure is designed so that nearby curve comparisons isolate the
effect of each ingredient.
Comparing $U$ with $A$ shows the effect of learned allocation within
the same GLS framework.
Comparing $U$ with $U_{\rm rec}$, or $A$ with $A_{\rm rec}$, shows
the effect of reconstruction on the same underlying measurement
records.
Comparing $U_{\rm all}$ with $A_{\rm rec}$ gives the overall change
from the simplest uniform estimator to the full method, but that
comparison combines several changes at once and should not be
interpreted as the effect of allocation alone.

Figure~\ref{fig:budget-curves} shows that learned allocation gives
a lower empirical GLS MSE in all 21 tested case--budget combinations.
The size of the improvement varies by problem, but the direction is
consistent across the tested budgets.
Reconstruction provides a further reduction in the sparse four-spin
and Hubbard examples, and the additional gain is especially pronounced
for the six-spin case with $m=4$.
By contrast, reconstruction has little visible effect for six-spin
$m=3$, and for dense four-spin the reconstructed and unreconstructed
curves are nearly indistinguishable at the two larger budgets.
Thus the observed reconstruction gains vary substantially across
the tested problems.

\section{Conclusion and Future Work}
\label{sec:limits}
s
Even when a quantum circuit is shallow enough to run, estimating
an energy accurately can require many repetitions. In real-time
quantum Krylov diagonalization, this raises a practical question:
with a fixed number of shots, which circuits should we repeat more
often to estimate the lowest Ritz energy in the chosen subspace?
We use a small pilot sample to guide that choice. It tells us which
measurement errors matter most for the energy and whether errors
from the same shot tend to reinforce or cancel each other. Under
our assumptions, as the shot budget grows, the resulting energy
estimate approaches the accuracy of the best
allocation we could choose if we knew these quantities in advance.
This comparison includes the shots spent on the pilot.

A second contribution concerns a structural relation between the overlap matrix \(S\) and the projected Hamiltonian \(H\) that is not enforced by the usual procedure of assembling the measured matrices and solving the resulting generalized eigenvalue problem. Both matrices are determined by the spectral measure of the initial state: at each time lag, the overlap moment is the average of the evolution phase with respect to this measure, while the corresponding Hamiltonian moment is the average of the same phase weighted by energy. Consequently, two measured matrices can each have the required Hermitian Toeplitz structure and yet be incompatible with any single spectral measure. Our reconstruction enforces this consistency by fitting both families of moments to a common measure, with the discrepancy weighted by the inverse covariance of the moment estimates. The relevant question is whether this fit improves the energy, since a reduction in moment error alone does not establish a reduction in energy error. We give a precise answer when the moment errors have a Gaussian limit and two linearly independent inequalities characterize the local feasible directions. In this setting, reconstruction cannot increase the leading energy MSE, and the improvement is strict precisely when at least one of the two covariance-weighted constraint normals has a nonzero first-order effect on the Ritz energy.

These results suggest a broader question for near-term quantum simulation: how should a fixed measurement budget be distributed when the reported observable depends on several measured quantities linked by physical constraints? Even in the Krylov setting, an allocation that is optimal before reconstruction need not remain optimal afterward, because the fit suppresses some errors that would otherwise affect the energy. A natural next step is therefore to design the allocation for the reconstructed estimator itself. The same principle may be useful for other tasks in which the total number of shots limits the attainable precision. For example, the uncertainty in an excitation gap depends on how measurement errors in two energy estimates cancel, while the uncertainty in a frequency-resolved response depends on how errors at different time lags combine. In each case, the quantity ultimately reported, together with the physical relations used to estimate it, provides a basis for deciding which circuits merit additional shots.

\section*{Acknowledgments}
\paragraph{Disclosure of AI assistance.}
We used GPT-6 Astra (OpenAI) to assist with implementing
and debugging numerical code, discussing mathematical
derivations, and refining the exposition. The authors
reviewed and revised the AI-assisted material and take
full responsibility for the mathematical claims,
computational results, interpretations, and conclusions
presented in this work.

\paragraph{Funding support.} The work is supported by NSF Grant DMS-2552687, and a Superseed grant from Penn State Institute for Computational and Data Sciences (ICDS).

\bibliographystyle{unsrt}
\bibliography{references}

@article{abramslloyd1999,
  author        = {Abrams, Daniel S. and Lloyd, Seth},
  title         = {Quantum Algorithm Providing Exponential Speed Increase for Finding Eigenvalues and Eigenvectors},
  journal       = {Physical Review Letters},
  volume        = {83},
  number        = {24},
  pages         = {5162--5165},
  year          = {1999},
  doi           = {10.1103/PhysRevLett.83.5162},
  url           = {https://doi.org/10.1103/PhysRevLett.83.5162},
  note          = {\href{https://doi.org/10.1103/PhysRevLett.83.5162}{10.1103/PhysRevLett.83.5162}}
}

@misc{perrin2026variational,
  author        = {Perrin, Alexandre and Messud, J{\'e}r{\'e}mie and
                   Burdin, S{\'e}bastien and Sauban{\`e}re, Matthieu},
  title         = {Exploring Variational Regularization of {Krylov} Subspaces},
  year          = {2026},
  howpublished  = {arXiv preprint},
  eprint        = {2611.09229},
  archivePrefix = {arXiv},
  doi           = {10.48550/arXiv.2611.09229},
  url           = {https://arxiv.org/abs/2611.09229},
  note          = {\href{https://arxiv.org/abs/2611.09229}{arXiv:2611.09229}}
}

@article{krilov2001,
  author  = {Krilov, Goran and Sim, Eunji and Berne, B. J.},
  title   = {Quantum time correlation functions from complex time {Monte Carlo} simulations: A maximum entropy approach},
  journal = {The Journal of Chemical Physics},
  year    = {2001},
  volume  = {114},
  number  = {3},
  pages   = {1075--1088},
  doi     = {10.1063/1.1331613},
  note    = {\href{https://doi.org/10.1063/1.1331613}{10.1063/1.1331613}}
}

@article{shen2026modmd,
  author  = {Shen, Yizhi and Buzali, Alex and Hu, Hong-Ye and Klymko, Katherine and Camps, Daan and Yelin, Susanne F. and {Van Beeumen}, Roel},
  title   = {Efficient Measurement-Driven Eigenenergy Estimation with Classical Shadows},
  journal = {PRX Quantum},
  year    = {2026},
  volume  = {7},
  number  = {1},
  pages   = {010328},
  doi     = {10.1103/74s6-3jsz},
  note    = {\href{https://doi.org/10.1103/74s6-3jsz}{10.1103/74s6-3jsz}}
}

@article{reiher2017,
  author        = {Reiher, Markus and Wiebe, Nathan and Svore, Krysta M. and Wecker, Dave and Troyer, Matthias},
  title         = {Elucidating reaction mechanisms on quantum computers},
  journal       = {Proceedings of the National Academy of Sciences},
  volume        = {114},
  number        = {29},
  pages         = {7555--7560},
  year          = {2017},
  doi           = {10.1073/pnas.1619152114},
  url           = {https://doi.org/10.1073/pnas.1619152114},
  note          = {\href{https://doi.org/10.1073/pnas.1619152114}{10.1073/pnas.1619152114}}
}

@article{lintong2022,
  author        = {Lin, Lin and Tong, Yu},
  title         = {{Heisenberg}-Limited Ground-State Energy Estimation for Early Fault-Tolerant Quantum Computers},
  journal       = {PRX Quantum},
  volume        = {3},
  number        = {1},
  pages         = {010318},
  year          = {2022},
  doi           = {10.1103/PRXQuantum.3.010318},
  url           = {https://doi.org/10.1103/PRXQuantum.3.010318},
  note          = {\href{https://doi.org/10.1103/PRXQuantum.3.010318}{10.1103/PRXQuantum.3.010318}}
}

@article{epperly,
  author        = {Epperly, Ethan N. and Lin, Lin and Nakatsukasa, Yuji},
  title         = {A theory of quantum subspace diagonalization},
  journal       = {SIAM Journal on Matrix Analysis and Applications},
  volume        = {43},
  pages         = {1263--1290},
  year          = {2022},
  doi           = {10.1137/21M145954X},
  url           = {https://doi.org/10.1137/21M145954X},
  note          = {\href{https://doi.org/10.1137/21M145954X}{10.1137/21M145954X}}
}

@article{sampling,
  author        = {Lee, G. and Lee, D. and Huh, J.},
  title         = {Sampling error analysis in quantum {Krylov} subspace diagonalization},
  journal       = {Quantum},
  volume        = {8},
  pages         = {1477},
  year          = {2024},
  doi           = {10.22331/q-2024-09-19-1477},
  url           = {https://doi.org/10.22331/q-2024-09-19-1477},
  note          = {\href{https://doi.org/10.22331/q-2024-09-19-1477}{10.22331/q-2024-09-19-1477}}
}

@article{zhang,
  author        = {Zhang, Z. and Wang, A. and Xu, X. and Li, Y.},
  title         = {Measurement-efficient quantum {Krylov} subspace diagonalisation},
  journal       = {Quantum},
  volume        = {8},
  pages         = {1438},
  year          = {2024},
  doi           = {10.22331/q-2024-08-13-1438},
  url           = {https://doi.org/10.22331/q-2024-08-13-1438},
  note          = {\href{https://doi.org/10.22331/q-2024-08-13-1438}{10.22331/q-2024-08-13-1438}}
}

@article{kirby,
  author        = {Kirby, W.},
  title         = {Analysis of quantum {Krylov} algorithms with errors},
  journal       = {Quantum},
  volume        = {8},
  pages         = {1457},
  year          = {2024},
  doi           = {10.22331/q-2024-08-29-1457},
  url           = {https://doi.org/10.22331/q-2024-08-29-1457},
  note          = {\href{https://doi.org/10.22331/q-2024-08-29-1457}{10.22331/q-2024-08-29-1457}}
}

@article{lee,
  author        = {Lee, G. and Choi, S. and Huh, J. and Izmaylov, A. F.},
  title         = {Efficient strategies for reducing sampling error in quantum {Krylov} subspace diagonalization},
  journal       = {Digital Discovery},
  volume        = {4},
  pages         = {954--969},
  year          = {2025},
  doi           = {10.1039/D4DD00321G},
  url           = {https://doi.org/10.1039/D4DD00321G},
  note          = {\href{https://doi.org/10.1039/D4DD00321G}{10.1039/D4DD00321G}}
}

@article{nakamura,
  author        = {Nakamura, Y. and Yano, Y. and Yoshioka, N.},
  title         = {Adaptive measurement strategy for quantum subspace methods},
  journal       = {New Journal of Physics},
  volume        = {26},
  pages         = {033028},
  year          = {2024},
  doi           = {10.1088/1367-2630/ad2c3b},
  url           = {https://doi.org/10.1088/1367-2630/ad2c3b},
  note          = {\href{https://doi.org/10.1088/1367-2630/ad2c3b}{10.1088/1367-2630/ad2c3b}}
}

@misc{rotated,
  author        = {{De Vriendt}, Xeno and Bringewatt, Jacob and Gjonbalaj, Nik O. and Ostermann, Stefan and Vodola, Davide and Borregaard, Johannes and K{\"u}hn, Michael and Yelin, Susanne F.},
  title         = {A new angle on quantum subspace diagonalization for quantum chemistry},
  year          = {2026},
  howpublished  = {arXiv preprint},
  eprint        = {2602.11985v1},
  archivePrefix = {arXiv},
  doi           = {10.48550/arXiv.2602.11985},
  url           = {https://arxiv.org/abs/2602.11985v1},
  note          = {\href{https://doi.org/10.48550/arXiv.2602.11985}{10.48550/arXiv.2602.11985}}
}

@article{mcclean,
  author        = {McClean, Jarrod R. and Kimchi-Schwartz, Mollie E. and Carter, Jonathan and de Jong, Wibe A.},
  title         = {Hybrid quantum-classical hierarchy for mitigation of decoherence and determination of excited states},
  journal       = {Physical Review A},
  volume        = {95},
  number        = {4},
  pages         = {042308},
  year          = {2017},
  doi           = {10.1103/PhysRevA.95.042308},
  url           = {https://doi.org/10.1103/PhysRevA.95.042308},
  note          = {\href{https://doi.org/10.1103/PhysRevA.95.042308}{10.1103/PhysRevA.95.042308}}
}

@misc{parrish,
  author        = {Parrish, Robert M. and McMahon, Peter L.},
  title         = {Quantum Filter Diagonalization: Quantum Eigendecomposition without Full Quantum Phase Estimation},
  year          = {2019},
  howpublished  = {arXiv preprint},
  eprint        = {1909.08925},
  archivePrefix = {arXiv},
  primaryClass  = {quant-ph},
  doi           = {10.48550/arXiv.1909.08925},
  url           = {https://arxiv.org/abs/1909.08925},
  note          = {\href{https://doi.org/10.48550/arXiv.1909.08925}{10.48550/arXiv.1909.08925}}
}

@article{stair,
  author        = {Stair, Nicholas H. and Huang, Renke and Evangelista, Francesco A.},
  title         = {A Multireference Quantum {Krylov} Algorithm for Strongly Correlated Electrons},
  journal       = {Journal of Chemical Theory and Computation},
  volume        = {16},
  number        = {4},
  pages         = {2236--2245},
  year          = {2020},
  doi           = {10.1021/acs.jctc.9b01125},
  url           = {https://doi.org/10.1021/acs.jctc.9b01125},
  note          = {\href{https://doi.org/10.1021/acs.jctc.9b01125}{10.1021/acs.jctc.9b01125}}
}

@article{ibm,
  author        = {Yoshioka, Nobuyuki and Amico, Mirko and Kirby, William and Jurcevic, Petar and Dutt, Arkopal and Fuller, Bryce and Garion, Shelly and Haas, Holger and Hamamura, Ikko and Ivrii, Alexander and Majumdar, Ritajit and Minev, Zlatko and Motta, Mario and Pokharel, Bibek and Rivero, Pedro and Sharma, Kunal and Wood, Christopher J. and Javadi-Abhari, Ali and Mezzacapo, Antonio},
  title         = {{Krylov} diagonalization of large many-body {Hamiltonians} on a quantum processor},
  journal       = {Nature Communications},
  volume        = {16},
  pages         = {5014},
  year          = {2025},
  doi           = {10.1038/s41467-025-59716-z},
  url           = {https://doi.org/10.1038/s41467-025-59716-z},
  note          = {\href{https://doi.org/10.1038/s41467-025-59716-z}{10.1038/s41467-025-59716-z}}
}

@article{cao,
  author        = {Cao, Yudong and Romero, Jonathan and Olson, Jonathan P. and Degroote, Matthias and Johnson, Peter D. and Kieferov{\'a}, M{\'a}ria and Kivlichan, Ian D. and Menke, Tim and Peropadre, Borja and Sawaya, Nicolas P. D. and Sim, Sukin and Veis, Libor and Aspuru-Guzik, Al{\'a}n},
  title         = {Quantum chemistry in the age of quantum computing},
  journal       = {Chemical Reviews},
  volume        = {119},
  number        = {19},
  pages         = {10856--10915},
  year          = {2019},
  doi           = {10.1021/acs.chemrev.8b00803},
  url           = {https://doi.org/10.1021/acs.chemrev.8b00803},
  note          = {\href{https://doi.org/10.1021/acs.chemrev.8b00803}{10.1021/acs.chemrev.8b00803}}
}

@article{properties,
  author        = {Oumarou, Oumarou and Ollitrault, Pauline J. and Cortes, Cristian L. and Scheurer, Maximilian and Parrish, Robert M. and Gogolin, Christian},
  title         = {Molecular properties from quantum {Krylov} subspace diagonalization},
  journal       = {Journal of Chemical Theory and Computation},
  volume        = {21},
  number        = {9},
  pages         = {4543--4552},
  year          = {2025},
  doi           = {10.1021/acs.jctc.5c00194},
  url           = {https://doi.org/10.1021/acs.jctc.5c00194},
  note          = {\href{https://doi.org/10.1021/acs.jctc.5c00194}{10.1021/acs.jctc.5c00194}}
}

@article{krylov,
  author        = {Krylov, A. N.},
  title         = {De la r\'esolution num\'erique de l'\'equation servant \`a d\'eterminer dans des questions de m\'ecanique appliqu\'ee les fr\'equences de petites oscillations des syst\`emes mat\'eriels},
  journal       = {Bulletin de l'Acad\'emie des Sciences de l'URSS. Classe des sciences math\'ematiques et naturelles},
  pages         = {491--539},
  year          = {1931},
  url           = {https://www.mathnet.ru/eng/im5215},
  note          = {No. 4, in Russian. \href{https://www.mathnet.ru/eng/im5215}{MathNet: im5215}}
}

@article{lanczos,
  author        = {Lanczos, Cornelius},
  title         = {An iteration method for the solution of the eigenvalue problem of linear differential and integral operators},
  journal       = {Journal of Research of the National Bureau of Standards},
  volume        = {45},
  pages         = {255--282},
  year          = {1950},
  doi           = {10.6028/jres.045.026},
  url           = {https://doi.org/10.6028/jres.045.026},
  note          = {\href{https://doi.org/10.6028/jres.045.026}{10.6028/jres.045.026}}
}

@book{parlett,
  author        = {Parlett, Beresford N.},
  title         = {The Symmetric Eigenvalue Problem},
  publisher     = {Society for Industrial and Applied Mathematics},
  series        = {Classics in Applied Mathematics},
  volume        = {20},
  year          = {1998},
  doi           = {10.1137/1.9781611971163},
  url           = {https://doi.org/10.1137/1.9781611971163},
  note          = {\href{https://doi.org/10.1137/1.9781611971163}{10.1137/1.9781611971163}}
}

@article{silverman1984,
  author        = {Silverman, Jeremiah N.},
  title         = {Generalized {Hellmann--Feynman} and curvature theorems in classical and quantum mechanics},
  journal       = {International Journal of Quantum Chemistry},
  volume        = {25},
  number        = {5},
  pages         = {915--928},
  year          = {1984},
  doi           = {10.1002/qua.560250512},
  url           = {https://doi.org/10.1002/qua.560250512},
  note          = {\href{https://doi.org/10.1002/qua.560250512}{10.1002/qua.560250512}}
}

@article{cohn,
  author        = {Cohn, Jeffrey and Motta, Mario and Parrish, Robert M.},
  title         = {Quantum filter diagonalization with compressed double-factorized {Hamiltonians}},
  journal       = {PRX Quantum},
  volume        = {2},
  number        = {4},
  pages         = {040352},
  year          = {2021},
  doi           = {10.1103/PRXQuantum.2.040352},
  url           = {https://doi.org/10.1103/PRXQuantum.2.040352},
  note          = {\href{https://doi.org/10.1103/PRXQuantum.2.040352}{10.1103/PRXQuantum.2.040352}}
}

@article{klymko,
  author        = {Klymko, Katherine and Mejuto-Zaera, Carlos and Cotton, Stephen J. and Wudarski, Filip and Urbanek, Miroslav and Hait, Diptarka and Head-Gordon, Martin and Whaley, K. Birgitta and Moussa, Jonathan and Wiebe, Nathan and de Jong, Wibe A. and Tubman, Norm M.},
  title         = {Real-time evolution for ultracompact {Hamiltonian} eigenstates on quantum hardware},
  journal       = {PRX Quantum},
  volume        = {3},
  number        = {2},
  pages         = {020323},
  year          = {2022},
  doi           = {10.1103/PRXQuantum.3.020323},
  url           = {https://doi.org/10.1103/PRXQuantum.3.020323},
  note          = {\href{https://doi.org/10.1103/PRXQuantum.3.020323}{10.1103/PRXQuantum.3.020323}}
}

@article{kemper,
  author        = {Kemper, Alexander F. and Yang, Chao and Gull, Emanuel},
  title         = {Denoising and extension of response functions in the time domain},
  journal       = {Physical Review Letters},
  volume        = {132},
  number        = {16},
  pages         = {160403},
  year          = {2024},
  eprint        = {2309.02566},
  archivePrefix = {arXiv},
  doi           = {10.1103/PhysRevLett.132.160403},
  url           = {https://doi.org/10.1103/PhysRevLett.132.160403},
  note          = {\href{https://doi.org/10.1103/PhysRevLett.132.160403}{10.1103/PhysRevLett.132.160403}}
}

@article{odmd,
  author        = {Shen, Yizhi and Camps, Daan and Szasz, Aaron and Darbha, Siva and Klymko, Katherine and Williams-Young, David B. and Tubman, Norm M. and {Van Beeumen}, Roel},
  title         = {Estimating eigenenergies from quantum dynamics: A unified noise-resilient measurement-driven approach},
  journal       = {Quantum},
  volume        = {9},
  pages         = {1836},
  year          = {2025},
  doi           = {10.22331/q-2025-08-27-1836},
  url           = {https://doi.org/10.22331/q-2025-08-27-1836},
  note          = {\href{https://doi.org/10.22331/q-2025-08-27-1836}{10.22331/q-2025-08-27-1836}}
}

@misc{fdodmd,
  author        = {Bassi, Hardeep and Shen, Yizhi and Bhat, Harish S. and {Van Beeumen}, Roel},
  title         = {From noisy observables to accurate ground state energies: A quantum classical signal subspace approach with denoising},
  year          = {2025},
  howpublished  = {arXiv preprint},
  eprint        = {2505.10735},
  archivePrefix = {arXiv},
  doi           = {10.48550/arXiv.2505.10735},
  url           = {https://arxiv.org/abs/2505.10735},
  note          = {\href{https://doi.org/10.48550/arXiv.2505.10735}{10.48550/arXiv.2505.10735}}
}

@article{partitioned,
  author        = {O'Leary, Tom and Anderson, Lewis W. and Jaksch, Dieter and Kiffner, Martin},
  title         = {Partitioned quantum subspace expansion},
  journal       = {Quantum},
  volume        = {9},
  pages         = {1726},
  year          = {2025},
  doi           = {10.22331/q-2025-05-05-1726},
  url           = {https://doi.org/10.22331/q-2025-05-05-1726},
  note          = {\href{https://doi.org/10.22331/q-2025-05-05-1726}{10.22331/q-2025-05-05-1726}}
}

@misc{superkrylov,
  author        = {Byrne, Adam and Kirby, William and Soodhalter, Kirk M. and Zhuk, Sergiy},
  title         = {The quantum super-{Krylov} method},
  year          = {2024},
  howpublished  = {arXiv preprint},
  eprint        = {2412.17289},
  archivePrefix = {arXiv},
  doi           = {10.48550/arXiv.2412.17289},
  url           = {https://arxiv.org/abs/2412.17289},
  note          = {Revised 2025. \href{https://doi.org/10.48550/arXiv.2412.17289}{10.48550/arXiv.2412.17289}}
}

@misc{samplekrylov,
  author        = {Yu, Jeffery and {Robledo Moreno}, Javier and Iosue, Joseph T. and Bertels, Luke and Claudino, Daniel and Fuller, Bryce and Groszkowski, Peter and Humble, Travis S. and Jurcevic, Petar and Kirby, William and Maier, Thomas A. and Motta, Mario and Pokharel, Bibek and Seif, Alireza and Shehata, Amir and Sung, Kevin J. and Tran, Minh C. and Tripathi, Vinay and Mezzacapo, Antonio and Sharma, Kunal},
  title         = {Quantum-centric algorithm for sample-based {Krylov} diagonalization},
  year          = {2025},
  howpublished  = {arXiv preprint},
  eprint        = {2501.09702},
  archivePrefix = {arXiv},
  doi           = {10.48550/arXiv.2501.09702},
  url           = {https://arxiv.org/abs/2501.09702},
  note          = {\href{https://doi.org/10.48550/arXiv.2501.09702}{10.48550/arXiv.2501.09702}}
}

@misc{filters,
  author        = {{Jord\~ao Oliveira}, Maria Gabriela and Ziems, Karl Michael and Glaser, Nina},
  title         = {Tackling instabilities of quantum {Krylov} subspace methods: An analysis of the numerical and statistical errors},
  year          = {2026},
  howpublished  = {arXiv preprint},
  eprint        = {2604.11532},
  archivePrefix = {arXiv},
  doi           = {10.48550/arXiv.2604.11532},
  url           = {https://arxiv.org/abs/2604.11532},
  note          = {\href{https://doi.org/10.48550/arXiv.2604.11532}{10.48550/arXiv.2604.11532}}
}

@article{thermal,
  author        = {Gentinetta, Gian and Metz, Friederike and Kirby, William and Carleo, Giuseppe},
  title         = {Quantum finite-temperature {Lanczos} method},
  journal       = {Physical Review Research},
  volume        = {8},
  number        = {2},
  pages         = {023285},
  year          = {2026},
  doi           = {10.1103/f3q5-cm1n},
  url           = {https://doi.org/10.1103/f3q5-cm1n},
  note          = {\href{https://doi.org/10.1103/f3q5-cm1n}{10.1103/f3q5-cm1n}}
}

@article{mirror,
  author        = {Kanasugi, Shota and Nakagawa, Yuya O. and Matsumoto, Norifumi and Hidaka, Yuichiro and Maruyama, Kazunori and Oshima, Hirotaka},
  title         = {Mirror subspace diagonalization: A real-time quantum {Krylov} algorithm with near-optimal sampling cost},
  journal       = {npj Quantum Information},
  year          = {2026},
  doi           = {10.1038/s41534-026-01384-y},
  url           = {https://doi.org/10.1038/s41534-026-01384-y},
  note          = {Advance online publication. \href{https://doi.org/10.1038/s41534-026-01384-y}{10.1038/s41534-026-01384-y}}
}

@misc{qkl,
  author        = {Matsuura, Shunji and Kawamura, Yoji and Salfi, Joseph and Iso, Satoshi},
  title         = {Quantum {Krylov} learning},
  year          = {2026},
  howpublished  = {arXiv preprint},
  eprint        = {2610.00537},
  archivePrefix = {arXiv},
  doi           = {10.48550/arXiv.2610.00537},
  url           = {https://arxiv.org/abs/2610.00537},
  note          = {\href{https://doi.org/10.48550/arXiv.2610.00537}{10.48550/arXiv.2610.00537}}
}

@book{boyd,
  author        = {Boyd, Stephen and Vandenberghe, Lieven},
  title         = {Convex Optimization},
  publisher     = {Cambridge University Press},
  year          = {2004},
  doi           = {10.1017/CBO9780511804441},
  url           = {https://web.stanford.edu/~boyd/cvxbook/},
  note          = {\href{https://doi.org/10.1017/CBO9780511804441}{10.1017/CBO9780511804441}}
}

@article{aldaz,
  author        = {Aldaz, J. M. and Kounchev, O. and Render, H.},
  title         = {{Bernstein} operators for exponential polynomials},
  journal       = {Constructive Approximation},
  volume        = {29},
  number        = {3},
  pages         = {345--367},
  year          = {2009},
  eprint        = {0805.1618},
  archivePrefix = {arXiv},
  doi           = {10.1007/s00365-008-9010-6},
  url           = {https://doi.org/10.1007/s00365-008-9010-6},
  note          = {\href{https://doi.org/10.1007/s00365-008-9010-6}{10.1007/s00365-008-9010-6}}
}

\appendix

\tableofcontents

\section{Measurement Identities and GLS}
\label{app:measurements}

\subsection{Circuit readout and joint probabilities}
\label{app:measurement-details}
At lag $\ell$, the interference circuit prepares
\[
 |\bm{\Psi}_\ell\rangle
 =\frac{|0\rangle|\bm{\psi}_0\rangle+
 |1\rangle U_\ell|\bm{\psi}_0\rangle}{\sqrt2},
 \qquad U_\ell=e^{-i\ell\tau\HH}.
\]
We measure the ancilla Pauli $Q\in\{X,Y\}$ and the system Pauli
string $P_k$, obtaining signs $A,B\in\{-1,1\}$.
The basis changes before computational readout are
$R_X=\mathsf H$, $R_Y=\mathsf H\mathsf S^\dagger$, and $R_Z=I$,
where $\mathsf S=\operatorname{diag}(1,i)$ and
$R_Q^\dagger ZR_Q=Q$. The system sign $B$ is the product of the
single-qubit signs for the nonidentity factors of $P_k$.

For any Hermitian system operator $P$, direct evaluation gives
\[
 \langle X\otimes P\rangle=\Re\langle\bm{\psi}_0|PU_\ell|\bm{\psi}_0\rangle,
 \qquad
 \langle Y\otimes P\rangle=\Im\langle\bm{\psi}_0|PU_\ell|\bm{\psi}_0\rangle.
\]
Taking $P=I$ gives the overlap components from $A$; taking
$P=P_k$ gives the Pauli contributions to $h_\ell$ from $AB$.
The system mean is
\[
 \E B=\tfrac12\left(
 \langle\bm{\psi}_0|P_k|\bm{\psi}_0\rangle+
 \langle\bm{\psi}_0|U_\ell^\dagger P_kU_\ell|\bm{\psi}_0\rangle\right),
\]
which does not depend on the ancilla basis.

Write $u=\E A$, $b=\E B$, and $t=\E(AB)$.
Since $(1+\alpha A)(1+\beta B)/4$ is the indicator of
$(A,B)=(\alpha,\beta)$,
\[
 \Prb(A=\alpha,B=\beta)
 =\tfrac14(1+\alpha u+\beta b+\alpha\beta t),
 \qquad \alpha,\beta\in\{-1,1\}.
\]
These four probabilities define the exact-outcome sampling model.
The measurement construction uses the ancilla-interference principle
of \cite{ibm}.

\subsection{Toeplitz assembly and a three-state example}
\label{app:lag-example}
\label{app:matrices}
For $|\bm{\phi}_j\rangle=U_j|\bm{\psi}_0\rangle$, exact evolution gives
$U_i^\dagger U_j=U_{j-i}$ and $[\HH,U_i]=0$. Consequently,
\[
 S_{ij}=s_{j-i},\qquad H_{ij}=h_{j-i},\qquad
 s_{-\ell}=\overline{s_\ell},\quad
 h_{-\ell}=\overline{h_\ell}.
\]
Each positive-lag estimate fills $m-\ell$ entries above the
diagonal. Repeated entries therefore have the same measurement error.

For example, take
\[
 \HH=(Z\otimes I+I\otimes Z)/2,\qquad
 |\bm{\psi}_0\rangle=(|00\rangle+|01\rangle+|11\rangle)/\sqrt3,
 \qquad \tau=\pi/2,\quad m=3.
\]
The occupied energies are $1,0,-1$, each with probability $1/3$, so
$s_\ell=(1+2\cos(\ell\tau))/3$ and
$h_\ell=-2i\sin(\ell\tau)/3$. Hence
\[
 S=\frac13\begin{pmatrix}3&1&-1\\1&3&1\\-1&1&3\end{pmatrix},
 \qquad
 H=\frac23\begin{pmatrix}0&-i&0\\i&0&-i\\0&i&0\end{pmatrix}.
\]
The pairs $(\bm{\phi}_0,\bm{\phi}_1)$ and $(\bm{\phi}_1,\bm{\phi}_2)$ both use
lag one; $(\bm{\phi}_0,\bm{\phi}_2)$ uses lag two. Measurements at these
two lags supply all off-diagonal entries. Direct Pauli measurements
supply $h_0$, while $s_0=1$ is fixed.

For any nonzero trial combination, its Rayleigh quotient is
$\bm{c}^\dagger H\bm{c}/(\bm{c}^\dagger S\bm{c})$. If $S\succ0$, the substitution
$\bm{c}=S^{-1/2}\bm{u}$ converts the generalized problem into the Hermitian
problem $S^{-1/2}HS^{-1/2}\bm{u}=E\bm{u}$ \cite{epperly,parlett}.
For an exactly singular $S$, restriction to its support removes
null trial combinations. Additional thresholding changes the retained
space and must be declared separately.

The spectral representation gives
\[
 S=V\operatorname{diag}(\bm{p})V^\dagger,
 \qquad V_{ja}=e^{ij\tau E_a},\quad 0\le j<m.
\]
Combining equal phases, let $r_\phi$ be the number of distinct
$e^{i\tau E_a}$ with positive weight. The Vandermonde rank formula
gives $\operatorname{rank}S=\min(m,r_\phi)$, so $S\succ0$ precisely
when $r_\phi\ge m$.

\subsection{Direct pooling}
\label{app:pooling}
For $\HH=\sum_{k=1}^J\alpha_kP_k$, let $n_{\ell k}^Q$ be the
count at setting $(\ell,k,Q)$ and $n_\ell^Q=\sum_kn_{\ell k}^Q$.
Overbars denote sample means within one setting. Direct pooling uses
\begin{equation}
 \widehat s_\ell^Q
 =\frac{\sum_kn_{\ell k}^Q\overline A_{\ell k}^Q}{n_\ell^Q},
 \qquad
 \widehat h_\ell^Q=\sum_k\alpha_k\overline{AB}_{\ell k}^Q.
 \label{eq:pooling}
\end{equation}
The $X$ and $Y$ components form the real and imaginary parts.
At zero lag, $\widehat h_0=\sum_k\alpha_k\overline B_{0k}$.
With independent settings and fixed counts, $A(AB)=B$ gives
\[
 \operatorname{Cov}(\widehat s_\ell^Q,\widehat h_\ell^Q)
 =\frac1{n_\ell^Q}\sum_k\alpha_k
 \bigl[\E B_{\ell k}^Q-\E A_{\ell k}^Q\E(A_{\ell k}^QB_{\ell k}^Q)\bigr].
\]
The count in the pooling weight cancels the inverse count in
the covariance of the two setting means.

\subsection{Single-shot covariances}

For the joint record $\bm{X}_q=(A,B,AB)^T$, write
$u_q=\E A$, $b_q=\E B$, and $t_q=\E(AB)$.
Because $A,B\in\{-1,1\}$, we have
$A^2=B^2=(AB)^2=1$, $A(AB)=B$, and $B(AB)=A$.
These identities give
\begin{equation}
 \Gamma_q=\Cov(\bm{X}_q)=
 \begin{pmatrix}
 1-u_q^2&t_q-u_qb_q&b_q-u_qt_q\\
 t_q-u_qb_q&1-b_q^2&u_q-b_qt_q\\
 b_q-u_qt_q&u_q-b_qt_q&1-t_q^2
 \end{pmatrix}.
 \label{eq:joint-covariance}
\end{equation}
For the scalar zero-lag record $X_q=B$, the variance is
$1-(\E B)^2$. In either case, averaging $n_q$ independent
records gives
\[
 \Cov(\overline{\bm{X}}_q)=\frac{\Gamma_q}{n_q}.
\]
Appendix~\ref{app:allocation-gls-details} derives the GLS
estimator and propagates these covariances to the estimated
moment vector.

\section{Proofs and Supporting Bounds for Shot Allocation}
\label{app:allocation}

\subsection{Covariance clipping and numerical allocation}
\label{app:learned-policy}

\paragraph{Choosing the clipping bounds.}
For $\bm{X}_q=(A,B,AB)^T$, the covariance is positive semidefinite,
so its largest eigenvalue is at most its trace.
Since $A^2=B^2=(AB)^2=1$,
\[
 \begin{aligned}
 \lambda_{\max}(\Gamma_q)
 &\le\tr(\Gamma_q)\\
 &=\Var(A)+\Var(B)+\Var(AB)\\
 &=3-(\E A)^2-(\E B)^2-\bigl(\E[AB]\bigr)^2
 \le3.
 \end{aligned}
\]
Thus $b=3$ is valid for every joint record.
Scalar sign records have variance at most one, so the
same bound covers them.

To choose the lower cutoff, prescribe a maximum covariance
condition number $K_{\max}>1$ and set $\kappa=b/K_{\max}$
before collecting the pilot.
Clip each empirical covariance eigenvalue to $[\kappa,b]$,
retaining its eigenvector. Then
\[
 \frac{\lambda_{\max}(\widehat\Gamma_q)}
      {\lambda_{\min}(\widehat\Gamma_q)}
 \le\frac b\kappa=K_{\max}.
\]
This controls the covariance weights used by the algorithm.
Theorem~\ref{thm:pilot} separately requires
$2\kappa\le\min_q\lambda_{\min}(\Gamma_q)$.
If a positive lower bound on the true eigenvalues is known,
choose $K_{\max}$ so that $\kappa=b/K_{\max}$ is no larger
than half that bound.
Otherwise, the theorem remains conditional on this inequality;
bounded sign outcomes alone do not give a positive lower bound.

\paragraph{Computing the allocation.}
For an accepted pilot, we minimize $\widehat V$ over
$\mathcal W_\epsilon$ to the tolerance specified in
Section~\ref{sec:learned-allocation}.
The pilot gradient and covariance weights remain fixed.
Start from $\bm{w}^{(0)}=\boldsymbol1/Q_{\rm tot}$.
At the current allocation $\bm{w}$, solve
\[
 \widehat{\mathcal J}(\bm{w})\bm{x}=F^T\widehat{\bm{g}}.
\]
Equation~\eqref{eq:gls-design-covariance-derivation} then gives
\begin{equation}
 \widehat V(\bm{w})=(F^T\widehat{\bm{g}})^T\bm{x},\qquad
 \partial_{w_q}\widehat V(\bm{w})
 =-(D_q\bm{x})^T\widehat\Gamma_q^{-1}(D_q\bm{x}).
 \label{eq:allocation-objective-derivative}
\end{equation}
The objective is convex, since
\begin{equation}
 \widehat V(\bm{w})
 =\max_{\bm{x}}\left\{
 2(F^T\widehat{\bm{g}})^T\bm{x}
 -\bm{x}^T\widehat{\mathcal J}(\bm{w})\bm{x}
 \right\}.
 \label{eq:allocation-convex-objective}
\end{equation}
The maximizing vector solves the linear system above.
For fixed $\bm{x}$, the expression is affine in $\bm{w}$, so its
maximum is convex.

At the current allocation $\bm{w}$, let $\bm{w}^{\rm lin}$ denote the allocation
that minimizes the linear approximation to $\widehat V$. For a
candidate allocation $\bm{v}$, the predicted change is
$\nabla\widehat V(\bm{w})^T(\bm{v}-\bm{w})$. This is minimized by giving each
setting its required fraction $\epsilon/Q_{\rm tot}$ and assigning
the remaining fraction $1-\epsilon$ to a setting with the smallest
derivative in \eqref{eq:allocation-objective-derivative}. Let $q_*$
be the smallest such index. Then
\[
 \bm{w}^{\rm lin}
 =\frac{\epsilon}{Q_{\rm tot}}\boldsymbol1
  +(1-\epsilon)\bm{e}_{q_*},
\]
where $\bm{e}_{q_*}$ is the $q_*$th coordinate vector.
This keeps the required fraction at every setting and
assigns the remaining mass to $q_*$.
It minimizes $\nabla\widehat V(\bm{w})^T\bm{v}$ over
$\bm{v}\in\mathcal W_\epsilon$, so convexity gives
\begin{equation}
 \begin{aligned}
 0
 &\le\widehat V(\bm{w})
      -\min_{\bm{v}\in\mathcal W_\epsilon}\widehat V(\bm{v})\\
 &\le\nabla\widehat V(\bm{w})^T(\bm{w}-\bm{w}^{\rm lin})
 =:\mathcal G_p(\bm{w}).
 \end{aligned}
 \label{eq:allocation-certificate}
\end{equation}
The computable quantity $\mathcal G_p(\bm{w})$ therefore bounds
the remaining optimization error.
Stop and return $\widehat{\bm{w}}=\bm{w}$ when
$\mathcal G_p(\bm{w})\le N_p^{-1/2}$.
Otherwise, update
\[
 \bm{w}^{(j+1)}
 =(1-\gamma_j)\bm{w}^{(j)}
  +\gamma_j\bm{w}^{{\rm lin},(j)},\qquad
 \gamma_j=\frac2{j+2},\qquad j=0,1,\ldots.
\]
The update stays in $\mathcal W_\epsilon$ because it is
a convex combination of admissible allocations.

The stopping rule terminates in exact arithmetic.
Identifiability, clipping, and the positive exploration
floor make $\widehat{\mathcal J}(\bm{w})$ uniformly positive
definite on the compact set $\mathcal W_\epsilon$.
Hence $\widehat V$ has a Lipschitz gradient there.
For a finite pilot-dependent constant $L_p$,
\[
 \widehat V(\bm{w}^{(j+1)})
 \le\widehat V(\bm{w}^{(j)})
 -\gamma_j\mathcal G_p(\bm{w}^{(j)})
 +L_p\gamma_j^2.
\]
If every gap exceeded $N_p^{-1/2}$, summing this inequality
would contradict $\widehat V\ge0$, because
$\sum_j\gamma_j=\infty$ and $\sum_j\gamma_j^2<\infty$.
Thus a finite iterate meets the tolerance.

If numerical optimization stops before meeting it, use
uniform fractions and record the solver failure.
Theorem~\ref{thm:pilot} requires this failure probability
to tend to zero. A fixed iteration limit alone does not
ensure that condition.

\paragraph{Converting fractions to counts.}
For the selected fractions $\widehat{\bm{w}}$, set
$n_q=\lfloor n\widehat w_q\rfloor$ and distribute the remaining
shots to the largest fractional remainders, with a fixed
index order for ties. Then
\[
 \sum_qn_q=n,\qquad
 |n_q-n\widehat w_q|<1,\qquad
 n_q\ge\lfloor n\epsilon/Q_{\rm tot}\rfloor.
\]
Require the last quantity to be at least one.
Recompute $R_q(\boldsymbol n)$ and
$\widehat\Sigma(\boldsymbol n)$ using these counts.

\subsection{Construction of the GLS weights}
\label{app:allocation-gls-details}
\label{app:gls}

To construct the weights in \eqref{eq:gls-production},
we fit the distinct outcome means shared across settings.
The auxiliary vector $\bm{\theta}$ contains these means;
$\bm{z}$ contains the moments used to assemble $H$ and $S$.

For zero-lag measurements, write
$\mu_k=\langle\bm{\psi}_0|P_k|\bm{\psi}_0\rangle$.
At positive lag, let $b_{\ell k}$ be the system mean,
which is independent of the ancilla basis.
Using $s_\ell,t_{\ell k}$ from \eqref{eq:moments}, collect
the true means as
\begin{equation}
 \begin{aligned}
 \bm{\theta}_*
 &=(\mu_1,\ldots,\mu_J,
    \bm{\theta}_{*,1}^T,\ldots,\bm{\theta}_{*,m-1}^T)^T,\\
 \bm{\theta}_{*,\ell}
 &=(b_{\ell1},\ldots,b_{\ell J},
    \Re s_\ell,\Im s_\ell,
    \Re t_{\ell1},\ldots,\Re t_{\ell J},
    \Im t_{\ell1},\ldots,\Im t_{\ell J})^T.
 \end{aligned}
 \label{eq:gls-lag-coordinate-block}
\end{equation}
This vector has $p=J+(m-1)(3J+2)$ coordinates.
A trial vector is denoted by $\bm{\theta}$ and its fitted value
by $\widehat{\bm{\theta}}$.

The matrix $D_q$ selects the means observed at setting $q$:
\begin{equation}
 \begin{aligned}
 D_{(\ell,k,X)}\bm{\theta}_*
 &=(\Re s_\ell,b_{\ell k},\Re t_{\ell k})^T,\\
 D_{(\ell,k,Y)}\bm{\theta}_*
 &=(\Im s_\ell,b_{\ell k},\Im t_{\ell k})^T.
 \end{aligned}
 \label{eq:gls-selection-example}
\end{equation}
For a zero-lag measurement of $P_k$, it selects $\mu_k$.
The matrix $F$ copies the overlap coordinates and forms the
Hamiltonian coordinates through
\[
 h_0=\sum_k\alpha_k\mu_k,\qquad
 \Re h_\ell=\sum_k\alpha_k\Re t_{\ell k},\qquad
 \Im h_\ell=\sum_k\alpha_k\Im t_{\ell k}.
\]
These selections and sums define $D_q$ and $F$ on every
trial vector. In particular,
\begin{equation}
 \E \bm{X}_q=D_q\bm{\theta}_*,\qquad \bm{z}_*=F\bm{\theta}_*.
 \label{eq:gls-shared-model}
\end{equation}
The system means $b_{\ell k}$ do not enter $\bm{z}_*$ directly.
Their shared values and correlations with the other outcomes
can nevertheless improve its estimation.

Known contributions may be subtracted before fitting and
restored afterward. The following formulas then apply to
the remaining uncertain coordinates.
We require these coordinates to be identifiable:
\begin{equation}
 D_q\bm{v}=\bm{0}\ \text{for all sampled }q
 \quad\Longrightarrow\quad \bm{v}=\bm{0},
 \qquad\text{equivalently}\qquad
 \sum_qD_q^TD_q\succ0.
 \label{eq:gls-identifiability}
\end{equation}
Any reduction of singular records must preserve the
shared-mean model and satisfy this condition.

For positive counts and fixed pilot covariance estimates,
the GLS fit is
\begin{equation}
 \widehat{\bm{\theta}}
 =\underset{\bm{\theta}}{\operatorname{argmin}}
 \sum_qn_q(\overline{\bm{X}}_q-D_q\bm{\theta})^T
          \widehat\Gamma_q^{-1}
          (\overline{\bm{X}}_q-D_q\bm{\theta}),
 \qquad \widehat{\bm{z}}=F\widehat{\bm{\theta}}.
 \label{eq:gls-fit}
\end{equation}
Differentiation gives the normal equations
\begin{equation}
 \begin{aligned}
 \widehat{\mathcal J}(\boldsymbol n)\widehat{\bm{\theta}}
 &=\sum_qn_qD_q^T\widehat\Gamma_q^{-1}\overline{\bm{X}}_q,\\
 \widehat{\mathcal J}(\boldsymbol n)
 &=\sum_qn_qD_q^T\widehat\Gamma_q^{-1}D_q.
 \end{aligned}
 \label{eq:gls-normal-system}
\end{equation}
The normal matrix is positive definite because, for $\bm{v}\ne\bm{0}$,
\[
 \bm{v}^T\widehat{\mathcal J}(\boldsymbol n)\bm{v}
 =\sum_qn_q\|\widehat\Gamma_q^{-1/2}D_q\bm{v}\|^2>0.
\]
Solving and applying $F$ gives the weights in the main text:
\begin{equation}
 R_q(\boldsymbol n)
 =F\widehat{\mathcal J}(\boldsymbol n)^{-1}
       n_qD_q^T\widehat\Gamma_q^{-1}.
 \label{eq:gls-normal}
\end{equation}

These weights reproduce the shared-mean model:
\begin{equation}
 \sum_qR_q(\boldsymbol n)D_q
 =F\widehat{\mathcal J}(\boldsymbol n)^{-1}
       \widehat{\mathcal J}(\boldsymbol n)
 =F.
 \label{eq:gls-unbiased-map}
\end{equation}
Since the production weights are fixed conditional on
the pilot,
\[
 \E[\widehat{\bm{z}}\mid\text{pilot}]
 =\sum_qR_qD_q\bm{\theta}_*=\bm{z}_*.
\]
Independence across settings gives \eqref{eq:sandwich}.
Substituting the GLS weights into
\eqref{eq:gls-fit-covariance} also gives
\begin{equation}
 \begin{aligned}
 \widehat\Sigma(\boldsymbol n)
 &=F\widehat{\mathcal J}(\boldsymbol n)^{-1}
   \left(\sum_qn_qD_q^T\widehat\Gamma_q^{-1}D_q\right)
   \widehat{\mathcal J}(\boldsymbol n)^{-1}F^T\\
 &=F\widehat{\mathcal J}(\boldsymbol n)^{-1}F^T.
 \end{aligned}
 \label{eq:gls-design-covariance-derivation}
\end{equation}
This simplification applies to $\widehat\Sigma$.
The actual covariance uses $\Gamma_q$ inside the sum
in \eqref{eq:sandwich}.

If $F$ has full row rank, then $\widehat\Sigma\succ0$.
For the unreduced model, this holds when at least one
$\alpha_k$ is nonzero: the overlap coordinates are copied
directly, and each Hamiltonian coordinate uses a nonzero
combination of a separate group of Pauli means.
After imposing known values or additional identities,
use independent uncertain moment coordinates and check
the corresponding rank.

For positive allocation fractions, the same formulas give
\[
 \widehat{\mathcal J}(n\bm{w})=n\widehat{\mathcal J}(\bm{w}),
 \qquad R_q(n\bm{w})=R_q(\bm{w}).
\]
Writing $\Sigma(\bm{w})$ and $\widehat\Sigma(\bm{w})$ for the covariance
formulas evaluated at $\bm{w}$, we obtain
\begin{equation}
 \Sigma(n\bm{w})=\frac{\Sigma(\bm{w})}n,\qquad
 \widehat\Sigma(n\bm{w})=\frac{\widehat\Sigma(\bm{w})}n.
 \label{eq:allocation-fraction-covariances}
\end{equation}
The pilot affects $\Sigma(\bm{w})$ through the implemented
weights $R_q(\bm{w})$.
If those weights instead use the true covariances, the
energy-variance coefficient becomes
\begin{equation}
 V(\bm{w})
 =\bm{g}^TF
 \left(\sum_qw_qD_q^T\Gamma_q^{-1}D_q\right)^{-1}
 F^T\bm{g}.
 \label{eq:allocation-oracle-coefficient}
\end{equation}
This is the oracle coefficient in Theorem~\ref{thm:pilot}.

\subsection{Proof of the learned-allocation guarantee}
\label{app:allocation-proof}

The reporting rule in Theorem~\ref{thm:pilot} must agree
with $E$ near $\bm{z}_*$.
For the cutoff-and-clipping rule, sufficient conditions are
$\lambda_{\min}(S)>s_{\rm cut}$ and $L<E_*<U$.
Continuity then makes both safeguards inactive near the target.

\begin{proof}[Proof of Theorem~\ref{thm:pilot}]
Let $\widehat{\bm{w}}$ be the selected fractions before rounding
and $\overline w_q=n_q/n$ the fractions afterward.
Write $\boldsymbol n_p$ for the pilot counts and
$\overline{\bm{X}}_{p,q}$ for the pilot sample mean at setting $q$.
All constants below may depend on the fixed model,
regularity margins, and $\epsilon,\kappa,b$, but not on
$N$ or the realized pilot.

\emph{1. Consistency of the pilot estimates.}
For $\bm{w}\in\mathcal W_{\epsilon/2}$,
\[
 \widehat{\mathcal J}(\bm{w})
 \succeq
 \frac{\epsilon}{2Q_{\rm tot}b}\sum_qD_q^TD_q,
 \qquad
 \|\widehat\Gamma_q^{-1}\|_{\rm op}\le\kappa^{-1}.
\]
Identifiability therefore bounds
$\widehat{\mathcal J}(\bm{w})^{-1}$ and $R_q(\bm{w})$ uniformly.
For sufficiently large budgets, both the balanced pilot
fractions and the rounded production fractions belong
to $\mathcal W_{\epsilon/2}$.

Every pilot count tends to infinity. The laws of large
numbers give convergence of the sample means and covariances.
Because the true covariance eigenvalues lie in $[\kappa,b]$,
continuity of clipping gives
\[
 \overline{\bm{X}}_{p,q}\longrightarrow D_q\bm{\theta}_*,
 \qquad
 \widehat\Gamma_q\longrightarrow\Gamma_q
 \quad\text{in probability}.
\]
The pilot means and weights use the same records, but
\eqref{eq:gls-unbiased-map} is an algebraic identity and yields
\[
 \widehat{\bm{z}}_p-\bm{z}_*
 =\sum_qR_q(\boldsymbol n_p)
       (\overline{\bm{X}}_{p,q}-D_q\bm{\theta}_*).
\]
The weights are uniformly bounded and the number of settings
is fixed, so $\widehat{\bm{z}}_p\to \bm{z}_*$ in probability.

Near $\bm{z}_*$, the overlap minimum and lowest Ritz gap are
continuous with positive limits, and the energy gradient
is continuous with finite limit $\bm{g}$.
Since $t_p\to0$, the pilot passes the checks with probability
tending to one and $\widehat{\bm{g}}\to \bm{g}$ on accepted pilots.
For $m=1$, the gap check is omitted.
Let $\mathcal A_p$ be the event that the pilot is accepted
and the allocation meets the optimization tolerance.
The assumption on numerical failures gives
$\Prb(\mathcal A_p)\to1$.

\emph{2. Convergence of the selected variance coefficient.}
On $\mathcal A_p$, define
\[
 \zeta_p
 =\sup_{\bm{w}\in\mathcal W_\epsilon}
       |\widehat V(\bm{w})-V(\bm{w})|,
\]
and set $\zeta_p=0$ outside this event.
The consistency from Step~1 and the uniform inverse bounds
imply $\zeta_p\to0$ in probability.
Uniform continuity of inversion follows from
$A^{-1}-B^{-1}=A^{-1}(B-A)B^{-1}$.

The stopping rule gives
\begin{equation}
 \widehat V(\widehat{\bm{w}})
 -\min_{\bm{w}\in\mathcal W_\epsilon}\widehat V(\bm{w})
 \le N_p^{-1/2}
 \qquad\text{on }\mathcal A_p.
 \label{eq:learned-design-tolerance}
\end{equation}
The continuous function $V$ attains its minimum at some
$\bm{w}_*\in\mathcal W_\epsilon$. On $\mathcal A_p$,
\[
 \begin{aligned}
 V_*\le V(\widehat{\bm{w}})
 &\le\widehat V(\widehat{\bm{w}})+\zeta_p\\
 &\le\widehat V(\bm{w}_*)+N_p^{-1/2}+\zeta_p\\
 &\le V_*+N_p^{-1/2}+2\zeta_p.
 \end{aligned}
\]
Outside $\mathcal A_p$, the procedure uses uniform fractions.
Since this event has vanishing probability,
$V(\widehat{\bm{w}})\to V_*$ in probability.

Rounding changes each fraction by less than $1/n$.
The derivative of $V$ is bounded on
$\mathcal W_{\epsilon/2}$, so
\[
 |V(\overline{\bm{w}})-V(\widehat{\bm{w}})|=O(n^{-1}),\qquad
 V(\overline{\bm{w}})\longrightarrow V_*
 \quad\text{in probability}.
\]

\emph{3. Convergence of the actual production variance.}
The implemented GLS weights use estimated covariances, while
their actual covariance is
\[
 \Cov(\widehat{\bm{z}}\mid\text{pilot})
 =\frac{\Sigma(\overline{\bm{w}})}n.
\]
With true-covariance weights, \eqref{eq:allocation-oracle-coefficient}
gives $\bm{g}^T\Sigma(\bm{w})\bm{g}=V(\bm{w})$.
Consistency of the fitted weights and the uniform inverse
bounds therefore imply
\[
 \sup_{\bm{w}\in\mathcal W_{\epsilon/2}}
 |\bm{g}^T\Sigma(\bm{w})\bm{g}-V(\bm{w})|
 \longrightarrow0
 \quad\text{in probability}.
\]
Together with Step~2, this gives
$\bm{g}^T\Sigma(\overline{\bm{w}})\bm{g}\to V_*$ in probability.
The bounded weights, bounded covariances, and positive
fraction floor bound this scalar uniformly over all pilots.
Consequently,
\[
 \E[\bm{g}^T\Sigma(\overline{\bm{w}})\bm{g}]\longrightarrow V_*.
\]

\emph{4. Control of the nonlinear energy error.}
Conditional on the pilot, the centered moment error is
\[
 \bm{\varepsilon}
 =\sum_q\sum_{r=1}^{n_q}\widehat{\bm{Y}}_{qr},\qquad
 \widehat{\bm{z}}_{qr}
 =\frac{R_q(\boldsymbol n)}{n_q}
       (\bm{X}_{q,r}-D_q\bm{\theta}_*).
\]
The summands are independent, centered, and uniformly
$O(n^{-1})$, since every count is bounded below by a fixed
multiple of $n$ for sufficiently large budgets.
Thus
$\E[\|\bm{\varepsilon}\|^2\mid\text{pilot}]=O(n^{-1})$.
Enumerating the summands by $j$, independence and centering
give, for each coordinate $i$,
\[
 \begin{aligned}
 \E[\varepsilon_i^4\mid\text{pilot}]
 &=\sum_j\E[Y_{j,i}^4\mid\text{pilot}]\\
 &\quad+6\sum_{j<k}
   \E[Y_{j,i}^2\mid\text{pilot}]
   \E[Y_{k,i}^2\mid\text{pilot}]
 =O(n^{-2}).
 \end{aligned}
\]
The dimension is fixed, so
$\E[\|\bm{\varepsilon}\|^4\mid\text{pilot}]=O(n^{-2})$.
Coordinatewise Hoeffding bounds also give, for each fixed
$r_0>0$,
\[
 \Prb\{\|\bm{\varepsilon}\|>r_0\mid\text{pilot}\}
 \le C_0e^{-c_0n},
\]
with positive constants uniform over pilots.

Since $S\succ0$ and $E_*$ is simple, the lowest generalized
eigenvalue is smooth near $\bm{z}_*$.
This follows from the implicit function theorem applied
to its simple root in $\det(H(\bm{z})-ES(\bm{z}))=0$.
Choose $r_0$ within a neighborhood where $E$ is smooth
and $E_{\rm rep}=E$, and define
\[
 r(\bm{\varepsilon})
 =E_{\rm rep}(\bm{z}_*+\bm{\varepsilon})-E_*-\bm{g}^T\bm{\varepsilon}.
\]
Taylor's theorem gives
$|r(\bm{\varepsilon})|\le C_E\|\bm{\varepsilon}\|^2$ inside this
neighborhood.
Outside it, $r$ remains bounded because the report and
the GLS outputs are uniformly bounded.
Hence
\[
 \E[r(\bm{\varepsilon})^2\mid\text{pilot}]
 \le C_E^2\E[\|\bm{\varepsilon}\|^4\mid\text{pilot}]
      +O(e^{-c_0n})
 =O(n^{-2}).
\]
By Cauchy--Schwarz, the cross term with
$\bm{g}^T\bm{\varepsilon}$ is $O(n^{-3/2})$.
Expanding the squared energy error yields
\[
 \E[(\widehat E-E_*)^2\mid\text{pilot}]
 =\frac{\bm{g}^T\Sigma(\overline{\bm{w}})\bm{g}}{n}
  +O(n^{-3/2}).
\]
Taking expectations and applying Step~3 gives
$n\,\MSE(\widehat E;E_*)\to V_*$.
Finally, $N_p/N\to0$ implies $N/n\to1$, proving
\[
 \MSE(\widehat E;E_*)
 =\frac{V_*}{N}+o(N^{-1}).
\]
\end{proof}

The proof treats the reported energy as the specified map
$E_{\rm rep}$. Additional numerical errors in evaluating
that map require separate control.

\section{Proofs and Supporting Results for Spectral Reconstruction}
\label{app:projection}

\subsection{Proof of Lemma~\ref{lem:projection-bound}}
\label{app:proj_ineq}

\begin{proof}
Fix $\bm{z}_*\in\C$. Since $\C$ is convex, the segment from
$\widehat{\bm{z}}_{\rm rec}$ to $\bm{z}_*$ lies in $\C$.
The fitting objective along this segment is
\[
 \phi(t)=\tfrac12
 \bigl\|\widehat{\bm{z}}-\widehat{\bm{z}}_{\rm rec}
       -t(\bm{z}_*-\widehat{\bm{z}}_{\rm rec})\bigr\|_M^2,
 \qquad 0\le t\le1.
\]
The projection minimizes the objective at $t=0$, so its
right derivative satisfies
\[
 \phi'(0^+)
 =(\widehat{\bm{z}}-\widehat{\bm{z}}_{\rm rec})^T
   M(\widehat{\bm{z}}_{\rm rec}-\bm{z}_*)\ge0.
\]
This derivative is the cross term, before multiplication by
two, in the expansion
\begin{equation}
 \begin{aligned}
 \|\widehat{\bm{z}}-\bm{z}_*\|_M^2
 &=
 \|\widehat{\bm{z}}-\widehat{\bm{z}}_{\rm rec}\|_M^2
 +\|\widehat{\bm{z}}_{\rm rec}-\bm{z}_*\|_M^2\\
 &\quad
 +2(\widehat{\bm{z}}-\widehat{\bm{z}}_{\rm rec})^T
    M(\widehat{\bm{z}}_{\rm rec}-\bm{z}_*).
 \end{aligned}
\end{equation}
Dropping the nonnegative cross term and rearranging gives
\eqref{eq:contraction}. The bound holds for every measured
vector $\widehat{\bm{z}}$, including one affected by circuit bias.
\end{proof}

\subsection{Proof of Theorem~\ref{thm:wedge}}
\label{app:wedge}

\begin{proof}
Our goal is to compare the leading energy errors before
and after reconstruction. We first identify their limiting
distributions, then compare their second moments.

\medskip
\noindent\textit{1. We identify the part of the moment error
that affects the energy.}

The assumptions on $S(\bm{z}_*)$ and $E_*$ make $E$ smooth
near $\bm{z}_*$. Therefore,
\[
 E(\bm{z}_*+\bm{h})-E_*=\bm{g}^T\bm{h}+o(\|\bm{h}\|),
 \qquad \bm{g}=\nabla E(\bm{z}_*).
\]
Thus, to first order, the energy error is the scalar
$\bm{g}^T\bm{h}$, rather than the full moment error $\bm{h}$.
Since the measurement errors have size $n^{-1/2}$,
we need the limits of
\[
 \sqrt n\,\bm{g}^T(\widehat{\bm{z}}_n-\bm{z}_*),
 \qquad
 \sqrt n\,\bm{g}^T(\widehat{\bm{z}}_{{\rm rec},n}-\bm{z}_*).
\]
Their second moments will determine the coefficients
of $1/n$ in the two energy MSEs.

\medskip
\noindent\textit{2. We choose coordinates that simplify
both the noise and the projection.}

Define
\[
 \bm{W}_n=\sqrt n\,\Omega^{-1/2}(\widehat{\bm{z}}_n-\bm{z}_*),
 \qquad \bm{b}=\Omega^{1/2}\bm{g}.
\]
The Gaussian assumption becomes
$\bm{W}_n\Rightarrow \bm{W}\sim N(\bm{0},I_d)$, and the scaled
first-order production energy error becomes
\[
 \sqrt n\,\bm{g}^T(\widehat{\bm{z}}_n-\bm{z}_*)=\bm{b}^T\bm{W}_n.
\]

The same coordinates turn the weighted fit into
Euclidean projection. Indeed, for any candidate
moment vector $\bm{z}$,
\[
 (\widehat{\bm{z}}_n-\bm{z})^T\Omega^{-1}(\widehat{\bm{z}}_n-\bm{z})
 =
 \frac1n
 \left\|\bm{W}_n-\sqrt n\,\Omega^{-1/2}(\bm{z}-\bm{z}_*)\right\|^2.
\]
Hence, with
\[
 D_n=\sqrt n\,\Omega^{-1/2}(\mathcal C-\bm{z}_*),
\]
the reconstructed error satisfies the exact identity
\[
 \sqrt n\,\Omega^{-1/2}(\widehat{\bm{z}}_{{\rm rec},n}-\bm{z}_*)
 =\Pi_{D_n}\bm{W}_n.
\]
We have therefore reduced the problem to understanding
Euclidean projections of approximately standard
Gaussian errors.

To analyze this projection as $n$ grows, we use the
tangent cone $T$, which describes the feasible directions
near $\bm{z}_*$. Set $K=\Omega^{-1/2}T$.
The convex sets $D_n$ increase with
$\overline{\bigcup_nD_n}=K$, so their Euclidean
projections converge to $\Pi_K$ uniformly on compact
sets. Together with $\bm{W}_n\Rightarrow \bm{W}$, this gives
\[
 \sqrt n\,\Omega^{-1/2}(\widehat{\bm{z}}_{{\rm rec},n}-\bm{z}_*)
 =\Pi_{D_n}\bm{W}_n
 \Rightarrow\Pi_K\bm{W}.
\]
Thus the limiting first-order energy errors are
$\bm{b}^T\bm{W}$ before reconstruction and $\bm{b}^T\Pi_K\bm{W}$ afterward.
We can now compare their second moments using the
geometry of $K$.

The set $D_n$ describes the feasible errors after
enlarging them by $\sqrt n$.
As $n$ increases, this enlargement reveals the
tangent cone
\[
 K=\Omega^{-1/2}T.
\]
More precisely, convexity and $\bm{z}_*\in\mathcal C$
imply that the sets $D_n$ are nested and that
\[
 \overline{\bigcup_{n\ge1}D_n}=K.
\]
For any fixed $\bm{w}$, the distances from $\bm{w}$ to $D_n$
decrease to its distance to $K$. The closest points
are bounded, and every limit point must be the
unique closest point in $K$. Thus
$\Pi_{D_n}\bm{w}\to\Pi_K\bm{w}$.
Because projections are 1-Lipschitz, this convergence
is uniform on compact sets.

The sequence $\bm{W}_n$ is tight, so the same limit applies
to these random inputs:
\[
 \sqrt n\,\Omega^{-1/2}(\widehat{\bm{z}}_{{\rm rec},n}-\bm{z}_*)
 =\Pi_{D_n}\bm{W}_n
 \Rightarrow\Pi_K\bm{W}.
\]
Together with the energy expansion in Step 1, this gives
\[
 \begin{aligned}
 \sqrt n(\widehat E-E_*)
 &\Rightarrow \bm{b}^T\bm{W},\\
 \sqrt n(\widehat E_{\rm rec}-E_*)
 &\Rightarrow \bm{b}^T\Pi_K\bm{W}.
 \end{aligned}
\]
We now compare the second moments of these two limits.
Step 5 uses Lemma~\ref{lem:projection-bound} to justify
passing from these distributional limits to the energy MSEs.

\medskip
\noindent\textit{3. We separate the directions that
reconstruction changes from those it leaves unchanged.}

In the transformed coordinates, the constraints are
\[
 (\Omega^{1/2}\bm{a}_i)^T\bm{\xi}\ge0,
 \qquad i=1,2.
\]

The transformed tangent cone is defined by
\[
 K=\{\bm{\xi}:(\Omega^{1/2}\bm{a}_1)^T\bm{\xi}\ge0,\;
          (\Omega^{1/2}\bm{a}_2)^T\bm{\xi}\ge0\}.
\]
These two inequalities depend only on the component
of $\bm{\xi}$ in the plane
\[
 \mathcal F
 =\operatorname{span}\{\Omega^{1/2}\bm{a}_1,\Omega^{1/2}\bm{a}_2\}.
\]
Indeed, writing $\bm{\xi}=\bm{\xi}_{\mathcal F}+\bm{\xi}_\perp$,
orthogonality gives
\[
 (\Omega^{1/2}\bm{a}_i)^T\bm{\xi}
 =(\Omega^{1/2}\bm{a}_i)^T\bm{\xi}_{\mathcal F},
 \qquad i=1,2.
\]
Thus the perpendicular component $\bm{\xi}_\perp$ does not
affect whether either inequality holds. Projection
onto $K$ leaves this component unchanged and adjusts
only the component in $\mathcal F$.
Consequently,
\[
 K=\mathcal F^\perp\oplus K_{\mathcal F},
 \qquad K_{\mathcal F}=K\cap\mathcal F,
\]
where $K_{\mathcal F}$ is a wedge.

Recall that
\[
 \bm{W}_n=\sqrt n\,\Omega^{-1/2}(\widehat{\bm{z}}_n-\bm{z}_*)
 \Rightarrow \bm{W}\sim N(\bm{0},I_d).
\]
Thus $\bm{W}$ represents the limiting measurement error
after scaling and changing coordinates.
The vector
\[
 \bm{b}=\Omega^{1/2}\bm{g},\qquad \bm{g}=\nabla E(\bm{z}_*),
\]
describes how errors in these coordinates affect
the energy, since
\[
 \sqrt n\,\bm{g}^T(\widehat{\bm{z}}_n-\bm{z}_*)=\bm{b}^T\bm{W}_n.
\]

We now separate both vectors into their components
within $\mathcal F$ and perpendicular to $\mathcal F$:
\[
 \bm{W}=\bm{W}_\perp+\bm{Z},
 \qquad
 \bm{b}=\bm{b}_\perp+\bm{b}_{\mathcal F},
\]
where $\bm{Z},\bm{b}_{\mathcal F}\in\mathcal F$ and
$\bm{W}_\perp,\bm{b}_\perp\in\mathcal F^\perp$.
This separates the contribution to the energy error
that projection can change from the contribution
it leaves unchanged.
Projection leaves $\bm{W}_\perp$ unchanged and replaces $\bm{Z}$ by
\[
 \bm{Q}=\Pi_{K_{\mathcal F}}\bm{Z}.
\]
The Gaussian vectors $\bm{W}_\perp$ and $\bm{Z}$ are independent,
and $\bm{W}_\perp$ has mean zero. Therefore,
\[
 \begin{aligned}
 \mathbb E(\bm{b}^T\bm{W})^2
 &=\|\bm{b}_\perp\|^2+\|\bm{b}_{\mathcal F}\|^2,\\
 \mathbb E(\bm{b}^T\Pi_K\bm{W})^2
 &=\|\bm{b}_\perp\|^2+\mathbb E(\bm{b}_{\mathcal F}^T\bm{Q})^2.
 \end{aligned}
\]
The unchanged contribution cancels. Thus the entire
comparison reduces to
\begin{equation}
 \Delta
 =\|\bm{b}_{\mathcal F}\|^2
  -\mathbb E(\bm{b}_{\mathcal F}^T\bm{Q})^2.
 \label{eq:wedge-gain-coefficient}
\end{equation}

\medskip
\noindent\textit{4. We conclude that projection reduces the
remaining second moment.}

If $\bm{b}_{\mathcal F}=\bm{0}$, then $\Delta=0$.
Otherwise, put $\bm{v}=\bm{b}_{\mathcal F}$.
We must show
\[
 \mathbb E(\bm{v}^T\bm{Q})^2<\|\bm{v}\|^2.
\]
Let $\bm{r}_1,\bm{r}_2$ be unit vectors along the two boundary
rays of the wedge. We bound the second moment by
separating the three possible projection outcomes.

If $\bm{Z}$ lies inside the wedge, then $\bm{Q}=\bm{Z}$.
To bound this contribution, enlarge the wedge to a
half-plane $H$ through the origin. Symmetry under
$\bm{Z}\mapsto-\bm{Z}$ gives
\[
 \begin{aligned}
 \mathbb E\!\left[(\bm{v}^T\bm{Z})^2
       \mathbf1_{\{\bm{Z}\in K_{\mathcal F}\}}\right]
 &\le
 \mathbb E\!\left[(\bm{v}^T\bm{Z})^2\mathbf1_{\{\bm{Z}\in H\}}\right]\\
 &=\frac12\mathbb E(\bm{v}^T\bm{Z})^2
 =\frac12\|\bm{v}\|^2.
 \end{aligned}
\]

Next consider points that project onto the positive
part of a boundary ray $\bm{r}_i$. Let $\bm{\nu}_i$ be the
outward unit normal to that ray and write
\[
 \bm{Z}=U\bm{r}_i+V\bm{\nu}_i.
\]
Here $U,V$ are independent standard Gaussians.
The corresponding projection region is $U>0,V>0$:
$U$ gives the position along the ray and $V$ gives
the outward distance from it. Projection removes
the outward component, leaving $\bm{Q}=U\bm{r}_i$.
This region therefore contributes
\[
 (\bm{v}^T\bm{r}_i)^2
 \mathbb E[U^2\mathbf1_{\{U>0\}}]\,
 \mathbb P(V>0)
 =\frac14(\bm{v}^T\bm{r}_i)^2.
\]

All remaining points project to the origin and
contribute zero. Adding the contributions gives
\[
 \mathbb E(\bm{v}^T\bm{Q})^2
 \le
 \frac12\|\bm{v}\|^2
 +\frac14\left[(\bm{v}^T\bm{r}_1)^2+(\bm{v}^T\bm{r}_2)^2\right]
 <\|\bm{v}\|^2.
\]
For the strict inequality, each squared inner product
is at most $\|\bm{v}\|^2$, and equality in both would require
$\bm{v}$ to be parallel to both boundary rays.
This is impossible because the rays are nonparallel.

We have proved that $\Delta>0$ exactly when
$\bm{b}_{\mathcal F}\ne\bm{0}$. Since $\mathcal F$ is spanned by
$\Omega^{1/2}\bm{a}_1,\Omega^{1/2}\bm{a}_2$ and $\bm{b}=\Omega^{1/2}\bm{g}$,
this is equivalent to
\[
 (\bm{a}_1^T\Omega \bm{g},\ \bm{a}_2^T\Omega \bm{g})\ne(0,0).
\]

\medskip
\noindent\textit{5. We transfer the comparison to the
energy MSEs.}

Convergence in distribution alone does not imply
convergence of second moments, so we must also
control large errors.
Lemma~\ref{lem:projection-bound} with $M=\Omega^{-1}$ gives
\[
 \|\widehat{\bm{z}}_{{\rm rec},n}-\bm{z}_*\|_{\Omega^{-1}}
 \le \|\widehat{\bm{z}}_n-\bm{z}_*\|_{\Omega^{-1}}.
\]
Since $\Omega$ is fixed and positive definite, the same bound
holds in the Euclidean norm up to a fixed factor.
Thus, whenever $\widehat{\bm{z}}_n$ is sufficiently close to
$\bm{z}_*$, both moment estimates lie in the neighborhood
where $E_{\rm rep}=E$ and $E$ is Lipschitz. In that case,
both absolute energy errors are bounded by a constant times
$\|\widehat{\bm{z}}_n-\bm{z}_*\|$. Outside this neighborhood
of measured estimates, $\|\widehat{\bm{z}}_n-\bm{z}_*\|$ is
bounded away from zero, so boundedness of $E_{\rm rep}$ gives
the same bound after increasing the constant.
Hence, for some constant $C$,
\[
 n\left[
   (\widehat E-E_*)^2
   +(\widehat E_{\rm rec}-E_*)^2
 \right]
 \le C\|\sqrt n(\widehat{\bm{z}}_n-\bm{z}_*)\|^2.
\]
The assumed $(2+\eta)$-moment bound makes these
squared scaled errors uniformly integrable.
The distributional limits from Step 2 therefore imply
\[
 \begin{aligned}
 n\,\operatorname{MSE}(\widehat E;E_*)
 &\longrightarrow \mathbb E(\bm{b}^T\bm{W})^2,\\
 n\,\operatorname{MSE}(\widehat E_{\rm rec};E_*)
 &\longrightarrow \mathbb E(\bm{b}^T\Pi_K\bm{W})^2.
 \end{aligned}
\]
Their difference converges to $\Delta$, whose sign
and strictness condition were established above.
This proves \eqref{eq:gain}.
\end{proof}

\subsection{Proof of Corollary~\ref{cor:computed-mse}}
\label{app:metric}
\label{app:finite}

We prove that the computed reconstruction retains the
MSE reduction when the covariance weight, grid, and
solver satisfy the conditions in
Corollary~\ref{cor:computed-mse}.

\begin{proof}
Set
\[
 \begin{aligned}
 M&=\Omega^{-1},
 &M_n&=(n\widehat\Sigma_{{\rm fit},n})^{-1},\\
 \widehat{\bm{z}}_{{\rm rec},n}
 &=\Pi_\C^M(\widehat{\bm{z}}_n),
 &\widehat{\bm{z}}_{{\rm rec},n}^{\,p}
 &=\Pi_\C^{M_n}(\widehat{\bm{z}}_n).
 \end{aligned}
\]
Both projections use the same data and continuous
feasible set. The first uses the true covariance
weight; the second uses the pilot estimate.
The computed fit $\widehat{\bm{z}}_{{\rm rec},n}^{\,{\rm num}}$ additionally uses
a finite grid and an approximate solver.

We show that these changes affect the reported
energy by $o(n^{-1})$ in mean square, which is too
small to change the leading MSE coefficient.

\medskip
\noindent\textit{1. Estimating the weight has a
negligible effect.}

The assumptions give
\[
 M_n\xrightarrow{p}M,\qquad
 C_0^{-1}I\preceq M_n\preceq c_0^{-1}I.
\]
Both fits belong to $\C$, but they minimize different
objectives. With weight $M$, the minimizer is
$\widehat{\bm{z}}_{{\rm rec},n}$, so
\[
 \|\widehat{\bm{z}}_n-\widehat{\bm{z}}_{{\rm rec},n}\|_M^2
 =\min_{\bm{z}\in\C}\|\widehat{\bm{z}}_n-\bm{z}\|_M^2
 \le\|\widehat{\bm{z}}_n-\widehat{\bm{z}}_{{\rm rec},n}^{\,p}\|_M^2.
\]
With weight $M_n$, the minimizer is
$\widehat{\bm{z}}_{{\rm rec},n}^{\,p}$, so
\[
 \|\widehat{\bm{z}}_n-\widehat{\bm{z}}_{{\rm rec},n}^{\,p}\|_{M_n}^2
 =\min_{\bm{z}\in\C}\|\widehat{\bm{z}}_n-\bm{z}\|_{M_n}^2
 \le\|\widehat{\bm{z}}_n-\widehat{\bm{z}}_{{\rm rec},n}\|_{M_n}^2.
\]
Write the two residuals as
\[
 \bm{r}_n=\widehat{\bm{z}}_n-\widehat{\bm{z}}_{{\rm rec},n},\qquad
 \bm{r}_n^{\,p}=\widehat{\bm{z}}_n-\widehat{\bm{z}}_{{\rm rec},n}^{\,p}.
\]
Since $\widehat{\bm{z}}_{{\rm rec},n}^{\,p}\in\C$, we can use it
as the comparison point in Lemma~\ref{lem:projection-bound}
for the projection with weight $M$. This gives
\[
 \|\widehat{\bm{z}}_{{\rm rec},n}^{\,p}-\widehat{\bm{z}}_{{\rm rec},n}\|_M^2
 \le \|\bm{r}_n^{\,p}\|_M^2-\|\bm{r}_n\|_M^2.
\]
To bound the right-hand side, add and subtract
the corresponding terms with weight $M_n$:
\[
 \begin{aligned}
 \|\bm{r}_n^{\,p}\|_M^2-\|\bm{r}_n\|_M^2
 &=
 \underbrace{
 \|\bm{r}_n^{\,p}\|_{M_n}^2-\|\bm{r}_n\|_{M_n}^2
 }_{\le0}\\
 &\quad
 +(\bm{r}_n^{\,p})^T(M-M_n)\bm{r}_n^{\,p}
 -\bm{r}_n^T(M-M_n)\bm{r}_n.
 \end{aligned}
\]
The underbraced term is nonpositive because
$\widehat{\bm{z}}_{{\rm rec},n}^{\,p}$ minimizes the $M_n$-weighted
objective. For the remaining terms, use
\[
 |\bm{u}^T(M-M_n)\bm{u}|
 \le\|M-M_n\|_{\rm op}\|\bm{u}\|^2.
\]
Dropping the nonpositive term and bounding each
remaining term by its absolute value therefore gives
\[
 \|\widehat{\bm{z}}_{{\rm rec},n}^{\,p}-\widehat{\bm{z}}_{{\rm rec},n}\|_M^2
 \le
 \|M_n-M\|_{\rm op}
 \left(\|\bm{r}_n^{\,p}\|^2+\|\bm{r}_n\|^2\right).
\]
Each residual is bounded by a constant times
$\|\widehat{\bm{z}}_n-\bm{z}_*\|$, since $\bm{z}_*$ is feasible and the
weights have uniform eigenvalue bounds. Therefore,
\[
 \|\widehat{\bm{z}}_{{\rm rec},n}^{\,p}-\widehat{\bm{z}}_{{\rm rec},n}\|^2
 \le C\|M_n-M\|_{\rm op}\|\widehat{\bm{z}}_n-\bm{z}_*\|^2.
\]
$\|M_n-M\|_{\rm op}$ is uniformly bounded
by the eigenvalue assumptions and converges to zero
in probability, while the moment assumption makes
$n\|\widehat{\bm{z}}_n-\bm{z}_*\|^2$ uniformly integrable. Hence
\begin{equation}
 n\E\|\widehat{\bm{z}}_{{\rm rec},n}^{\,p}-\widehat{\bm{z}}_{{\rm rec},n}\|^2
 \longrightarrow0.
 \label{eq:metric-mean-square}
\end{equation}

\medskip
\noindent\textit{2. The grid and solver errors are
also negligible.}

For fixed $m,\tau,[L,U]$, the vector of real moments
$\bm{v}(x)$ has bounded second derivative. Linear
interpolation therefore approximates each vector
column with error $O(h_n^2)$. Averaging these
approximations gives the same bound for every
feasible moment vector. With the uniform bounds
on $M_n$, this yields
\[
 \sup_{\bm{z}\in\C}
 \operatorname{dist}_{M_n}(\bm{z},\C_{G_n})
 \le\eta_n,\qquad \eta_n=Ch_n^2.
\]
Interpolation also preserves a fixed mean energy,
if that constraint is imposed.

Choose $\bm{q}_n\in\C_{G_n}$ within $\eta_n$ of
$\widehat{\bm{z}}_{{\rm rec},n}^{\,p}$ in the $M_n$ norm, and write
\[
 f_n(\bm{z})=\tfrac12\|\widehat{\bm{z}}_n-\bm{z}\|_{M_n}^2.
\]
The solver-gap bound gives
$f_n(\widehat{\bm{z}}_{{\rm rec},n}^{\,{\rm num}})\le f_n(\bm{q}_n)+\epsilon_n$.
Applying Lemma~\ref{lem:projection-bound} with weight $M_n$
and feasible comparison point
$\widehat{\bm{z}}_{{\rm rec},n}^{\,{\rm num}}$ gives the first
inequality below. Expanding $f_n(\bm{q}_n)$ about
$\widehat{\bm{z}}_{{\rm rec},n}^{\,p}$ gives the second:
\[
 \begin{aligned}
 \tfrac12
 \|\widehat{\bm{z}}_{{\rm rec},n}^{\,{\rm num}}-\widehat{\bm{z}}_{{\rm rec},n}^{\,p}\|_{M_n}^2
 &\le f_n(\widehat{\bm{z}}_{{\rm rec},n}^{\,{\rm num}})-f_n(\widehat{\bm{z}}_{{\rm rec},n}^{\,p})\\
 &\le
 \eta_n\|\widehat{\bm{z}}_n-\widehat{\bm{z}}_{{\rm rec},n}^{\,p}\|_{M_n}
 +\tfrac12\eta_n^2+\epsilon_n.
 \end{aligned}
\]
Since $\bm{z}_*\in\C$, the expected residual is at most
$\E\|\widehat{\bm{z}}_n-\bm{z}_*\|_{M_n}=O(n^{-1/2})$. Thus
\[
 n\E\|\widehat{\bm{z}}_{{\rm rec},n}^{\,{\rm num}}-\widehat{\bm{z}}_{{\rm rec},n}^{\,p}\|^2
 \le C\left(
 \sqrt n\,\eta_n+n\eta_n^2+n\epsilon_n
 \right)\longrightarrow0.
\]
Here we used
$h_n=o(n^{-1/4})$, so $\eta_n=o(n^{-1/2})$,
and $\epsilon_n=o(n^{-1})$.

Combining this estimate with Step 1 gives
\begin{equation}
 n\E\|\widehat{\bm{z}}_{{\rm rec},n}^{\,{\rm num}}-\widehat{\bm{z}}_{{\rm rec},n}\|^2
 \longrightarrow0.
 \label{eq:computed-moment-equivalence}
\end{equation}

\medskip
\noindent\textit{3. Transfer the error bound to the energy MSE.}

Write
\[
 \widehat E_{\rm rec}=E_{\rm rep}(\widehat{\bm{z}}_{{\rm rec},n}),
 \qquad
 \widehat E_{\rm rec}^{\,{\rm num}}
 =E_{\rm rep}(\widehat{\bm{z}}_{{\rm rec},n}^{\,{\rm num}}).
\]
We already know from
\eqref{eq:computed-moment-equivalence} that
$\E\|\widehat{\bm{z}}_{{\rm rec},n}^{\,{\rm num}}-\widehat{\bm{z}}_{{\rm rec},n}\|^2=o(n^{-1})$.
We now transfer this bound to the energies.

Choose a fixed $r>0$ such that $E_{\rm rep}=E$ within
$2r$ of $\bm{z}_*$ and
\[
 |E(\bm{z})-E(\bm{z}')|\le L_E\|\bm{z}-\bm{z}'\|
\]
whenever both points are within that neighborhood.
Define
\[
 A_n=
 \{\|\widehat{\bm{z}}_{{\rm rec},n}-\bm{z}_*\|\le r,\;
   \|\widehat{\bm{z}}_{{\rm rec},n}^{\,{\rm num}}-\widehat{\bm{z}}_{{\rm rec},n}\|\le r\}.
\]
On $A_n$, both fits lie within $2r$ of $\bm{z}_*$.
Their energy difference is therefore at most
$L_E\|\widehat{\bm{z}}_{{\rm rec},n}^{\,{\rm num}}-\widehat{\bm{z}}_{{\rm rec},n}\|$.

To control the remaining outcomes, we bound the
probability that either condition fails.
Lemma~\ref{lem:projection-bound} bounds the exact fit's
error by the measurement error in the $M$-norm.
Since $M$ is fixed and positive definite, converting
to the Euclidean norm introduces only a fixed factor.
The moment assumption thus gives
\[
 \E\|\widehat{\bm{z}}_{{\rm rec},n}-\bm{z}_*\|^{2+\eta}
 \le C_1\E\|\widehat{\bm{z}}_n-\bm{z}_*\|^{2+\eta}
 \le C_2n^{-1-\eta/2}.
\]
Markov's inequality and the union bound imply
\[
 \begin{aligned}
 \Pr(A_n^c)
 &\le
 \frac{\E\|\widehat{\bm{z}}_{{\rm rec},n}-\bm{z}_*\|^{2+\eta}}{r^{2+\eta}}
 +\frac{\E\|\widehat{\bm{z}}_{{\rm rec},n}^{\,{\rm num}}-\widehat{\bm{z}}_{{\rm rec},n}\|^2}{r^2}\\
 &=O(n^{-1-\eta/2})+o(n^{-1})
 =o(n^{-1}),
 \end{aligned}
\]
where the last equality uses $\eta>0$.

Let $B_E=\max\{|L|,|U|\}$ bound the absolute reported
energy. The difference between the two reported energies
on $A_n^c$ is therefore at most $2B_E$.
Splitting the expectation over the two events gives
\[
 \E|\widehat E_{\rm rec}^{\,{\rm num}}-\widehat E_{\rm rec}|^2
 \le
 L_E^2\E\|\widehat{\bm{z}}_{{\rm rec},n}^{\,{\rm num}}-\widehat{\bm{z}}_{{\rm rec},n}\|^2
 +4B_E^2\Pr(A_n^c)
 =o(n^{-1}).
\]

Finally,
\[
 \widehat E_{\rm rec}^{\,{\rm num}}-E_*
 =
 (\widehat E_{\rm rec}-E_*)
 +(\widehat E_{\rm rec}^{\,{\rm num}}-\widehat E_{\rm rec}).
\]
Squaring and taking expectations introduces an
additional squared term and a cross term.
The squared term is $o(n^{-1})$ by the preceding
bound. Cauchy--Schwarz bounds the cross term by
\[
 2\sqrt{
 \MSE(\widehat E_{\rm rec};E_*)\,
 \E|\widehat E_{\rm rec}^{\,{\rm num}}-\widehat E_{\rm rec}|^2}
 =o(n^{-1}),
\]
because the exact reconstructed energy has MSE
$O(n^{-1})$.

The two reconstructed MSEs therefore differ by
$o(n^{-1})$. Combining this with
Theorem~\ref{thm:wedge} gives
\[
 \MSE(\widehat E;E_*)
 -
 \MSE(\widehat E_{\rm rec}^{\,{\rm num}};E_*)
 =\frac{\Delta}{n}+o(n^{-1}).
\]
When $\Delta>0$, the computed reconstruction has
smaller energy MSE for all sufficiently large $n$.
\end{proof}

\section{Numerical specifications}
\label{app:numerical-details}

\subsection{Hamiltonians and reference states}
\label{app:numerical-states}

The unnormalized Hamiltonians are
\[
\begin{aligned}
 \HH_{\rm spin}^{(0)}
 &=\sum_{j=0}^{K-2}(1+0.08j)
 (0.28X_jX_{j+1}+0.27Y_jY_{j+1}+0.23Z_jZ_{j+1})\\
 &\quad +\sum_{j=0}^{K-1}0.11(-1)^j(1+0.06j)Z_j,\\
 \HH_{\rm Hub}^{(0)}
 &=-0.34\sum_{j\in\{0,1,3,4\}}
 (X_jX_{j+1}+Y_jY_{j+1})
 +0.48\sum_{j=0}^{2}Z_jZ_{j+3}.
\end{aligned}
\]
Here $K=4,6$ is the chain length. In the Hubbard model, qubits
$0,1,2$ encode the up-spin orbitals and $3,4,5$ the down-spin orbitals.
For $\HH^{(0)}=\sum_\alpha h_\alpha P_\alpha$, we use
$\HH=\HH^{(0)}/W_0$ with $W_0=\sum_\alpha|h_\alpha|$, so that
$\|\HH\|_{\rm op}\le1$. In the seven-case comparison the time step
for $\HH$ is $\tau=W_0\tau_0$, with $\tau_0=2.55$, $2.90$, and
$1.18$ for four spins, six spins, and Hubbard, respectively.
The time steps in the two theorem tests refer directly to $\HH$.

For the eigenstate superpositions we use
$|\bm{\psi}_0\rangle=\sum_j\sqrt{p_j}\,e^{0.31\,i j}|\bm{u}_j\rangle$,
where $j$ indexes the saved eigenbasis in order of increasing energy,
starting at zero. In the seven-case comparison, the sparse four-spin
state has probabilities $(0.31,0.28,0.23,0.18)$ at indices
$(0,2,6,13)$. The sparse six-spin and Hubbard states have probabilities
$(0.30,0.26,0.25,0.19)$ at $(0,9,24,53)$. All remaining probabilities
are zero. The dense states use $p_j\propto1+0.4\cos(1.137j)$.

For both sparse Hubbard runs ($m=3$ and $m=4$), we used the same
six-qubit reference state. In the energy eigenbasis used for the
simulation, the eigenvectors with indices $(0,9,24,53)$ have the
four probabilities listed above, respectively. These probabilities
sum to one, and each of the other 60 eigenvectors has probability
zero.

To specify the reference state used in the Pauli measurements, we
give its amplitudes in the computational basis:
\[
  |\bm{\psi}_0\rangle=\sum_{k=0}^{63} c_k |k\rangle.
\]
Here $k$ labels a six-bit string, with qubit~0 on the left.
For example, $k=10$ denotes $|001010\rangle$. A measurement in
the computational basis returns bitstring $k$ with probability
$|c_k|^2$. Each occupied energy eigenvector can contribute to
several bitstrings.

The table gives the real and imaginary parts of the 39 amplitudes
that are nonzero to eight decimal places. The first listed index
is $10$ and the last is $55$. The amplitudes at all 25 unlisted
indices, including those skipped between $10$ and $55$, round to zero.

\begin{center}
\small
\setlength{\tabcolsep}{3pt}
\begin{tabular}{rrr@{\quad}rrr@{\quad}rrr}
\hline
$k$ & $\Re c_k$ & $\Im c_k$ &
$k$ & $\Re c_k$ & $\Im c_k$ &
$k$ & $\Re c_k$ & $\Im c_k$ \\
\hline
10 &  0.00348479 & -0.00248844 & 25 & -0.08398246 & -0.13190967 & 41 &  0.24938882 &  0.09327422 \\
11 & -0.06569903 & -0.05785760 & 26 &  0.24938882 &  0.09327422 & 42 &  0.37327228 &  0          \\
12 &  0.00557578 & -0.00204539 & 28 &  0.18663614 &  0          & 43 & -0.00866579 &  0.16025032 \\
13 &  0.04644940 &  0.04091150 & 29 &  0.00866579 & -0.16025032 & 44 &  0.03756511 & -0.09327422 \\
14 & -0.00000670 &  0          & 30 &  0.09630711 & -0.03532876 & 46 &  0.12753304 &  0.11028790 \\
15 &  0.19514568 & -0.09722194 & 31 &  0.00096295 &  0.00084805 & 47 & -0.00136182 & -0.00119932 \\
17 & -0.00348479 &  0.00248844 & 33 & -0.00557578 &  0.00204539 & 48 & -0.03670665 & -0.28863845 \\
19 &  0.04644940 &  0.04091150 & 34 & -0.00440056 &  0.00040418 & 49 &  0.18663614 &  0          \\
20 &  0.00440056 & -0.00040418 & 35 & -0.00000670 &  0          & 50 &  0.03756511 & -0.09327422 \\
21 & -0.00001341 &  0          & 37 & -0.04645971 & -0.04091150 & 51 & -0.09630711 &  0.03532876 \\
22 & -0.04645971 & -0.04091150 & 38 &  0.06569430 &  0.05785760 & 52 &  0.21558149 &  0.13190967 \\
23 &  0.28969585 & -0.10627041 & 39 &  0.21454612 & -0.05306711 & 53 & -0.12753304 & -0.11028790 \\
24 &  0.19191699 &  0.23170196 & 40 &  0.10975029 & -0.04026018 & 55 &  0.00096295 &  0.00084805 \\
\hline
\end{tabular}
\end{center}
\subsection{Pilot and production settings}
\label{sec:numerical-policies}

\subsubsection{Product-state allocation test}
\label{app:numerical-product}

The four-spin test in Section~\ref{sec:new-design} uses $m=3$,
$\tau=6$, and the product state
\[
 |\bm{\psi}_0\rangle=\bigotimes_{j=0}^{3}
 \left[\cos(\theta_j/2)|0\rangle+
 e^{i\phi_j}\sin(\theta_j/2)|1\rangle\right],
 \qquad
 \bm{\theta}/\pi=(0.2,0.8,0.3,0.7),\quad \phi_j=0.19+0.23j.
\]
The smallest eigenvalue of the ideal overlap matrix is $0.158$,
the gap between the two lowest ideal Ritz values is $0.328$, and the
smallest eigenvalue among the single-shot covariance matrices is
$0.216$, all rounded.

For each total budget $N=80{,}000\cdot 2^k$, $k=0,\ldots,8$, we
use a pilot of $N_p$ shots to estimate the record covariances and
the energy gradient. We set
\[
 N_p=\min\!\left\{\lfloor N/2\rfloor,\;
 65\left\lceil \frac{N^{2/3}(\log N)^{1/3}}{65}\right\rceil\right\}.
\]
This choice gives each of the 65 measurement settings a growing
pilot sample while the pilot fraction $N_p/N$ tends to zero, as
required by \eqref{eq:learned-pilot-growth}. Rounding up to a
multiple of 65 gives equal pilot counts at every tested budget,
and the cap reserves at least half the shots for production.

As explained in Appendix~\ref{app:learned-policy}, we clip the
eigenvalues of the pilot covariance estimates to control their
condition numbers. For this experiment, we use the interval
$[\kappa,b]=[10^{-6},3]$. If the pilot fails the checks in
\eqref{eq:learned-pilot-acceptance}, we allocate production shots
uniformly. Otherwise, we choose the fraction $w_q$ assigned to each
setting $q$ by minimizing $\widehat V(\bm{w})$, the pilot estimate of the
leading variance coefficient of the GLS energy. We require
$\sum_q w_q=1$ and $w_q\ge 0.16/65$. Starting from uniform
fractions, we use sequential quadratic programming with a limit of
140 iterations. All 10,794 accepted pilots satisfied
$\mathcal G_p(\bm{w})\le N_p^{-1/2}$ at this stage. By
\eqref{eq:allocation-certificate}, no feasible allocation could
lower $\widehat V(\bm{w})$ by more than this tolerance.

To approximate $V_*$, we minimize $V(\bm{w})$ using the true covariances
and energy gradient. Starting from the allocation returned by
sequential quadratic programming, we repeatedly move toward
$\bm{w}^{\rm lin}$ from Appendix~\ref{app:learned-policy}, choosing
the step size by minimizing $V$ along that segment. After the
200-step limit, the remaining optimality gap was
$3.91\times10^{-4}$, placing the computed coefficient within
$4\times10^{-4}$ of $V_*$.

The learned and uniform GLS estimates use the same pilot. We convert
each allocation into integer counts for the $N-N_p$ production shots
by largest-remainder rounding.
\subsubsection{Seven-case comparison}

The two GLS estimators use the same balanced $20\%$ pilot. At
setting $q$, let $G_q$ be the empirical covariance of the $n_{p,q}$
pilot records. A small pilot can
make a linear combination of the recorded outcomes appear nearly
noiseless. Inverting $G_q$ could then give that combination
excessive weight. To stabilize the GLS weights, we shrink $G_q$
toward the identity and add a small diagonal term, giving
\[
 \begin{aligned}
 W_q&=(1-\rho_q)G_q+\rho_q I+\epsilon_q I,\\
 \rho_q&=\min\{1,2/\sqrt{n_{p,q}}\},\qquad
 \epsilon_q=\min\{10^{-5},1/n_{p,q}\}.
 \end{aligned}
\]
The shrinkage brings the estimated variances toward one and the
cross-covariances toward zero. Its weight decreases as the pilot
grows. We use $W_q$ to choose the learned shot counts and to
compute the GLS weights under both allocations.

Reconstruction needs the uncertainty of the resulting GLS moments.
For this purpose we use the ordinary sample covariance, with
divisor $n_{p,q}-1$, plus a small diagonal term. This gives
\[
 \widehat\Gamma_q
 =\frac{n_{p,q}}{n_{p,q}-1}G_q+\frac{I}{n_{p,q}}.
\]
This retains the sample cross-covariances. Equation
\eqref{eq:gls-fit-covariance} combines $\widehat\Gamma_q$ with
the GLS maps $R_q(\boldsymbol n)$ to estimate the covariance of
the GLS moments. Its inverse weights the reconstruction.

\subsection{Numerical validation of the two-normal reconstruction theorem}
\label{app:numerical-controlled}

Section~\ref{sec:theory-validation} tests whether the MSE reduction
observed in the four-spin experiment agrees with the prediction of
Theorem~\ref{thm:wedge}. The target $E_*$ is the lowest energy in
the fixed Krylov subspace. In each trial, we use the same measurement
records to compute two estimates: $\widehat E$ from the GLS moments
and $\widehat E_{\rm rec}$ from the reconstructed moments.

\paragraph{Experimental setup.}
We use the four-spin XYZ Hamiltonian $H$ with two Krylov basis
states ($m=2$), the reference state $\lvert\psi_0\rangle$ and its
time evolution $e^{-i\tau H}\lvert\psi_0\rangle$, with $\tau=2.5$.
The reference state assigns probabilities $0.6$ and $0.4$ to
eigenvectors $0$ and $13$, whose energies are
$e_1\simeq-0.649597$ and $e_2\simeq0.305110$, respectively.

To isolate the effect of reconstruction, we omit the pilot and
allocate shots uniformly across the 39 settings. The target budgets
are
\[
 N/10^3\in\{10,20,46,100,215,464,1000,2150,4640,10000\}.
\]
Each setting receives $n_q=\lfloor N/39\rfloor$ shots, giving
$n=39\lfloor N/39\rfloor$ shots per trial. We run 5,000 independent
trials at each budget.

Let $\Gamma_q$ be the exact covariance of one measurement record
at setting $q$. We use $\Gamma_q+10^{-8}I$ when computing the GLS
weights $R_q(\bm n)$ that combine the sample means into
$\widehat{\bm z}$. The diagonal shift is a numerical safeguard
for matrix inversion. Since the sample mean at setting $q$ has
covariance $\Gamma_q/n_q$, equation~\eqref{eq:sandwich} gives
\[
 \Cov(\widehat{\bm z})
 =\sum_q R_q(\bm n)\frac{\Gamma_q}{n_q}R_q(\bm n)^T
 =\frac{\Omega}{n}.
\]
Reconstruction fits the measured moments to a probability
distribution on $[L,U]=[-0.7,0.4]$, weighting the moment errors
by $(\Omega/n)^{-1}$. A grid of 401 energies initializes the fit.
The solver then varies the energy locations continuously over
$[L,U]$.

\paragraph{Prediction for the four-spin example.}
To predict how much the fit changes the energy MSE, we need to
know which moment errors reconstruction corrects. For $m=2$, a
distribution concentrated at energy $x$ gives the moment vector
from \eqref{eq:single-energy-vector},
\[
 \bm v(x)=\bigl(x,\cos(\tau x),x\cos(\tau x),
                 -\sin(\tau x),-x\sin(\tau x)\bigr)^T.
\]
The feasible moment set $\mathcal C$ consists of all
probability-weighted averages of these vectors over $[L,U]$.
The ideal moment vector in this experiment is
\[
 \bm z_*=0.6\bm v(e_1)+0.4\bm v(e_2).
\]
As the shot budget grows, the measured moments concentrate near
$\bm z_*$. To first order, reconstruction projects their errors
onto the tangent cone $T_{\mathcal C}(\bm z_*)$, the set of feasible
directions of change at $\bm z_*$. Theorem~\ref{thm:wedge} gives
the leading energy MSE when this cone is described by two
inequalities. The following lemma identifies those inequalities
for our choice of state and fitting interval.

\begin{lemma}[Tangent cone for the four-spin example]
\label{lem:four-spin-normals}
Define the matrix
\[
 M=
 \begin{pmatrix}
  \bigl(\bm v(e_2)-\bm v(e_1)\bigr)^T\\
  \bm v'(e_1)^T\\
  \bm v'(e_2)^T\\
  \bigl(\bm v(L)-\bm z_*\bigr)^T\\
  \bigl(\bm v(U)-\bm z_*\bigr)^T
 \end{pmatrix}.
\]
The two systems
\[
 M\bm a_1=(0,0,0,1,0)^T,\qquad
 M\bm a_2=(0,0,0,0,1)^T
\]
have unique solutions. These vectors are linearly independent
inward normals at $\bm z_*$, and the tangent cone is
\[
 T_{\mathcal C}(\bm z_*)
 =
 \bigl\{\delta\bm z\in\mathbb R^5 :
   \bm a_1^T\delta\bm z\ge0,\quad
   \bm a_2^T\delta\bm z\ge0\bigr\}.
\]
\end{lemma}

Lemma~\ref{lem:four-spin-normals} supplies the two normals needed
to evaluate the prediction. Let $\bm G\sim N(\bm 0,\Omega)$
denote the limiting moment error after scaling by $\sqrt n$, and
let $\bm g$ be the energy gradient at $\bm z_*$. Before
reconstruction, the first-order energy error is $\bm g^T\bm G$.
After reconstruction, it is $\bm g^T\bm G_T$, where $\bm G_T$ is
the projection of $\bm G$ onto the cone in the lemma using the
weight $\Omega^{-1}$. Thus the two leading MSE coefficients in
Theorem~\ref{thm:wedge} are
\[
 V=\mathbb E[(\bm g^T\bm G)^2],
 \qquad
 V-\Delta=\mathbb E[(\bm g^T\bm G_T)^2].
\]
For this example, the predicted relative reduction is
$\Delta/V\simeq45.9\%$. At the largest shot budget, the observed
reduction was $46.9\%$. The approximate $95\%$ Monte Carlo
interval at that budget, adjusted for simultaneous coverage across
all ten budgets, was $[44.1\%,49.6\%]$ and contains the prediction.

\begin{proof}[Proof of Lemma~\ref{lem:four-spin-normals}]
We first derive the rows of $M$. A vector $\bm a$ is an inward
normal at $\bm z_*$ if $\bm a^T(\bm z-\bm z_*)\ge0$ for every
$\bm z\in\mathcal C$. Since each $\bm v(x)$ is feasible and every
feasible $\bm z$ is an average of these vectors, it suffices to
check that
\[
 q_{\bm a}(x):=\bm a^T\bigl(\bm v(x)-\bm z_*\bigr)\ge0
 \qquad\text{for all }x\in[L,U].
\]
The definition of $\bm z_*$ gives
$0.6q_{\bm a}(e_1)+0.4q_{\bm a}(e_2)=0$. For an inward normal,
both values are nonnegative, so both vanish. Subtracting them
gives the first row of $M\bm a=\bigl(0,0,0,
q_{\bm a}(L),q_{\bm a}(U)\bigr)^T$. The next two rows follow
because $e_1$ and $e_2$ are interior minima of $q_{\bm a}$, so
its derivative vanishes at both energies. The last two rows give
the endpoint values. We choose these values to be $(1,0)$ for
$\bm a_1$ and $(0,1)$ for $\bm a_2$.

To show that the systems have unique solutions, suppose
$M\bm a=\bm 0$. The first row gives $q_{\bm a}(e_1)=q_{\bm a}(e_2)$.
Their weighted average is zero, so both values are zero. The other
rows give $q_{\bm a}'(e_1)=q_{\bm a}'(e_2)=0$ and
$q_{\bm a}(L)=q_{\bm a}(U)=0$. Thus the four distinct points
$L,e_1,e_2,U$ account for at least $1+2+2+1=6$ zeros counted
with multiplicity.

The function $q_{\bm a}$ lies in the six-dimensional space spanned
by the constant function $1$ and the five components of $\bm v(x)$.
Since $\tau(U-L)=2.75<\pi$, every nonzero function in this space
has at most five zeros counted with multiplicity
\cite[Theorem~9]{aldaz}. Hence $q_{\bm a}(x)=0$ for every
$x\in[L,U]$. In the expansion
$q_{\bm a}(x)=-\bm a^T\bm z_*+\bm a^T\bm v(x)$, the entries
of $\bm a$ are the coefficients of the five components of
$\bm v(x)$. Linear independence forces each coefficient to vanish.
Thus $\bm a=\bm 0$, and $M$ is invertible. The two solutions are
linearly independent because their right-hand sides are.

It remains to show that these two vectors give all the required
inequalities. Each function $q_{\bm a_1}$ and $q_{\bm a_2}$ equals
one at one endpoint and zero at the other. Its endpoint zero and
its zeros at $e_1,e_2$ already account for five zeros counted with
multiplicity. The interior zeros are therefore exactly double,
and there are no other zeros. The function keeps its sign across
each double zero. Its value one at the other endpoint makes it
nonnegative throughout $[L,U]$. Thus both vectors are inward
normals.

Finally, for any inward normal $\bm a$, the formula for $M\bm a$
and the two defining systems give
\[
 M\bm a=q_{\bm a}(L)\,M\bm a_1+q_{\bm a}(U)\,M\bm a_2.
\]
Since $M$ is invertible, $\bm a$ is a combination of $\bm a_1$
and $\bm a_2$ with coefficients $q_{\bm a}(L),q_{\bm a}(U)\ge0$.
For a convex set, a direction belongs to the tangent cone exactly
when it has nonnegative inner product with every inward normal.
It therefore suffices to check the two inequalities in the lemma.
\end{proof}

\end{document}